\documentstyle[12pt]{article}
\input psfig.tex
\font\twelvebf=cmbx10 scaled\magstep 2
\font\twelverm=cmr10 scaled\magstep 2
\font\twelveit=cmti10 scaled\magstep 2
 1
\font\elevenrm=cmr10 scaled\magstep 1
 1

\renewcommand{\thefootnote}{\fnsymbol{footnote}}
\begin{document}
\begin{flushright}
{\elevenrm MSUHEP-50228}
\end{flushright}
\begin{flushright}
{\elevenrm February 1995}
\end{flushright}
\baselineskip=22pt
\centerline{\twelvebf TOP QUARK PHYSICS}
\baselineskip=22pt
\vspace{0.8cm}
\centerline{\twelverm C.--P. Yuan}
\vspace{0.3cm}
\centerline{\twelveit Department of Physics and Astronomy}
\baselineskip=14pt
\centerline{\twelveit Michigan State University}
\baselineskip=14pt
\centerline{\twelveit East Lansing, MI 48824, USA}
\vspace{0.9cm}
\begin{abstract}
\noindent
In these three lectures, I review the physics of top quark.
Each of the lectures is self-contained.
\end{abstract}

\vspace{0.8cm}
\rm\baselineskip=14pt

\renewcommand{\thefootnote}{\arabic{footnote}}
\setcounter{footnote}{0}


\centerline{\twelvebf LECTURE ONE:}
\vspace{0.2cm}
\centerline{\twelvebf Global Analysis of the Top Quark}
\baselineskip=14pt
\centerline{\twelvebf Couplings to Gauge Bosons}
\vspace{0.8cm}

%
\def\kln{\kappa_{L}^{\rm{NC}}}
\def\krn{\kappa_{R}^{\rm{NC}}}
\def\klc{\kappa_{L}^{\rm{CC}}}
\def\krc{\kappa_{R}^{\rm{CC}}}
\def\ttz{{\mbox {$t$-$t$-$Z$}\,}}
\def\bbz{{\mbox {$b$-$b$-$Z$}\,}}
\def\tta{{\mbox {$t$-$t$-$A$}\,}}
\def\bba{{\mbox {$b$-$b$-$A$}\,}}
\def\tbw{{\mbox {$t$-$b$-$W$}\,}}
\def\tltlz{{\mbox {$t_L$-$t_L$-$Z$}\,}}
\def\blblz{{\mbox {$b_L$-$b_L$-$Z$}\,}}
\def\brbrz{{\mbox {$b_R$-$b_R$-$Z$}\,}}
\def\tlblw{{\mbox {$t_L$-$b_L$-$W$}\,}}
\def\ppbar{\bar{{\rm p}}{\rm p}}
\def\beq{\begin{equation}}
\def\enq{\end{equation}}
\def\ra{\rightarrow}

\section{\twelvebf Introduction and Motivations}

    Despite the success of the Standard Model
(SM) \cite{mele,alta,lep94,sld94,alt},
there is little faith that the SM is the final theory.
For instance, the SM contains many
arbitrary parameters with no apparent connections \cite{pc}.
In addition,  the SM provides no
satisfactory explanation for the symmetry-breaking mechanism which takes
place and gives rise to the observed mass spectrum of the gauge bosons
and fermions.
In this lecture, we study how to use the top quark to probe the origin of
the spontaneous symmetry-breaking and the generation of fermion masses.

There are strong experimental and theoretical arguments
suggesting the top quark must exist \cite{kane}; {\it e.g.}, from the
measurement of the weak isospin quantum number of the left-handed $b$ quark
we know the top quark has to exist.
{}From the direct search at the Tevatron, assuming SM top quark,
$m_t$ has to be larger than 131\,GeV \cite{D0}.
Recently, data were presented by the CDF group at
FNAL to support the existence of a heavy top quark with mass
$m_t = 174 \pm 10 \, {\rm (stat)} \pm 12 {\rm (syst)}$\,GeV
\cite{CDF}.
Furthermore, studies on radiative corrections concluded that
the mass ($m_t$) of a standard top quark
has to be less than $200$\,GeV \cite{mele}. However, there
are no compelling reasons to believe that the top quark couplings to light
particles should be of the SM nature.
Because the top quark is heavy
relative to other observed fundamental particles, one expects that any
underlying theory at high energy scale $\Lambda \gg m_t$ will
easily reveal itself at low energy through the effective interactions of
the top quark to other light particles.
Also because the top quark mass is of the order of the Fermi scale
$v={(\sqrt{2}G_F)}^{-1/2}=246$\,GeV, which characterizes the
electroweak symmetry-breaking scale, the top quark may be a useful tool
to probe the symmetry-breaking sector. Since the
fermion mass generation can be closely related to the
electroweak symmetry-breaking,
one expects some residual
effects of this breaking to appear in accordance with the generated mass
\cite{pczh,sekh}. This means new effects should be more apparent in the
top quark sector than any other light sector of the theory.
Therefore, it is important to study the top quark system as a direct tool
to probe new physics effects \cite{kane}.

Undoubtedly, any real analysis including the top quark cannot be completed
without actually discovering it. In the SM, which is a renormalizable theory,
the couplings of the top quark to gauge bosons are fixed by the
linear realization of the gauge symmetry $SU(2)_{L}\times U(1)_Y$,
although the top quark mass remains a free parameter.
However, through the non-linear realization of the gauge symmetry,
the couplings of the top quark to gauge bosons
can be different from the SM predictions.
The goal of this lecture is to study the couplings of the top quark to
gauge bosons from the precision data at LEP and examine how to improve
our knowledge about the top quark at the current and future colliders.
Also we will discuss how to use this knowledge to probe the
symmetry-breaking mechanism.

   Generally one studies a specific model
({\it e.g.}, a grand unified theory)
valid up to some high energy scale and evolves that theory down
to the electroweak scale to compare its predictions
with the low energy data \cite{sekh,lopez}.
In addition to such a model by model
study, one can incorporate new physics effects in a model-independent
way formulated in terms of either
a set of variables \cite{pesk,bar,bar1,bar2}
or an effective Lagrangian \cite{geor,gold,buch}. In this
lecture, we will adopt the latter approach. We simply address the
problem in the following way.
Assume there is an underlying theory at some high energy scale.
How does this theory {\it appreciably} manifest itself at low energy?
Because we do not know the shape of the underlying theory
and because a general treatment is usually very complicated,
we cannot provide a satisfying answer. Still, one can get some crude
answers to this question based on a few
{\it negotiable} arguments suggested by the status of low energy
data with the application of the electroweak chiral Lagrangian.

It is generally believed that new physics is likely to come in via processes
involving longitudinal gauge bosons (equivalent to Goldstone bosons)
and/or heavy fermions such as the top quark.
One commonly discussed method to probe the electroweak symmetry sector is
to study the interactions among the longitudinal
gauge bosons in the TeV
region. Tremendous work has been done in the literature \cite{wwww}.
However, this is not the subject of this lecture.
As we argued above, the top quark plays an important role
in the search for new physics. Because of its heavy mass, new physics
will feel its presence easily and eventually may
show up in its couplings to the gauge bosons. If the top quark is
a participant in a dynamical symmetry-breaking mechanism, {\it e.g.},
through the $\overline{t}t$ condensate (Top Mode Standard Model) \cite{tana}
which is suggested by the fact that its mass is of the order of the
Fermi scale $v$, then the top quark
is one of the best candidates for search of new physics.

An attempt to study the nonuniversal interactions of the top quark has
been carried out in Ref.~\cite{pczh} by Peccei {\it et al}. However,
in that study only the vertex \ttz was considered based on the
assumption that this is the only vertex which gains a significant
modification due to a speculated dependence of
the coupling strength on the fermion mass:
$\kappa_{ij} \leq {\cal O} \left ( \frac{\sqrt{m_{i}m_{j}}}{v} \right )$,
where $\kappa_{ij}$ parameterizes some new
dimensional-four interactions among gauge bosons and fermions $i$
and $j$. However, this is not the only possible pattern of interactions,
{\it e.g.}, in some extended technicolor models
\cite{sekh} one finds that the nonuniversal residual interactions
associated with the vertices \blblz, \tltlz, and \tlblw
to be of the same order.
In Sec. 4 we discuss the case of the SM with a heavy Higgs boson
($m_H > m_t$) in which we find the size of the nonuniversal effective
interactions \tltlz and \tlblw
to be of the same order but with a negligible \blblz effect.

Here is the outline of our approach.
First, we use the chiral Lagrangian approach \cite{wein,geor2,cole,fer}
to construct the most general $SU(2)_L \times U(1)_Y$ invariant
effective Lagrangian including up to dimension-four operators
for the top and bottom quarks. Then
we deduce the SM (with and without a scalar Higgs boson) from this
Lagrangian, and only consider new
physics effects which modify the top quark couplings to gauge
bosons and possibly the vertex \blblz.
With this in hand, we perform a comprehensive analysis using precision
data from LEP.
We include the contributions from the vertex \tbw in addition to the
vertex \ttz, and discuss the special
case of having a comparable size in \bbz as in \ttz.
Second, we build an  effective model
with an approximate custodial symmetry ($\rho \approx 1$)
connecting the \ttz and \tbw couplings.
This reduces the number of parameters in the effective
Lagrangian and strengthens its structure and predictability.
After examining what we have learned from the LEP data, we study how
to improve our knowledge on these couplings at
the electron colliders, such as the SLC and the
NLC (Next Linear Collider) \cite{nlc}.\footnote{
We use NLC to represent a generic $e^-e^+$ supercollider.}
We will discuss the physics of top quark at hadron colliders
, such as the Tevatron and the LHC (Large Hadron Collider),
in great details in the third lecture.

The rest of this lecture is organized as follows. In Sec. 2
we provide a brief introduction to the chiral Lagrangian
with an emphasis on the top quark sector. In Sec. 3
we present the complete analysis of the top quark interactions with gauge
bosons using LEP data for various scenarios of symmetry-breaking
mechanism.
In Sec. 4 we discuss the heavy Higgs limit ($m_{H} > m_t$)
in the SM model as an example of our proposed effective model at the top
quark mass scale.
In Sec. 5 we discuss how the electron colliders (such as the
SLC and the NLC)
can contribute to the measurement of these couplings.
Some discussion and conclusions for this lecture are given in Sec. 6.

\section{\twelvebf Introduction to the Chiral Lagrangian}

 The chiral Lagrangian approach has been used
in understanding the low energy strong interactions because it can
systematically describe the phenomenon of spontaneous symmetry-breaking
\cite{wein}. Recently, the chiral Lagrangian technique has been widely
used in studying the electroweak sector
\cite{pczh,gold,fer,pecc,hol,how1,fer1,app},
to which this work has been directed.

A chiral Lagrangian can be constructed solely
on the symmetry of the theory without assuming
any explicit underlying dynamics. Thus,
it is the most general effective Lagrangian
that can accommodate any truly fundamental theory possessing that
symmetry at low energy. Since one is interested in the low energy
behavior of such a theory, an expansion in powers of the external momentum is
performed in the chiral Lagrangian \cite{geor2}.

  In general one starts from a Lie group G which breaks down spontaneously
into a subgroup H, hence a Goldstone boson for every broken
 generator is to be introduced \cite{cole}.
 Consider, for example, G~$=SU(2)_L\times U(1)_Y$ and H~$=U(1)_{em}$.
There are three Goldstone bosons generated by this breakdown,
$\phi^{a},\,a=1,2,3$ which are eventually eaten by $W^{\pm}$ and $Z$
and become the longitudinal degree of freedom of these gauge bosons.

 The Goldstone bosons transform non-linearly under G but linearly
under the subgroup H. A convenient way to handle this is to introduce
the matrix field
\beq
\Sigma ={\rm{exp}}\left ( i\frac{\phi^{a}\tau^{a}}{v_{a}}\right )\, ,
\enq
where $\tau^{a},\, a=1,2,3, $ are the Pauli matrices normalized as
${\rm{Tr}}(\tau^a \tau^b)=2 \delta_{ab}$. Because of $U(1)_{em}$ invariance
$v_1=v_2=v$, but $v$ is not necessarily equal to $v_3$.
The matrix field $\Sigma$ transforms under G as
\beq
\Sigma\rightarrow {\Sigma}^{\prime}={\rm {exp}}\left ( i
\frac{\alpha^{a}\tau^{a}}{2}\right )\,
\Sigma \,{\rm {exp}}(-iy\frac{\tau^3}{2})\, ,
\enq
where $\alpha^{1,2,3}$ and $y$ are
the group parameters of G.

In the SM, being a special case of the chiral Lagrangian,
$v=246$\,GeV is the vacuum expectation value of the Higgs
boson field. Also
$v_3=v$ arises from the approximate custodial symmetry
in the SM.
It is this symmetry that is responsible for the
tree-level relation
\beq
\rho= \frac{{M^2}_W}{{M^2}_Z\,{\cos^2} \theta_W}=1\,
\label{yeq3}
\enq
in the SM, where $\theta_W$ is the electroweak mixing angle.
In this lecture, we assume the full theory guarantees that
$v_1=v_2=v_3=v$.

Out of the Goldstone bosons and the gauge boson fields one can construct
the bosonic gauge invariant terms in the chiral Lagrangian
\beq
{\cal L}_{B} =-\frac{1}{4} {W_{\mu \nu}^a} {W^{\mu \nu}}^{a}
 -\frac{1}{4} B_{\mu \nu}B^{\mu \nu}
 +\frac{1}{4}v^{2}\rm{Tr}({D_{\mu}\Sigma}^{\dagger}D^{\mu}
\Sigma)\, ,
\label{yeq4}
\enq
where the covariant derivative
\beq
D_{\mu}\Sigma =\partial_{\mu}\Sigma - ig{W_{\mu}^a}\frac{\tau^{a}}{2}\Sigma
+i{g}^{\prime}\Sigma B_{\mu}\frac{\tau^{3}}{2}\, .
\enq
In the unitary gauge $\Sigma =1$, one can easily see how the gauge bosons
acquire a mass.
In Eq.~(\ref{yeq3}), $M_W=gv/2$ is the mass of $W^\pm_\mu=
(W^1_\mu\mp i W^2_\mu)/\sqrt{2})$, $M_Z=gv/2/\cos \theta_W$
is the mass of $Z_\mu=\cos \theta_W W^3_\mu - \sin \theta_W B_\mu$.
The photon field will be denoted as
$A_\mu = \sin \theta_W W^3_\mu + \cos \theta_W B_\mu$.

Fermions can be included in this context by assuming that they
transform under G$ =SU(2)_L\times U(1)_{Y}$ as \cite{pecc}
\beq
f\rightarrow {f}^{\prime}=e^{iyQ_f}f \label{eq1} \, ,
\enq
where $Q_{f}$ is the electromagnetic charge of $f$.

 Out of the fermion fields $f_1$, $f_2$ and the Goldstone bosons matrix
field $\Sigma$
the usual linearly realized fields
$\Psi$ can be constructed. For example, the left-handed
fermions [$SU(2)_L$ doublet] are constructed as
\beq
\Psi_{L} = \Sigma F_{L} = \Sigma{f_1\choose {f_2}}_{L} \label{psi} \,
\enq
with $Q_{f_1}-Q_{{f_2}}=1$.
One can easily show that $\Psi_{L}$\,transforms under G linearly as
\beq
\Psi_{L}\rightarrow {{\Psi}^{\prime}}_{L}=g\Psi_{L}\, ,
\enq
where $ g=
exp(i\frac{\alpha^{a}\tau^{a}}{2})exp(i\frac{y}{2})
 \in {\rm{G}} $.
Linearly realized right-handed fermions
$\Psi_{R}$  [$SU(2)_L$ singlet] simply coincide with $F_{R}$: {\it i.e.},
\beq
\Psi_{R}= F_{R}={f_1\choose {f_2}}_{R}\, .\label{psr}
\enq
Out of those fields with the specified transformations it is straightforward
to construct
a Lagrangian which is invariant under $SU(2)_L\times U(1)_Y$.

Since the interactions among the light fermions and the gauge bosons have
been well tested to agree with the SM, we only consider
new interactions involving
the top and bottom quarks. We ignore all possible mixing of the
top quark with light fermions in these new interactions.
In case there exists a fourth generation with heavy
fermions, there can be a substantial
impact on the Cabibbo-Kobayashi-Maskawa (CKM)
matrix element $V_{tb}$.
To be discussed later, this effect is effectively included
in the new nonstandard couplings of \tbw.

Following Ref.~\cite{pecc}, we define
\beq
{\Sigma_{\mu}^a}=-\frac{i}{2}{\rm{Tr}}(\tau^{a}\Sigma^{\dagger}D_{\mu}\Sigma)
\, ,
\enq
which transforms under G as:
\beq
 {\Sigma_{\mu}^3}\rightarrow {{{\Sigma}^{\prime}}_{\mu}^3}
       ={\Sigma_{\mu}^3}\, ,
\enq
\beq
{\Sigma_{\mu}^\pm}\rightarrow {{{\Sigma}^{\prime}}_{\mu}^\pm}
   =e^{\pm iy}{\Sigma_{\mu}^\pm}\, ,
\enq
where
\beq
{\Sigma_{\mu}^\pm}=\frac{1}{\sqrt{2}}({\Sigma_{\mu}^1}\mp i
{\Sigma_{\mu}^2})\, .
\enq
In the unitary gauge, $\Sigma =1$, we have
\beq
{\Sigma_{\mu}^3}= -\frac{1}{2}\frac{gZ_{\mu}}{\cos{\theta_W}}\, ,
\enq
\beq
{\Sigma_{\mu}^\pm}= -\frac{1}{2}g{W_{\mu}^\pm}\, .
\enq

Consider the interaction terms up to dimension-four for the
$t$ and $b$ quarks.
{}From Eqs.~(\ref{psi}) and (\ref{psr}) we denote
\beq
F= {t\choose b} = F_L + F_R\, ,
\enq
with $f_1=t$ and $f_2=b$.
The SM Lagrangian can be deduced from
\begin{eqnarray}
{\cal L}_{0}&=&\overline{F}i\gamma^{\mu} \left ( \partial_{\mu} -ig^{\prime}
(\frac{Y}{2}+\frac{\tau^3}{2})
B_{\mu} \right ) F - \overline{F}\,M\,F \nonumber \\
&-& \overline{F_{L}}\gamma^{\mu}\tau^{a} F_{L}{\Sigma_{\mu}^a}
+{\cal L}_{B} \, \, ,
\label{eq17}
\end{eqnarray}
where $Y=1/3$ and $M$ is a diagonal mass matrix
\beq
 M=\pmatrix{m_t & 0 \cr 0 & m_b \cr}\, .
\enq
${\cal L}_{0}$ is invariant under G, and the electric charge of fermions is
given by $Y/2+T^3$, where $T^3$ is the weak isospin quantum number.
Taking advantage of the chiral Lagrangian approach,
additional nonstandard interaction terms, invariant under G, are allowed
\cite{pecc}
\begin{eqnarray}
{\cal L}&=& -\kln \overline{t_{L}}\gamma^{\mu} t_{L}{\Sigma_{\mu}^3}
  -\krn \overline{{t}_{R}}\gamma^{\mu} t_{R}{\Sigma_{\mu}^3} \nonumber \\
&-&\sqrt{2}\klc \overline{{t}_{L}}\gamma^{\mu} b_{L}{\Sigma_{\mu}^+}
-\sqrt{2}{\klc}^{\dagger}\overline{{b}_{L}}\gamma^{\mu}t_{L}
{\Sigma_{\mu}^-} \nonumber \\
&-&\sqrt{2}\krc \overline{{t}_{R}}\gamma^{\mu} b_{R}{\Sigma_{\mu}^+}
-\sqrt{2}{\krc}^{\dagger} \overline{{b}_{R}}\gamma^{\mu} t_{R}
{\Sigma_{\mu}^-} \label{eq2} \, ,
\end{eqnarray}
where $\kln$, $\krn$ are two arbitrary real parameters,
$\klc$, $\krc$ are two arbitrary complex parameters, and the superscripts
$NC$ and $CC$ denote neutral and charged currents, respectively.
In the unitary gauge we derive the following nonstandard terms
in the chiral Lagrangian with the symmetry
$ { {SU(2)_L \times U(1)_Y} \over {U(1)_{em}} }$
\begin{eqnarray}
{\cal L}&=&\frac{g}{4\cos \theta_W}\bar{t}\left ( \kln \gamma^{\mu}
(1-\gamma_{5})+\krn \gamma^{\mu}(1+\gamma_{5}) \right ) t\,Z_{\mu} \nonumber \\
 &+&\frac{g}{2\sqrt{2}}\bar{t}\left ( \klc\gamma^{\mu}(1-\gamma_{5})
+ \krc \gamma^{\mu}(1+\gamma_{5})\right ) b\,{W_{\mu}^+} \nonumber \\
&+&\frac{g}{2\sqrt{2}}\bar{b}\left ( {\klc}^{\dagger}\gamma^{\mu}
(1-\gamma_{5})+{\krc}^{\dagger}\gamma^{\mu}(1+\gamma_{5}) \right )
t\,{W_{\mu}^-} \label{eq3} \, .
\end{eqnarray}

A few remarks are in order regarding the Lagrangian
${\cal L}$ in Eqs. (\ref{eq2}) and (\ref{eq3}).
\begin{itemize}
\item[(1)]
In principle, ${\cal L}$ can include
nonstandard neutral currents $\overline{b_{L}}\gamma_{\mu}b_{L}$ and
$\overline{b_{R}}\gamma_{\mu}b_{R}$.
For the left-handed neutral current $\overline{b_{L}}\gamma_{\mu}b_{L}$
we discuss two cases: \\
(a) The effective left-handed vertices \tltlz,
\tlblw, and \blblz are comparable
in size as in some extended technicolor models \cite{sekh}.
In this case, the top quark contribution to low energy observables
is of higher order through radiative corrections;
therefore, its contribution will be suppressed by $1/16{\pi}^{2}$.
In this case, as we will discuss in the next section,
the constraints derived from low energy data on the
nonstandard couplings are so stringent (of the order of a few percent)
 that it would be a challenge to directly probe the nonstandard
top quark couplings at the Tevatron, the LHC, and the NLC.\\
(b) The effective left-handed vertex \blblz is small
as compared to the
\ttz and \tbw vertices.
We will devote most of this lecture to the case where the vertex \blblz
is not modified by the dynamics of
the symmetry-breaking.
This assumption leads to interesting conclusions to be seen
in the next section. In this case one needs to consider
the contributions of the top quark to low energy data through loop
effects. A specific model with such properties is given in Sec. 4.
\item[(2)]
We shall assume that \brbrz is not modified by the
dynamics of the electroweak symmetry-breaking. This is the case in
the Extended
Technicolor models discussed in Ref. \cite{sekh}.
The model discussed in Sec. 4 is another example.
\item[3)]
The right-handed charged current contribution
$\krc$ in Eqs.~(\ref{eq2}) and (\ref{eq3}) is expected to be
suppressed by the bottom quark mass.
This can be understood in the following way. If $b$ is
massless ($m_b=0$), then
the left- and right-handed $b$ fields can be
associated with different global $U(1)$ quantum numbers. ($U(1)$ is
a chiral group, not the
hypercharge group.) Since the underlying theory has an exact
$SU(2)_L\times U(1)_Y$ symmetry at high energy,
the charged currents are purely left-handed before the
symmetry is broken.
After the symmetry is spontaneously
broken and for a massless $b$ the $U(1)$ symmetry associated
with $b_{R}$ remains exact (chiral invariant) so it is not possible to
generate right-handed charged currents.
Thus $\krc$ is usually suppressed by the bottom quark mass although
it could be enhanced in some
models with a larger group G, {\it i.e.}, in models
containing additional right-handed gauge bosons.

We find that in the limit of ignoring the bottom quark mass,
$\krc$ does not contribute to low energy data through loop insertion
at the order ${m_t^2}\ln {\Lambda}^2$, therefore we cannot constrain
$\krc$ from the LEP data.
 However, at the Tevatron and the LHC $\krc$ can be measured
by studying the direct detection of the top quark
and its decays.
This will be discussed in the third lecture.
\end{itemize}

It is worth mentioning that the photon does not participate in
the new nonuniversal interactions as described in the chiral Lagrangian
${\cal L}$ in Eq.~(\ref{eq3}) because the
$U(1)_{em}$ symmetry remains an exact symmetry of the effective
theory.
Using Ward identities one can show
that such nonuniversal terms should not
appear. To be precise, any new physics can only contribute to the
universal interactions of the photon to charged fields. This effect
can simply be absorbed in redefining  the
electromagnetic fine structure constant $\alpha$, hence no new
\tta or \bba interaction terms
will appear in the effective Lagrangian after a proper renormalization of
$\alpha$.

Here is a final note regarding the physical Higgs boson.
It is known that the gauge bosons acquire masses through the spontaneous
symmetry-breaking mechanism. In the chiral Lagrangian
this can be seen from the last term in ${\cal L}_B$ (see Eq.~(\ref{yeq4})),
which only involves the gauge bosons and the unphysical Goldstone bosons.
This indicates that the chiral Lagrangian can account for the mass generation
of the gauge bosons without the actual details of the symmetry-breaking
mechanism. Furthermore,
 the fermion mass term is also allowed in the chiral Lagrangian,
\beq
 -m_{f_i}\overline{f_i}f_i \, , \nonumber
\enq
because it is invariant under G, where
the fermion field $f_i$ transforms as
in Eq. (\ref{eq1}).

{}From this it is clear the Higgs boson is not necessary in constructing the
low energy effective Lagrangian. Indicating that the SM Higgs mechanism
is just one example of the possible
spontaneous symmetry-breaking scenarios which might take place in nature.
Still, a Higgs boson
can be inserted in the chiral Lagrangian as an additional field
($SU(2)_{L}\times U(1)_Y$ singlet) with arbitrary
couplings to the rest of the fields. To retrieve the SM Higgs boson
contribution at tree level, one can
 simply substitute the fermion mass $m_f$
by $g_f v$ and $v$ by $v+H$, where $g_f$ is the Yukawa coupling for fermion
$f$ and $H$ is the Higgs boson field.
Hence, we get the scalar sector Lagrangian
\beq
{\cal L}_{H}=\frac{1}{2}\partial_{\mu} H \partial^{\mu} H
-\frac{1}{2}m_H^2 H^2 -V(H) +\frac{1}{2}vH{\rm{Tr}}
\left ( {D_{\mu}\Sigma}^{\dagger}D^{\mu} \Sigma \right )
+\frac{1}{4}H^2{\rm{Tr}}
\left ( {D_{\mu}\Sigma}^{\dagger}D^{\mu} \Sigma \right )\, ,\label{higg}
\enq
where $V(H)$ describes the Higgs boson self-interaction.
The coefficients of the last two terms in the above equation can be
arbitrary for a chiral Lagrangian with a scalar field other than
the SM Higgs boson.
Similarly, the coupling of $f_i$ and the scalar (Higgs) boson $H$ can
in general be written as
\beq
 - c_i { m_{f_i} \over v} H {\bar f_i} f_i \,\, ,
\enq
where $c_i=1$ for the SM Higgs boson.
In this analysis we will discuss models
with and without a Higgs boson.
In the case of a light Higgs boson ($m_H <  m_t$) we will include the
Higgs boson field in the chiral Lagrangian as a
part of the light fields with no new physics being associated with it.
In the case of a heavy Higgs boson ($m_H > m_t$) in the full theory,
we assume the Higgs boson field has been integrated out and its effect on low
energy physics can be thought of as a new heavy physics effect which is
already included in the effective couplings of the top quark
at the scale of $m_t$. Finally, we
will consider the possibility of a spontaneous symmetry-breaking
scenario without including a SM Higgs boson in the full theory.
In this case we consider the effects on low energy data from the
new physics parameterized by the nonstandard
interaction terms in ${\cal L}$ in Eq.~(\ref{eq3})
and contributions from the SM
without a Higgs boson.

\section{\twelvebf  the Top Quark Couplings to Gauge Bosons}

As we discussed in the previous section, one possibility of new physics
effects is the modification of the vertices \bbz, \ttz, and \tbw in the
effective Lagrangian
by the same order of magnitude
\cite{sekh}. In this case, only the vertex \bbz can have large
contributions to low
energy data while, based on the dimensional counting method,
the contributions from the other two vertices
\ttz and \tbw are suppressed by
$1/16{\pi}^2$ due to their insertion in loops.

In this case, one can use $\Gamma_b$ (the partial decay width of the
$Z$ boson to $\overline{b}b$) to constrain the \bbz coupling.
Denote the nonstandard \bbz vertex as
\beq
\frac{g}{4\cos \theta_W}\kappa \gamma_{\mu}(1-\gamma_5)\, ,
\enq
which is purely left-handed.
In some Extended
technicolor models, discussed in Ref.~\cite{sekh},
this nonstandard effect arises from the same source as the mass generation
of the top quark, therefore
 $\kappa$ depends on the top quark mass.

As we will discuss later, the nonuniversal contribution to $\Gamma_b$
is parameterized by a measurable parameter denoted as
$\epsilon_b$ \cite{bar,bar1,bar2} which is measured to be \cite{bar}
\beq
\epsilon_b\,(10^{3}) = 4.4 \pm 7.0 \,. \nonumber
\enq
The SM contribution to $\epsilon_b$ is calculated in Refs.~\cite{bar,bar1},
{\it e.g.}, for a 150\,GeV top quark
\beq
\epsilon_b^{\rm{SM}}\, (10^{3}) =-4.88 \,\,. \nonumber
\enq
The contribution from $\kappa$ to $\epsilon_b$ is
\beq
\epsilon_b=-\kappa \, .
\enq
Within a 95\% confidence level (CL), from  $\epsilon_b$ we find that
\beq
-22.9 \leq \kappa\, (10^{3}) \leq 4.4 \, .
\enq

As an example, the simple commuting extended technicolor model
presented in {\mbox Ref. \cite{sekh}} predicts that
\beq
\kappa \approx  \frac{1}{2}{\xi}^2 \frac{m_t}{4\pi v} \, ,
\enq
where $\xi$ is of order 1.
Also in that model the top quark couplings
$\kln$, $\krn$, and $\klc$, as defined in Eqs.~(\ref{eq2}) and (\ref{eq3}),
are of the same order as $\kappa$.
For a 150\,GeV top quark, this model predicts
\beq
\kappa \, (10^{3})\approx 24.3\, {\xi}^2 \, \nonumber .
\enq
Hence, such a model is likely to be excluded using low energy data.

We will devote the subsequent discussion to models in which the
nonstandard \bbz coupling can be ignored relative to the
\ttz and \tbw couplings. In this
case one needs to study their effects at the quantum level, {\it i.e.},
through loop insertion. We will first discuss the general case where no
relations between the couplings are assumed. Later we will
impose a relation between $\kln$ and $\klc$ which are defined in
Eqs.~(\ref{eq2}) and (\ref{eq3}) using an effective model with an
approximate custodial symmetry.

\subsection{\twelveit General case}

  The chiral Lagrangian in general has a complicated structure and many
arbitrary coefficients which weaken its predictive power. Still, with
a few further assumptions, based on the status of
present low energy data, the chiral Lagrangian can provide a useful approach to
confine the coefficients parameterizing new physics effects.

In this subsection, we provide a general treatment for
the case under study with
minimal imposed assumptions in the chiral Lagrangian. In this case, we
only impose the assumption that the vertex \bbz is not modified by
the dynamics. In the chiral Lagrangian
${\cal L}$, as defined in Eqs.~(\ref{eq2}) and (\ref{eq3}),
there are six independent parameters ({\mbox {$\kappa$'s}})
which need to be constrained
using precision data.
Throughout this lecture we will only consider the
insertion of {\mbox {$\kappa$'s}} once in one-loop diagrams by assuming that
these nonstandard couplings are small; $\kappa_{L,R}^{NC,CC} < 1$.
At the one-loop level
the imaginary parts of the couplings do not contribute to those
LEP observables of interest.
Thus, hereafter we drop the imaginary pieces from the effective couplings,
which reduces the number of relevant parameters to four.
Since the bottom quark
mass is small relative to the top quark mass,
we find that  $\krc$ does not contribute
to low energy data up to the order ${m_t^2}\ln {\Lambda}^2$
in the $m_{b}\ra 0$ limit.
With these observations we conclude that only the three parameters
$\kln$, $\krn$ and $\klc$ can be constrained.

A systematic approach can be implemented for such an
analysis based on the scheme used in Refs.~\cite{bar,bar1,bar2}, where the
radiative corrections can be parameterized by 4 independent parameters,
three of those parameters $\epsilon_1$, $\epsilon_2$, and $\epsilon_3$
are proportional to the variables $T$, $U$ and $S$ \cite{pesk},
and the fourth  one; $\epsilon_b$ is due to the Glashow-Iliopoulos-Miani- (GIM)
violating contribution in $Z\rightarrow b \overline{b}$ \cite{bar}.

  These parameters are derived from four basic measured \mbox{observables},
$\Gamma_{\ell}$\,(the partial width of $Z$ to a
charged lepton pair),
$A_{FB}^{\ell}$\,(the forward-backward asymmetry at the $Z$ peak for
the charged lepton $\ell$), $M_{W}/M_{Z}$,
and $\Gamma_{b}$\,(the partial width
of $Z$ to a $b\overline{b}$ pair).
The expressions of these observables in terms of
$\epsilon$'s were given in Refs.~\cite{bar,bar1}.
In this lecture we only give the relevant terms in $\epsilon$~'s
which might contain the leading effects from new physics.

We denote the vacuum polarization for the $W^1, W^2, W^3, B$ gauge bosons as
\beq
{{\Pi}^{ij}}_{\mu \nu}(q) = -ig_{\mu \nu}\left [A^{ij}(0) + q^{2}F^{ij}(
q^2)
\right ] + q_{\mu}q_{\nu}\,\rm{terms}\, ,
\enq
where $i,j=1,2,3,0$ for $W^1, W^2, W^3$ and $B$,  respectively. Therefore,
\beq
\epsilon_1 = e_1-e_5 \,,
\enq
\beq
\epsilon_2 = e_2 - c^{2}e_5\, ,
\enq
\beq
\epsilon_3 = e_3 - c^{2}e_5 \, ,
\enq
\beq
\epsilon_b = e_b \, ,
\enq
where
\beq
e_1 = \frac{A^{33}(0) - A^{11}(0)}{M_W^2}\, ,
\enq
\beq
e_2 = F^{11}(M_W^2) - F^{33}(M_Z^2)\, ,
\enq
\beq
e_3 = \frac{c}{s}F^{30}(M_Z^2)\, ,
\enq
\beq
e_5 = {M_Z^2}\frac{dF^{ZZ}}{dq^2}(M_Z^2)\, ,
\enq
and $c\equiv \cos \theta_W$.
\beq
c^{2} \equiv \frac{1}{2}\left [ 1+{\left ( 1-\frac{4\pi \alpha (M_Z)}
{\sqrt{2} G_f M_Z^2}\right )}^{1/2} \right ] \, ,
\enq
and $s^2=1-c^2$.
$e_b$ is defined through the GIM-violating $Z\rightarrow b\overline{b}$ vertex
\beq
{V_{\mu}}^{GIM}\left ( Z \rightarrow b\bar{b}\right ) = -\frac{g}{2c}e_b
\gamma_{\mu}\frac{1-\gamma_5}{2}\, .
\enq

$\epsilon_1$ depends quadratically on $m_t$ \cite{bar,bar1}
and has been measured to
good accuracy, therefore $\epsilon_1$ is sensitive to any
new physics coming through the top quark. On the contrary, $\epsilon_2$\, and
$\epsilon_3$ do not play any
significant role in our analysis because their dependence on the top mass is
only logarithmic.

 Non-renormalizability of the effective Lagrangian presents
a major issue of how to find a scheme to
handle both the divergent and the finite pieces in
loop calculations \cite{burg,mart}. Such a problem arises because one
does not know the underlying theory; hence, no matching can be performed
to extract the correct scheme to be used in the effective Lagrangian
\cite{geor}.
One approach is to associate the divergent piece in
loop calculations with a physical
cutoff $\Lambda$, the upper
scale at which the effective Lagrangian is
valid \cite{pecc}. In the chiral Lagrangian approach this cutoff
$\Lambda$ is taken to be
$4\pi v \sim 3$\,TeV \cite{geor}.\footnote{
This scale, $4\pi v \sim 3$\,TeV, is only meant to indicate
the typical cutoff scale. It is equally probable to have, say,
$\Lambda=1$\,TeV.} For the finite piece no
completely satisfactory approach is available \cite{burg}.

\begin{figure}
\caption{
Some of the relevant Feynman diagrams in the
't\,Hooft-Feynman gauge, which contribute
to the order ${\cal O}\,({m_t^2}\ln {\Lambda}^2)$.}
\label{fey}
\end{figure}

To perform calculations using the chiral Lagrangian,
one should arrange
the contributions in powers of $1/4\pi v$ and then include all
diagrams up to the desired power. In
the $R_{\xi}$ gauge ($\Sigma \neq 1$), the couplings of
the Goldstone bosons to the fermions should also be included
in Feynman diagram calculations.
These couplings can be easily found
by expanding the terms in ${\cal L}$ as given
in Eq.~(\ref{eq2}).
We will not give the explicit expressions for those terms here.
Some of the relevant Feynman diagrams are shown in Fig.~\ref{fey}.
Calculations were done in the
't Hooft-Feynman gauge.
We have also checked our calculations in both the Landau gauge and the
unitary gauge and found agreement as expected.

We calculate the  contribution to $\epsilon_1$ and
$\epsilon_b$ due to the new interaction terms in the chiral Lagrangian
(see Eqs.~(\ref{eq2}) and (\ref{eq3})) using the dimensional regularization
scheme and
taking the bottom mass to be zero.
At the end of the calculation, we
replace the divergent piece $1/\epsilon$ by
$\ln(\Lambda^2/{m_t^2})$ for $\epsilon = (4-n)/2$ where $n$ is the
space-time dimension.
We have assumed that the underlying full theory is renormalizable.
The cutoff scale $\Lambda$ serves as the infrared cutoff of the
operators in the effective Lagrangian. Due to the renormalizability
of the full theory, from renormalization group analysis, we conclude
that the same cutoff $\Lambda$ should also serve as the ultraviolet
cutoff of the effective Lagrangian in calculating Wilson coefficients.
Hence, in the dimensional regularization scheme,
$1/\epsilon$ is replaced by $\ln(\Lambda^2/{\mu^2})$.
Furthermore, the renormalization scale $\mu$ is set to be $m_t$,
the heaviest mass scale in the effective Lagrangian of interest.

Since we are mainly interested
in new physics associated with the top quark couplings to gauge bosons,
we shall restrict ourselves
to the {\it leading} contribution enhanced by the top quark mass, {\it i.e.},
of the oder of $m_t^2\ln {\Lambda}^{2}$.
We find
\beq
\epsilon_1=\frac{G_F}{2\sqrt{2}{\pi}^2}3{m_t^2}
 (-\kln+\krn+\klc)\ln{\frac{{\Lambda}^2}{m_t^2}}\,\, , \label{cal1}
\enq
\beq
\epsilon_b=\frac{G_F}{2\sqrt{2}{\pi}^2}{m_t^2}
\left ( -\frac{1}{4}\krn+\kln \right ) \ln{\frac{{\Lambda}^2}{m_t^2}}
\,\, . \label{cal2}
\enq
Note that $\epsilon_2$ and $\epsilon_3$ do not contribute at this order.
That $\klc$ does not contribute to $\epsilon_b$ up to this order
can be understood from Eq.~(\ref{eq3}). If $\klc=-1$
then there is no net \tbw coupling in the chiral Lagrangian
after including both the standard and nonstandard contributions. Hence,
no dependence on the top quark mass can be generated, {\it i.e.},
the non-standard $\klc$ contribution to $\epsilon_b$
must cancel the SM contribution when
$\klc=-1$, independently of the couplings of
the neutral current.
{}From this observation and because the SM contribution to
$\epsilon_b$ is finite,
we conclude that $\klc$ cannot contribute to $\epsilon_b$ at the order
of interest.

Note that we set the renormalization scale $\mu$ to be $m_t$, which is the
natural scale to be used in our study because the top quark is
considered to be the heaviest
mass scale in the effective Lagrangian.
We have assumed that all other
heavy fields have been integrated out
to modify the effective couplings of the top quark to gauge bosons
at the scale $m_t$ in the chiral Lagrangian.
Here we ignore the effect of
the running couplings from the top quark
mass scale down to the $Z$ boson mass scale which is a
reasonable approximation for our study.

To constrain these nonstandard couplings we need to have both the
experimental values\footnote{
We should first discuss the old (1993) data in this lecture, then discuss
the impact of new (1994) data in the next lecture for comparison.
This is useful to test the sensitivity of the constraints
on $\kappa$'s from the measurements of $\epsilon$'s.
} and the SM predictions of {\mbox {$\epsilon$'s}}.
First, we tabulate the numerical inputs,
taken from Ref.~\cite{bar}, used in our analysis:
\begin{eqnarray}
{\alpha}^{-1}(M_Z^2) &=&   128.87 \pm 0.12  \nonumber\, ,\\
G_F          & =  & 1.16637(2) \times {10}^{-5} \,\,\,\,
 {\rm{GeV}}^{-2} \nonumber\, ,\\
M_Z          & =  &  91.187 \pm 0.007  \,\,\, \rm{GeV} \nonumber \, ,\\
M_W/M_Z      & =  &  0.8798 \pm 0.0028   \nonumber \, ,\\
\Gamma_{\ell}& =  &  83.52 \pm 0.28  \,\,\, \rm{MeV} \nonumber \, ,\\
\Gamma_b     & =  &  383 \pm 6  \,\,\,\rm{MeV}   \nonumber \, ,\\
A^{\ell}_{FB}       & =  &  0.0164 \pm 0.0021 \nonumber \, ,\\
A^{b}_{FB}       & =  &  0.098 \pm 0.009  \nonumber \, ,\\
A_{LR} \, ({\rm SLC}) \,       & =  &  0.100 \pm 0.044 \nonumber \, .
\end{eqnarray}
These values yield \cite{bar}
\begin{eqnarray}
\epsilon_1 \,10^{3}  & =  &  -0.3 \pm 3.4 \nonumber \, ,\\
\epsilon_b \,10^{3}  & =  &  4.4 \pm 7.0 \nonumber \, ,
\end{eqnarray}
and, for completeness,
\begin{eqnarray}
\epsilon_2 \,10^{3}  & =  &  -7.6 \pm 7.6 \nonumber \, ,\\
\epsilon_3\, 10^{3}  & =  &  0.4 \pm 4.2 \nonumber \, .
\end{eqnarray}

The SM contribution to $\epsilon$'s have been calculated in
Refs.~\cite{bar,bar1}. We will include these contributions in
our analysis in accordance with the assumed Higgs boson mass.
In the light Higgs boson case ($m_H < m_t$), the
calculated values of the $\epsilon$'s include both the SM contribution
calculated in Refs.~\cite{bar,bar1}
and the new physics contribution derived from
the effective couplings of the top quark to gauge bosons.
In the heavy Higgs boson case
($m_H > m_t$) we subtract
the Higgs boson contribution from the SM
calculations of $\epsilon$'s given in {\mbox {Refs.~\cite{bar,bar1}}}.
In this case, the Higgs  boson
contribution is implicitly included in the effective couplings of the top
quark to gauge bosons after the heavy Higgs boson field is integrated
out.
Finally, in a spontaneous symmetry scenario without a Higgs boson
the calculations of {\mbox {$\epsilon$'s}} are exactly the same as those done
in the
heavy Higgs boson case except that the effective couplings of the top quark
to gauge bosons are not due to an assumed heavy Higgs boson in the full
theory.

Choosing $m_t=150$\,GeV and $m_H=100$\,GeV
we span the parameter space  defined by $-1 \leq \kln \leq 1 $,
$-1 \leq \krn \leq 1 $, and $-1 \leq \klc \leq 1 $.
Within $95$\%~CL and including both the SM and the new physics
contributions, the allowed region of these three parameters is found
to form a thin slice in the specified volume.
The two-dimensional projections of this slice were shown
in Figs. 2, 3, and 4 of Ref.~\cite{ehab}.\footnote{
We do not reproduce those figures here because they exhibit the same
shape as those obtained using new data,
to be discussed in the next lecture.
} These nonstandard couplings {\mbox {($\kappa$'s)}} do exhibit
some interesting features.
\begin{itemize}
\item [(1)]
As a function of the top quark mass,
the allowed volume for the top quark
couplings to gauge bosons shrinks as the top quark
becomes more massive.
\item [(2)]
 New physics prefers positive $\kln$.
$\kln$ is constrained within $-0.3$ to 0.6 ($-0.2$ to 0.5) for
a 150 (175)\,GeV top quark.
\item [(3)]
New physics prefers $\klc \approx -\krn$.
\end{itemize}

A similar analysis was carried out in Ref.~\cite{pczh},
in which, however, the authors did
not include the charged current contribution and assumed only
the vertex \ttz gives large nonstandard effects.
The allowed region they found  simply
corresponds, in our analysis, to the region defined by the intersection
of the allowed volume and the plane
$\klc = 0$. This gives a small area confined in the vicinity of the line
$\kln = \krn$. This can be understood from the
expression of $\epsilon_1$ derived
in Eq.~(\ref{cal1}).
After setting $\klc=0$, we find
\beq
\epsilon_1 \propto \left (\krn - \kln \right )\, .
\enq
In this case we note that the length of
the allowed area
is merely determined by the contribution from
$\epsilon_b$. We will
elaborate on a more quantitative comparison in the second part of this
section.

\subsection{\twelveit Special case}

The allowed region in the parameter space obtained in Figs.~2, 3 and 4
of Ref.~\cite{ehab}
contains all possible new physics (to the order ${m_t^2}\ln {\Lambda}^2$ )
which can modify the couplings of the top quark to gauge bosons
as described by $\kln$, $\krn$, and $\klc$.
In this section we would like to examine a special class of models
in which an approximate custodial symmetry is assumed
as suggested by low energy data.

 The SM has an additional (accidental) symmetry called the
custodial symmetry which is responsible for the tree-level relation:
$\rho=1$. This symmetry is slightly broken at the quantum
level by the $SU(2)_L$ doublet fermion mass splitting and the hypercharge
coupling $g^{\prime}$ \cite{velt}. Writing $\rho=1 +\delta \rho$,
$\delta\rho$ would vanish to all orders if this symmetry is exact.
 Because low energy data indicate that
$\delta\rho$ is very close to zero
we shall therefore assume an underlying theory with a custodial
symmetry.
In other words we require the global
group $SU(2)_V$ associated with the
custodial symmetry to be a subgroup of the full group characterizing
the full theory. We will assume that the custodial symmetry is broken
by the same factors
which break it in the SM, {\it i.e.}, by the fermion
mass splitting and the hypercharge coupling $g^{\prime}$.

In the chiral Lagrangian this assumption of a custodial
symmetry sets  $v_3=v$,
and forces the couplings of the top quark to
gauge bosons ${W_{\mu}^a}$
to be equal after turning off the hypercharge and
assuming $m_b = m_t$.
If the dynamics of the symmetry-breaking is such
that the masses of the two $SU(2)$ partners $t$ and $b$ remain degenerate
then  we expect new physics to contribute
to the couplings of \ttz and \tbw by the same amount.
However, in reality, $m_b \ll m_t$;
thus, the custodial symmetry has to be broken.
We will discuss how this symmetry is broken shortly.
Since we are mainly interested in the
{\it leading} contribution enhanced
by the top quark mass at
the order ${m_t^2}\ln {\Lambda}^2$, turning the hypercharge
coupling on and off will not affect the final result up to this order.

We can construct the two Hermitian operators
$J_{L}$ and $J_{R}$, which transform under G as
\beq
J_{L}^{\mu}=-i\Sigma {D_{\mu}\Sigma}^{\dagger}
 \rightarrow g_{L}J_{L}^{\mu}g_{L}^{\dagger}\,\, ,
\enq
\beq
J_{R}^{\mu}=i{\Sigma}^{\dagger} D_{\mu}\Sigma
 \rightarrow g_{R}J_{R}^{\mu}g_{R}^{\dagger}\,\, ,
\enq
where $g_{L}=e^{i\alpha^{a}\frac{\tau^{a}}{2}} \in SU(2)_{L} $\,and
$g_{R}=e^{iy\frac{\tau^3}{2}}$ (note that $v_3=v$ in $\Sigma$).
In fact, using either $J_{L}$ or $J_R$ will lead to the same result.
Hence, from now on we will only consider $J_R$.
The SM Lagrangian can be derived from\footnote{
This Lagrangian is equivalent to the one defined in Eq.~(\ref{eq17}).
Both of them give the same S-matrix for any physical process.
}
\begin{eqnarray}
{\cal L}_{0}=\overline{\Psi_{L}}i\gamma^{\mu}D_{\mu}^{L}\Psi_{L}
            +\overline{\Psi_{R}}i\gamma^{\mu}D_{\mu}^{R}\Psi_{R}
           -(\overline{\Psi_{L}}\Sigma M\Psi_{R}+h.c.) \nonumber \\
-\frac{1}{4}W_{\mu \nu}^{a}{W^{\mu \nu}}^{a}-\frac{1}{4}B_{\mu \nu}B^{\mu \nu}
  +\frac{v^{2}}{4}{\rm{Tr}}(J_{R}^{\mu}{J_{R}}_{\mu})\,\, ,
\label{yeq48}
\end{eqnarray}
where $M$ is a diagonal mass matrix. We have chosen
 the left-handed fermion fields to be
the ones defined in Eq.~(\ref{psi}):
\beq
\Psi_{L}\equiv \Sigma{t\choose b}_{L}\, .
\enq
The right-handed fermion fields $t_{R}$ and $b_{R}$ coincide with the
original right-handed fields (see Eq.~(\ref{psr})). Also
\beq
D_{\mu}^{L}=\partial_{\mu}-igW_{\mu}^{a}\frac{\tau^{a}}{2}
  -i{g}^{\prime}B_{\mu}\frac{Y}{2}\,\, ,
\enq
\beq
D_{\mu}^{R}=\partial_{\mu}-i{g}^{\prime}B_{\mu}
\left ( \frac{Y}{2} + \frac{\tau^3}{2} \right )\,\, .
\enq
Note that in the nonlinear realized effective theories
using either set of fields ( $\Psi_{L,R}$ or $F_{L,R}$)
to construct a chiral Lagrangian will lead to the
same S matrix \cite{cole}.

The Lagrangian ${\cal L}_0$ in Eq.~(\ref{yeq48})
is not the most general Lagrangian one can construct
based solely on the symmetry of $G/H$. Taking
advantage of the chiral Lagrangian approach we can
derive additional interaction terms which
deviate from the SM. This is so because in this formalism
the $SU(2)_L\times U(1)_Y$ symmetry is nonlinearly realized
and  only the
$U(1)_{em}$ is linearly realized.

Because the SM is so successful one can think of the SM (without
the top quark)
as being the leading term in the expansion of the effective Lagrangian.
Any possible deviation associated with
the light fields can only come through
higher dimensional operators in the Lagrangian.
However, this assumption is neither
necessary nor preferable when dealing with the top quark because no
precise data are available to lead to such a conclusion. In this
lecture we will include nonstandard dimension-four
operators for the couplings of the top quark to gauge bosons.
In fact this is all
we will deal with and  we will not consider operators with dimension
higher than four. Note that higher dimensional operators are
naturally suppressed by powers of $1/\Lambda$.

One can write $J_{R}$ as
\beq
J_{R}^{\mu}= {J_{R}^{\mu}}^{a}\frac{{\tau}^{a}}{2} \label{tra} \, ,
\enq
with
\beq
{J_{R}^{\mu}}^{a}={\rm{Tr}}\left ( {\tau}^{a}{J_R^\mu}\right )
=i{\rm{Tr}}\left ( \tau^{a}{\Sigma}^{\dagger}{\rm{D}}^{\mu}\Sigma \right )\, .
\enq
The full operator $J_{R}$ posses an explicit custodial symmetry
when $g^{\prime}=0$ as can easily be checked by expanding it in powers of
the Goldstone boson fields.

Consider first the left-handed sector. One can add
additional interaction terms to the Lagrangian
${\cal L}_0$
\beq
{\cal L}_1 ={\kappa}_{1}\overline{\Psi_{L}}\gamma_{\mu}\Sigma J_{R}^{\mu}
         {\Sigma}^{\dagger}\Psi_{L}
     +{\kappa}_{2}\overline{\Psi_{L}}\gamma_{\mu}\Sigma
     {\tau}^{3}J_{R}^{\mu}{\Sigma}^{\dagger}\Psi_{L}
     +{\kappa}_{2}^\dagger \overline{\Psi_{L}}
       \gamma_{\mu}\Sigma J_{R}^{\mu}{\tau}^{3}
     {\Sigma}^{\dagger}\Psi_{L}
\label{eehab}\, ,
\enq
where ${\kappa}_{1}$ is an arbitrary real parameter and $\kappa_2$ is
an arbitrary complex parameter.
Here we do not include interaction terms such as
\beq
 {\kappa}_{3}\overline{\Psi_{L}}\gamma_{\mu}\Sigma {\tau}^{3}J_{R}^{\mu}
         {\tau}^{3}{\Sigma}^{\dagger}\Psi_{L}\, ,
\enq
where ${\kappa}_{3}$ is real, because
it is simply a linear combination of the other two terms in ${\cal L}_1$.
This can be easily checked by
using Eq.~(\ref{tra}) and the commutation relations of the Pauli matrices.
Note that ${\cal L}_{1}$ still is not the most general Lagrangian one can
write for the left-handed sector, as compared to Eq.~(\ref{eq2}).
In fact, it is our insistence on using
the fermion doublet form and the full operator $J_{R}$ that lead us to this
form. For shorthand, ${\cal L}_{1}$ can be further rewritten as
\beq
{\cal L}_1 =\overline{\Psi_{L}}\gamma_{\mu}\Sigma K_{L}J_{R}^{\mu}
         {\Sigma}^{\dagger}\Psi_{L}
         +\overline{\Psi_{L}}\gamma_{\mu}\Sigma J_{R}^{\mu} K_{L}^\dagger
         {\Sigma}^{\dagger}\Psi_{L}
\label{cp} \,\, ,
\enq
where $K_{L}$ is a complex diagonal matrix.

These new terms can be generated either through some
electroweak symmetry-breaking scenario or
through some other new heavy physics effects.
If $m_b=m_t$ and $g^{\prime}=0$, then we
require the effective Lagrangian to respect fully the custodial
symmetry to all orders. In this limit,
${\kappa}_{2}=0$ in Eq.~(\ref{eehab}) and
{\mbox {$K_{L} =\kappa_{1}${\bf 1}}}, where {\bf 1} is
the unit matrix and $\kappa_1$ is real.

Since $m_b \ll m_t$, we can think of $\kappa_{2}$
as generated through the evolution from $m_b=m_t$ to
$m_b=0$. In the
matrix notation this implies $K_{L}$ is not proportional
to the unit matrix and can be parameterized by
\beq
 K_{L}=\pmatrix{\kappa_L^t & 0 \cr 0 & \kappa_L^b\cr}\, ,
\enq
with
\beq
{\kappa_L^t}=\frac{\kappa_1}{2} + \kappa_2\, ,
\enq
and
\beq
{\kappa_L^b}=\frac{\kappa_1}{2} - \kappa_2\,
\label{me}.
\enq
In the unitary gauge we get the terms
\begin{eqnarray}
+\frac{g}{2c}2{\rm{Re}} ({\kappa_L^t})\overline{t_{L}}
\gamma^{\mu}t_{L}Z_{\mu}
+\frac{g}{\sqrt{2}}({\kappa_L^t}+{\kappa_L^b}^{\dagger})
\overline{t_{L}}\gamma^{\mu}b_{L}{W_\mu^+} \nonumber \\
+\frac{g}{\sqrt{2}}({\kappa_L^b}+{\kappa_L^t}^{\dagger})
\overline{b_{L}}\gamma^{\mu}t_{L}{W_\mu^-}
-\frac{g}{2c}2 {\rm{Re}} ({\kappa_L^b})
\overline{b_{L}}\gamma^{\mu}b_{L}Z_{\mu}\, .
\end{eqnarray}
As discussed in the previous section,
we will assume that new physics effects will not modify
the $b_{L}$-$b_{L}$-$Z$ vertex. This can be achieved by
choosing $\kappa_1=2{\rm{Re}}(\kappa_2)$ such that
${\rm{Re}}(\kappa_{L}^{b})$ vanishes
in Eq.~(\ref{me}). Later, in
{\mbox {Sec. 4}}, we will consider a specific model to support this
assumption.

 Since the imaginary parts of the couplings do not contribute to LEP
physics of interest, we simply drop them hereafter.
With this assumption we are left with
one real parameter $\kappa_L^t$ which will be
denoted from now on by $\kappa_L/2$. The left-handed top quark couplings to
the gauge bosons are
\beq
 t_{L}-t_{L}-Z :\,\, \frac{g}{4c}\kappa_{L}\gamma_{\mu}(1-\gamma_{5})\, ,
\enq
\beq
 t_{L}-b_{L}-W :\,\, \frac{g}{2\sqrt{2}}\frac{\kappa_{L}}{2}
 \gamma_{\mu}(1-\gamma_{5})\, .
\enq
Notice the connection between the neutral and the
charged current, as compared to Eq.~(\ref{eq3}):
\beq
\kln = 2\klc = \kappa_{L}\, .
\enq
This conclusion holds for any underlying theory with an approximate
custodial symmetry such that the vertex \bbz is not modified as discussed
above.

For the right-handed sector, the situation is different because the
right-handed fermion
fields are $SU(2)$ singlet, hence the induced interactions do not see
the full operator $J_{R}$ but its components individually.
Therefore, we cannot impose the
previous connection between the neutral
and charged current couplings.

The additional allowed interaction terms in
the right-handed sector are given by
\begin{eqnarray}
{\cal L}_{2}=
\frac{g}{2c}{\kappa_R^t}^{\rm{NC}} \overline{t_{R}}\gamma^{\mu}
t_{R} {J_{R}}_{\mu}^3
+\frac{g}{\sqrt{2}}{\kappa_R^{\rm{CC}}}\overline{t_{R}}\gamma^{\mu}b_{R}
{J_{R}}_{\mu}^+ \nonumber \\
+\frac{g}{\sqrt{2}}{\kappa_R^{\rm{CC}}}^{\dagger}\overline{b_{R}}\gamma^{\mu}
t_{R}{J_{R}}_{\mu}^-
-\frac{g}{2c} {\kappa_R^b}^{\rm{NC}} \overline{b_{R}}\gamma^{\mu}
b_{R} {J_{R}}_{\mu}^3 \label{neh}\,\, ,
\end{eqnarray}
where ${\kappa_R^t}^{\rm{NC}}$
and  ${\kappa_R^b}^{\rm{NC}}$ are two arbitrary
real parameters and $\kappa_R^{\rm{CC}}$ is an arbitrary complex parameter.
Note that in ${\cal L}_{2}$ we have one more additional coefficient than
we have in ${\cal L}_{1}$ (in Eq.~(\ref{eehab})),
this is due to our previous assumption of
using the full operator $J_{R}$ in constructing the left-handed interactions.
We assume that the \brbrz vertex just as the \blblz vertex
is not modified, then
the coefficient ${\kappa_R^b}^{\rm{NC}}$ vanishes.
Because $\krc$ does not contribute to LEP physics in the limit of
$m_b=0$ and at the order ${m_t^2} \ln {\Lambda}^2$ we are left with
one real parameter ${\kappa_R^t}^{\rm{NC}}$ which will be
denoted hereafter as $\kappa_{R}$.
The right-handed top quark coupling to $Z$ boson is
\beq
t_{R}-t_{R}-Z :\,\,
\frac{g}{4c}  \kappa_{R}\gamma_{\mu}(1+\gamma_{5}) \,\, .
\enq

In the rest of this section we consider models described by
${\cal L}_{1}$ and ${\cal L}_{2}$ with only two relevant parameters
$\kappa_L$ and $\kappa_R$. Performing the calculations as we discussed in
the previous subsection we find
\beq
\epsilon_{1} = \frac{G_F}{2\sqrt{2}{\pi}^2} 3 {m_t^2} \left (
\kappa_{R}-\frac{\kappa_{L}}{2}\right )\ln(\frac{{\Lambda}^2}{m_t^2})\,\, ,
\label{fn}
\enq
\beq
\epsilon_{b} = \frac{G_F}{2\sqrt{2}{\pi}^2}{m_t^2}
\left (-\frac{1}{4} \kappa_{R} + \kappa_{L}\right )
\ln(\frac{{\Lambda}^2}{m_t^2})
 \label{sb} \,\, .
\enq
These results simply correspond to those
in Eqs.~(\ref{cal1}) and (\ref{cal2})
after substituting $\kln=2\klc=\kappa_{L}$ and $\krn=\kappa_R$.

The constraints on $\kappa_L$ and $\kappa_R$
for models with a light Higgs boson
or a heavy Higgs boson,
or without a physical scalar (such as a
Higgs boson) are presented here in order.
Let us first consider a standard light Higgs boson
with mass $m_H=100$\,GeV. Including
the SM contribution from Ref.~\cite{bar} we span the plane
defined by $\kappa_{L}$ and $\kappa_{R}$ for top mass 150 and 175\,GeV,
respectively.
Figs. 5 and 6 of Ref.~\cite{ehab}
showed the allowed range for those parameters
within 95\% CL.
As a general feature one observes that the allowed range is a
narrow area aligned close to the line $\kappa_{L} = 2 \kappa_{R}$ where for
$m_t=150$\,GeV the maximum  range
for $\kappa_{L}$ is between $-0.1 $ and $0.5$. As the
top mass increases this range shrinks and moves downward and to the
right away form the origin ($\kappa_{L}$,$\kappa_{R}$) = {\mbox {(0,0)}}.
The deviation from the relation $\kappa_{L} = 2 \kappa_{R}$ for
various top quark masses was given in Fig.~7 of Ref.~\cite{ehab}
by calculating
$\kappa_{L} - 2 \kappa_{R}$ as a function of $m_t$. Note that the
SM has the solution $\kappa_{L}=\kappa_R=0$.
This solution ceases to exist for $m_{t}\geq 200$\,GeV.
The special relation $\kappa_{L}=2\kappa_{R}$ is
a consequence of the assumption we imposed in connecting the left-handed
neutral and charged current.

It is worth mentioning
that the SM contribution to $\epsilon_{b}$ is lower
than the experimental central value \cite{bar,bar1}.
This is reflected in the behavior of
$\kappa_{L}$ which prefers being positive to compensate this difference as
can be seen from Eq.~(\ref{sb}).
This means in models of electroweak symmetry-breaking with an approximate
custodial symmetry, a positive $\kappa_L$ is preferred.
In Fig.~8 of Ref.~\cite{ehab}
 we showed the allowed $\klc=\kln/2=\kappa_{L}/2$ as a
function of $m_t$.
With new physics effects ($\kappa_L \neq 0$) $m_t$ can be as large as
300\,GeV, although in the SM ($\kappa_L=0$) $m_t$ is bounded below
200\,GeV.

Now, we would like to discuss the effect of a light SM Higgs boson
($ m_H < m_t$) on the allowed range of these parameters.
It is easy to anticipate the effect; since
$\epsilon_{b}$ is not sensitive to the Higgs boson contribution up to one loop
\cite{bar},
the allowed range is only affected by the Higgs boson contribution to
$\epsilon_{1}$ which affects slightly the
width of the allowed area and its location relative to the line
$\kappa_{L}=2\kappa_{R}$.
One expects that as the Higgs boson mass increases the allowed area moves
upward. The reason simply lies in the fact that the standard Higgs boson
contribution to $\epsilon_{1}$ up to one loop becomes more negative
for heavier Higgs boson,
hence $2\kappa_{R}$ prefers to be larger than $\kappa_{L}$
to compensate this effect. However, this
modification is not significant because $\epsilon_{1}$
depends on the
Higgs boson mass only logarithmically \cite{bar1}.

If there is a heavy
Higgs boson ($m_H > m_t$), then it should be integrated out
from the full theory  and its effect in the chiral Lagrangian is
manifested through the effective couplings of the top quark to gauge
bosons.
In this case we simply
subtract the Higgs boson contribution from the
SM results obtained in Refs.~\cite{bar,bar1}.
Fig.~9 of Ref.~\cite{ehab}
showed the allowed area in the $\kappa_L$ and $\kappa_R$ plane
for a 175\,GeV top quark in such models.
Again we find no noticeable difference between the results from these models
and those with a light Higgs boson. That is because
up to one loop level neither $\epsilon_1$
nor $\epsilon_b$ is sensitive
to the Higgs boson contribution \cite{bar,bar1}.

\begin{figure}
\caption{
A comparison between our model and the model in Ref.~7.
The allowed regions in both models are shown on the plane of
$\kln$ and $\krn$, for $m_{t}=150$\,GeV.}
\label{f10o}
\end{figure}

If we consider a new
symmetry-breaking scenario without a fundamental scalar such
as a SM Higgs boson,
following the previous discussions we again find
negligible effects on the allowed range of $\kappa_L$ and $\kappa_R$.

What we learned is that
to infer a bound on the Higgs boson mass
from the measurement of the effective couplings of the top quark to gauge
bosons, we need very precise measurement of the parameters $\kappa_L$
and $\kappa_R$.
However, from the correlations between the effective
couplings {\mbox {($\kappa$'s)}}
of the top quark to gauge bosons, we can infer if
the symmetry-breaking sector is due to a Higgs boson or not, {\it i.e.},
we may be able
to probe the symmetry-breaking mechanism in the top quark system.
Further discussion will be given in the next section.

Finally, we would like to compare our results with those in
Ref.~\cite{pczh}. Fig.~\ref{f10o} shows the most general allowed region for
the couplings $\kln$ and $\krn$, {\it i.e.}, without imposing any relation
between $\kln$ and $\klc$. This region is for top mass 150\,GeV and is
covering the parameter space $-1.0 \leq \kln \, , \krn \leq 1.0 $. We find
\begin{eqnarray}
-0.3 &\leq& \kln \leq 0.6 \, ,\nonumber \\
-1.0 &\leq& \krn \leq 1.0 \nonumber \, .
\end{eqnarray}
Also shown on Fig.~\ref{f10o} the allowed regions from our model and
the model in Ref.~\cite{pczh}.
The two regions overlap in the vicinity of the origin (0, 0) which
corresponds to the SM case.
As $\kln \geq 0.1$, these two regions diverge and become separable.
One notices that the allowed range predicted in Ref.~\cite{pczh} lies
along the line $\kln=\krn$ whereas in our case the slope is different
$\kln =2\krn$. This difference comes in because of the assumed dependence
of $\klc$ on the other two couplings $\kln$ and $\krn$. In our case
$\klc =\kln/2$, and in Ref.~\cite{pczh} $\klc=0$.

Note that for $m_{t}\leq 200$\,GeV the allowed region of
{\mbox {$\kappa$'s}}
in all models of symmetry-breaking should overlap
near the origin because the SM is consistent with low energy data at
the 95\% CL If we imagine that any prescribed dependence between the
couplings corresponds to a symmetry-breaking scenario, then, given the
present status of low energy data, it is
possible to distinguish between different scenarios if $\kln$, $\krn$ and
$\klc$ are larger than 10\%. Better future measurements of
{\mbox {$\epsilon$'s}} can further discriminate between different
symmetry-breaking scenarios. We will discuss how the SLC
and the NLC can contribute to these measurements in Sec. 5,
and defer the discussions on hadron colliders (such as the Tevatron and
the LHC) in the third lecture.
Before that, let us examine a specific model that predicts certain relations
among the coefficients $\klc$, $\krc$, $\kln$ and $\krn$
of the effective couplings of the top quark to gauge bosons.

\section{\twelvebf  Heavy Higgs Limit in the SM }

The goal of this study
is to probe new physics effects, particularly the effects due
to the symmetry-breaking sector, in the top quark system by examining
the couplings of top quark to gauge bosons.
To illustrate how a specific symmetry-breaking mechanism might affect
these couplings, we consider in this section the Standard Model with a
heavy Higgs boson ($m_H > m_t$) as the full theory,
and derive the effective couplings
$\kln$, $\krn$, $\klc$, and $\krc$ at the top quark mass scale in the effective
Lagrangian after integrating out the heavy Higgs boson field.

Given the full theory (SM in this case), we can perform matching between
the underlying theory and the effective Lagrangian.
In this case, the heavy Higgs boson mass
acts as a regulator (cutoff) of the effective theory\cite{long}.

While setting $m_{b}=0 $, and
only keeping the leading terms of the order
${m_t^2}\ln{m_H^2}$,
we find the following effective couplings
\beq
 t-t-Z:\,\,
\frac{g}{4c}\frac{G_F}{2\sqrt{2}{\pi}^2}\left ( \frac{-1}{8}{m_t^2}
 \gamma_{\mu}(1-\gamma_{5})
  +\frac{1}{8}{m_t^2}\gamma_{\mu}(1+\gamma_{5}) \right )
  \ln \left(\frac{m_H^2}{m_t^2}\right)\, ,
\enq
\beq
 t-b-W:\,\,\frac{g}{2\sqrt{2}}
\frac{G_F}{2\sqrt{2}{\pi}^2}\left(\frac{-1}{16}\right){m_t^2}
\gamma_{\mu}(1-\gamma_{5})\ln\left(\frac{{m_H^2}}{m_t^2}\right)\, .
\enq
{}From this we conclude
\beq
\kappa_L^{\rm {NC}}=2\kappa_L^{\rm{CC}}= \frac {G_F}{2\sqrt{2}{\pi}^2}
 \left(\frac{-1}{8}\right){m_t^2}
 \ln\left(\frac{m_H^2}{m_t^2}\right)\, ,
\enq
\beq
\kappa_R^{\rm{NC}}=\frac{G_F}{2\sqrt{2}{\pi}^2}
\frac{1}{8}{m_t^2}
 \ln\left(\frac{m_H^2}{m_t^2}\right)\, ,
\enq
\beq
\kappa_{R}^{CC}=0\, .
\enq

 Note that the relation between the left-handed currents
($\kln=2\klc$) agree with our
prediction because of the approximate custodial symmetry
in the full theory (SM) and the fact that vertex \bbz is not modified. The
right-handed currents $\krc$ and $\krn$ are not correlated,
and $\krc$ vanishes for a massless $b$.
Also note an additional relation in the effective Lagrangian
between the left- and right-handed effective
couplings of the top quark to $Z$ boson, {\it i.e.},
\beq
\kln =-\krn \, .
\enq
This means only the axial vector current of \ttz acquires a nonuniversal
contribution but its vector current is not modified.

As discussed in Sec. 2, due to the Ward identities
associated with the photon field
there can be no nonuniversal contribution to either
the \bba or \tta vertex
after renormalizing the fine structure constant $\alpha$.
This can be explicitly checked in this model.
Furthermore, up to the order of $m_t^2\ln{m_H^2}$, the
vertex \bbz is not modified which agrees with the assumption we made in
Sec. 2 that there exist dynamics of electroweak symmetry-breaking so
that neither \brbrz nor \blblz in the effective Lagrangian
is modified at the scale of $m_t$.

{}From this example we learn that the effective couplings of the
top quark to gauge bosons arising from
a heavy Higgs boson are correlated in a specific way: namely,
\beq
\kln=2 \klc=-\krn \, .
\enq
(This relation in general
also holds for models with a heavy scalar which is not necessarily a SM
Higgs boson, {\it i.e.}, the coefficients of the last two terms in
Eq.~(\ref{higg})
can be arbitrary, and are not necessarily $1/2$ and $1/4$, respectively).
In other words, if the couplings of a heavy
top quark to the gauge bosons are measured
and exhibit large deviations from these
relations, then it is likely that the electroweak
symmetry-breaking is not due to the standard Higgs mechanism
which contains a heavy SM Higgs boson.
This illustrates how the symmetry-breaking sector can be probed
by measuring the effective couplings of the top quark to gauge bosons.

\section{\twelvebf  Direct Measurement of the Top Quark Couplings}

In Sec. 3 we concluded that the precision LEP
data can constrain the couplings $\kln$, $\krn$ and $\klc$,
but not $\krc$ (the right-handed charged current).
In this section we examine how to
improve our knowledge on these couplings at the current and future
electron colliders.
(We defer the discussions on hadron colliders in Lecture Three.)

\subsection{\twelveit  At the SLC }

The measurement of the left-right cross section asymmetry $A_{LR}$
in $Z$ production with a longitudinally polarized electron beam
at the SLC provides a stringent test of the SM and is sensitive to new
physics.

Additional constraints on the couplings $\kln$, $\krn$
and $\klc$ can be inferred from  $A_{LR}$ which can be written as
\cite{bar}
\beq
A_{LR} = \frac{2x}{1+x^2} \, ,
\enq
with
\beq
x=1-4s^2(1+\Delta k^{\prime} ) \,\, ,
\enq
\beq
\Delta k^{\prime} = \frac{\epsilon_{3}- c^2\epsilon_{1}}{c^2 - s^2} \, .
\enq
Up to the order ${m_t^2}\ln {\Lambda}^{2}$, only
$\epsilon_1$ contributes. In our model with the approximate custodial
symmetry, {\it i.e.}, $\kln= 2 \klc=\kappa_L$,
the SLC $A_{LR}$ measurement will have a significant influence on the
precise measurement of the nonuniversal couplings of the top quark.
This will decrease the width of the allowed area in the
parameter space of $\kappa_L$ versus $\kappa_R$.
However, SLC data will have no effect on
the length of the allowed region which in our approximation is solely
determined by $\epsilon_b$.
Hence, a more accurate measurement of $\epsilon_b$, {\it i.e.},
$\Gamma (Z \rightarrow b \bar b) $,
is required to further confine the nonuniversal interactions of
the top quark to gauge bosons to probe new physics.
(More details will be given in the next lecture.)

\subsection{\twelveit  At the NLC }

The best place to probe $\kln$ and $\krn$ associated with the
\ttz coupling is at the NLC through $e^- e^+ \ra A, Z \ra t \bar{t}$.
(We use NLC to represent a generic $e^-e^+$ supercollider \cite{nlc}.)
A detailed Monte Carlo study on the measurement of these couplings at the NLC
including detector effects and initial state radiation
can be found in Ref.~\cite{gal}.
The bounds were obtained by studying the
angular distribution and the polarization of the top quark
produced in $e^- e^+$ collisions.
Assuming a 50 $\rm{fb}^{-1}$
luminosity at $\sqrt{S}=500$\,GeV, we concluded that
within a 90\% confidence level, it should be possible to measure
$\kln$ to within about 8\%, while $\krn$ can be known
to within about 18\%.
A $1\,$TeV machine can do better than a $500\,$GeV machine in
determining $\kln$ and $\krn$ because the relative sizes of the
$t_R {(\overline{t})}_R$  and $t_L {(\overline{t})}_L$
production rates become small and the polarization of the $t \bar t$ pair
is purer. Namely, it is more likely to produce either
a $t_L {(\overline{t})}_R$ or a $t_R {(\overline{t})}_L$ pair.
A purer polarization of the $t \bar t$ pair makes $\kln$ and $\krn$
better determined. (The purity of the $t \bar t$ polarization
can be further improved by polarizing the electron beam.)
Furthermore, the top quark is
boosted more in a $1\,$TeV machine thereby allowing a better
determination of its polar angle in the $t \bar t$ system
because it is easier to find
the right $b$ associated with the lepton to reconstruct the
top quark moving direction.

Finally, we remark that at the NLC $\klc$ and $\krc$ can be studied
either from the decay of the top quark pair or from the single-top
quark production process, $W$-photon fusion
process $e^{-}e^{+}(W\gamma) \ra t X $,
or $ e^{-}\gamma (W\gamma) \ra {\bar t} X$, which is similar to the
$W$-gluon fusion process in hadron collisions.

\section{\twelvebf Discussions and Conclusions}

In this lecture we have applied the electroweak chiral Lagrangian to probe
new physics beyond the SM through studying the couplings of the top quark
to gauge bosons.
We first examined the precision LEP data to extract the information on these
couplings, then we discussed how to improve our knowledge at current and
future electron colliders such as at the SLC and the NLC.

Because of the non-renormalizability of the electroweak chiral Lagrangian we
can only estimate the size of these nonstandard couplings by studying the
contributions to LEP observables at the order of
${m_t^2}\ln{\Lambda}^{2}$, where $\Lambda$ ($= 4 \pi v \sim 3$\,TeV)
is the cutoff scale of the effective Lagrangian.
Already we found interesting constraints on these couplings.

Assuming \bbz vertex is not modified, we found that $\kln$ is already
constrained to be $-0.3 < \kln < 0.6$ ($-0.2 < \kln < 0.5$)
by LEP data at the 95\% CL.
for a 150 (175)\,GeV top quark.
Although $\krn$ and $\klc$ are allowed to be in the full
range of $\pm 1$, the precision LEP data do impose some correlations among
$\kln$, $\krn$, and $\klc$. ($\krc$ does not contribute to the LEP
observables of interest in the limit of $m_b=0$.)
In our calculations, these nonstandard couplings are only inserted once
in loop diagrams using dimensional regularization.

Inspired by the experimental fact $\rho \approx 1$, reflecting
the existence of an approximate custodial symmetry, we
proposed an effective model to relate $\kln$ and $\klc$.
We found that the nonuniversal
interactions of the top quark to
gauge bosons parameterized by $\kln$, $\krn$, and $\klc$
are well constrained by LEP data, within 95\% CL.
Also, the two parameters $\kappa_L=\kln$ and $\kappa_R=\krn$ are strongly
correlated. (In our model, $\kappa_L \sim 2 \kappa_R$.)

We note that the relations among $\kappa$'s
can be used to test different models of
electroweak symmetry-breaking. For instance, a heavy SM Higgs boson
($m_H > m_t$) will
modify the couplings \ttz and \tbw of
a heavy top quark at the scale $m_t$ such that
$\kln = 2\klc$, $\kln =-\krn$, and $\krc=0$.

It is also interesting to note that the upper bound on the top quark mass
can be raised from the SM bound $m_t < 200$\,GeV to as large as
300\,GeV if new physics occurs.
That is to say, if there is new physics
associated with the top quark, it is
possible that the top quark is heavier
than what the SM predicts, a similar
conclusion was reached in Ref.~\cite{pczh}.

With a better measurement of $A_{LR}$ at the SLC,
more constraint can be set on the correlation between
$\kappa_L$ and $\kappa_R$. To constrain the size of
$\kappa_L$ and $\kappa_R$, we need
a more precise measurement on the partial decay
width $\Gamma (Z \ra b \bar b)$.

Undoubtedly, direct detection of the top quark at
the NLC is crucial to measuring the couplings of
\tbw and \ttz.
The NLC shall be the best machine to measure $\kln$ and $\krn$
from studying the angular distribution and the polarization
of the top quark produced in $e^- e^+$ collision \cite{gal}.

\newpage

\setcounter{section}{0}
\setcounter{subsection}{0}
\centerline{\twelvebf LECTURE TWO:}
\vspace{0.2cm}
\centerline{\twelvebf Heavy Top Quark Effects To Low Energy Data}
\baselineskip=14pt
\centerline{\twelvebf In The EW Chiral Lagrangian}
\vspace{0.8cm}

\def\SUU{$SU(2)_{L}\times U(1)_{Y}$}
\def\beq{\begin{equation}}
\def\enq{\end{equation}}
\def\zz{{\cal Z}}
\def\ww{{\cal W}}
\def\ww3{{\cal W}^{3}}
\def\bb{{\cal B}}
\def\aa{{\cal A}}
\def\gblblzz{{{\overline{b_L}\gamma_{\mu} b_{L}{\cal Z}^{\mu}}\,}}
\def\gbrbrzz{{{\overline{b_R}\gamma_{\mu} b_{R}{\cal Z}^{\mu}}\,}}
\def\gblblaa{{{\overline{b_L}\gamma_{\mu} b_{L}{\cal A}^{\mu}}\,}}
\def\gbrbraa{{{\overline{b_R}\gamma_{\mu} b_{R}{\cal A}^{\mu}}\,}}
\def\gblblbb{{{\overline{b_L}\gamma_{\mu} b_{L}{\cal B}^{\mu}}\,}}
\def\gbrbrbb{{{\overline{b_R}\gamma_{\mu} b_{R}{\cal B}^{\mu}}\,}}
\def\gblblz{{{\overline{b_L}\gamma_{\mu} b_{L}{Z}^{\mu}}\,}}
\def\gblbl{{{\overline{b_L}\gamma_{\mu}\partial^{\mu} b_{L}}\,}}
\def\gbrbr{{{\overline{b_R}\gamma_{\mu}\partial^{\mu} b_{R}}\,}}
\def\zzzz{{\cal Z}_{\mu} {\cal Z}^{\mu}}
\def\wwww{{\cal W}^{+}_{\mu} {{\cal W}^{-}}^{\mu}}
\def\ttzz{{\mbox {$t$-${t}$-${\cal Z}$}\,}}
\def\bbzz{{\mbox {$b$-${b}$-${\cal Z}$}\,}}
\def\kln{\kappa_{L}^{\rm {NC}}}
\def\krn{\kappa_{R}^{\rm {NC}}}
\def\klc{\kappa_{L}^{\rm {CC}}}
\def\krc{\kappa_{R}^{\rm {CC}}}
\def\ttz{{\mbox {$t$-${t}$-$Z$}\,}}
\def\bbz{{\mbox {$b$-${b}$-$Z$}\,}}
\def\tta{{\mbox {$t$-${t}$-$A$}\,}}
\def\bba{{\mbox {$b$-${b}$-$A$}\,}}
\def\tbw{{\mbox {$t$-${b}$-$W$}\,}}
\def\tltlz{{\mbox {$t_L$-$\overline{t_L}$-$Z$}\,}}
\def\blblz{{\mbox {$b_L$-$\overline{b_L}$-$Z$}\,}}
\def\brbrz{{\mbox {$b_R$-$\overline{b_R}$-$Z$}\,}}
\def\tlblw{{\mbox {$t_L$-$\overline{b_L}$-$W$}\,}}
\def\ra{\rightarrow}
\def\centeron#1#2{{\setbox0=\hbox{#1}\setbox1=\hbox{#2}\ifdim
\wd1>\wd0\kern.5\wd1\kern-.5\wd0\fi
\copy0\kern-.5\wd0\kern-.5\wd1\copy1\ifdim\wd0>\wd1
\kern.5\wd0\kern-.5\wd1\fi}}
\def\ltap{\;\centeron{\raise.35ex\hbox{$<$}}{\lower.65ex\hbox{$\sim$}}\;}
\def\gtap{\;\centeron{\raise.35ex\hbox{$>$}}{\lower.65ex\hbox{$\sim$}}\;}
\def\gsim{\mathrel{\gtap}}
\def\lsim{\mathrel{\ltap}}
\def\D0{D\O}

\section{\twelvebf Introduction}
\indent

In Lecture One we studied the general (non-standard) couplings
of the top quark to the electroweak (EW) gauge bosons
in an effective chiral Lagrangian formulated electroweak theory with
the spontaneously broken symmetry $SU(2)_{L}\times U(1)_{Y}/U(1)_{em}$.
We found that from the previously announced LEP data \cite{mele} there were
still considerable rooms allowed to
accommodate such non-standard interactions \cite{ehab}.
The question regarding the origin of
such non-standard interactions is of a great importance,
and was discussed to some extent in the previous lecture.

In this lecture we will concentrate on two main points. The first point is
to take a different approach from that used in the previous lecture
to study the leading contributions of a heavy top quark (in powers
of $m_t$) to low energy observables.
In Lecture One we calculated the leading
corrections of ${\cal O}\left( m_t^2\ln \Lambda^2 \right)$
arising from some non-standard couplings of the top quark to
the EW gauge bosons, parameterized in the chiral Lagrangian.\footnote{
$\Lambda$ is the cutoff scale at which the effective
Lagrangian is valid.}
The set of Feynman diagrams calculated in  Lecture One contained
external gauge boson lines and its internal lines
could be gauge bosons and/or Goldstone bosons,
in addition to fermions, to form a gauge invariant set.
Because the
leading corrections (in powers of $m_t$) are closely related to the
spontaneously symmetry-breaking (SSB) sector,
we expect such corrections to be obtained from the interactions
of the top quark to the Goldstone bosons alone.
We shall develop a formalism to calculate these leading corrections
from the pure scalar sector in the chiral Lagrangian. We show how
to reproduce the results obtained in Lecture One from
a set of Feynman diagrams which only contain scalar boson
and fermion lines, gauge boson lines are however not needed.
The relation between these two sets of Green's functions,
one set is involving only the gauge bosons as external lines and
the other set with external scalar boson lines in the limit
of turning off the weak gauge coupling $g$, was derived
in Ref.~\cite{barxn2} for the SM.
As to be shown in sec. 3, the relation between these
two sets of Green's functions in the chiral Lagrangian
can be easily derived, and its derivation
is far more clear than that
in the SM in which the $SU(2)_L \times U(1)_Y$ symmetry
is linearly realized.
This is another example indicating the power and elegance of the
non-linearly realized chiral Lagrangian approach.

The second point is to update the constraints on the
non-standard couplings of the top quark to the EW gauge bosons using the
new LEP \cite{lep94} and SLC data \cite{sld94}.
The rest of this lecture is organized as follows. Sec. 2 is devoted to
study the large top quark mass contribution (in powers of $m_t$)
to low energy physics through the quantities $\rho$ and
$\tau$ \cite{barxn2}
in the chiral Lagrangian formulation.
In sec. 3 we update the constraints on the
non-standard couplings of the top quark to the EW gauge bosons.
Our previous constraints were given in
Lecture One. Sec. 4 contains some conclusions.

\section{\twelvebf Large Top Quark Mass Effects To Low Energy Physics}
\indent

In the Sec. 2 of Lecture One we have extensively discussed how to construct
a chiral Lagrangian formulated electroweak theory in which the gauge symmetry
$SU(2)_{L}\times U(1)_{Y}$ is non-linearly realized.
In this lecture, we will explore an equivalent but different formulation.

Define
\beq
{{\cal W}_{\mu}^a}=-i{\rm{Tr}}(\tau^{a}\Sigma^{\dagger}
D_{\mu}\Sigma)\,
\enq
and
\beq
{\cal B}_{\mu}=g^{\prime}B_{\mu}\,\, ,
\label{b}
\enq
where
\beq
D_{\mu}\Sigma=\left (\partial_{\mu}-ig\frac{\tau^a}{2}W_{\mu}^a\right )
\Sigma\,\,\, .\enq
$W_\mu^a$ and $B_\mu$ are the gauge bosons associated with the $SU(2)_L$
and $U(1)_Y$ groups, respectively. $g$ and $g'$ are the corresponding
gauge couplings.
These fields transform under G as
\beq
 {{\cal W}_{\mu}^3}\rightarrow {{{\cal W}^{\prime}}_{\mu}^3}
       ={{\cal W}_{\mu}^3}-\partial_{\mu}y\, ,
\enq
\beq
{{\cal W}_{\mu}^\pm}\rightarrow {{{\cal W}^{\prime}}_{\mu}^\pm}
  =e^{\pm iy}{{\cal W}_{\mu}^\pm}\, ,
\enq
\beq
{\cal B}_{\mu} \ra {\cal B}^{\prime}_{\mu} =
 {\cal B}_{\mu}+\partial_{\mu}y\,
\label{bb}\, ,
\enq
where
\beq
{\cal W}_{\mu}^{\pm}={\frac{{\cal W}_{\mu}^{1}\mp i{\cal W}_{\mu}^{2}}
{\sqrt{2}}}\, .
\enq

Introduce the fields ${\cal Z}_{\mu}$ and ${\cal A}_{\mu}$ as
\beq
\zz_\mu=\ww3_\mu +\bb_\mu
\label{b1}\,\, ,
\enq
\beq
s^2\aa_\mu = s^2\ww3_\mu - c^2 \bb_\mu\, ,
\label{b2}
\enq
where $s^2\equiv\sin^2\theta_W$, and $c^2=1-s^2$.
In the unitary gauge ($\Sigma =1$)
\beq
{\cal W}_{\mu}^a=-gW_{\mu}^a \,\, ,
\enq
\beq
{\zz}_{\mu} =-\frac{g}{c} Z_{\mu}\,\, ,
\enq
\beq
{\aa}_{\mu}=-\frac{e}{s^2}A_{\mu}\,\, ,
\enq
where we have used the relations $e=g s=g' c$,
$W_\mu^3= c Z_\mu + s A_\mu$, and
$B_\mu= -s Z_\mu + c A_\mu$.
The transformations of ${\cal Z}_{\mu}$ and
${\cal A}_{\mu}$ under G are
\beq
{\cal Z}_{\mu}\ra {\cal Z}_{\mu}^{\prime}={\cal Z}_{\mu}\,\, ,
\enq
\beq
{\cal A}_{\mu} \ra {\cal A}_{\mu}^{\prime} ={\cal A}_{\mu} -\frac{1}{s^2}
\partial_{\mu}y \,\, .
\enq
Hence, under G, the fields ${\cal W}_\mu^\pm$ and ${\cal Z}_\mu$
transform as vector fields, but ${\cal A}_\mu$ transforms as a gauge boson
field which plays the role of the photon field $A_\mu$.

Out of the fields defined as above, one may construct
the $SU(2)_L \times U(1)_Y$ gauge invariant interaction
terms in the chiral Lagrangian
\begin{eqnarray}
{\cal L}^B &=&-\frac{1}{4g^2} {{\cal W}_{\mu \nu}^a}
{{\cal W}^{a}}^{\mu \nu}
 -\frac{1}{4{g^\prime}^2} {\cal B}_{\mu \nu}{\cal B}^{\mu \nu}\nonumber \\
&+&\frac{v^2}{4}{\cal W}^{+}_{\mu}{{\cal W}^{-}}^{\mu}+\frac{v^2}{8}
\zz_{\mu}\zz^{\mu}+{\dots }\,\, ,
\label{eq4}
\end{eqnarray}
where
\beq
{\cal W}^{a}_{\mu \nu}=\partial_{\mu}{\cal W}^{a}_{\nu}
-\partial_{\nu}{\cal W}^{a}_{\mu}+\epsilon^{abc}{\cal W}^{b}_{\mu}
{\cal W}^{c}_{\nu} \,\, ,
\enq
\beq
{\cal B}_{\mu \nu}=\partial_{\mu}{\cal B}_{\nu}-\partial_{\nu}{\cal B}_{\mu}
\,\, ,\enq
and where ${\dots}$ denotes other possible four- or higher-
dimensional operators \cite{app,fer}.

It is easy to show that\footnote{
Use ${\cal W}_{\mu}^a \tau^a = -2 i \Sigma^{\dagger} D_\mu \Sigma ~$,
and $[\tau^a,\tau^b]=2 i \epsilon^{abc} \tau^c $.
}
\beq
{\cal W}_{\mu \nu}^a \tau^a=-g\Sigma^{\dagger}W^a_{\mu \nu}\tau^a
\Sigma\,\,
 \enq
and
\beq
{{\cal W}_{\mu \nu}^a} {{\cal W}^{a}}^{\mu \nu}=g^2 W_{\mu \nu}^{a}
{{W^a}^{\mu \nu}}\,\, .
\label{eq05}
\enq
This simply reflects the fact that this kinetic term is not related to
the Goldstone bosons sector, {\it i.e.}, it does not originate from the
symmetry-breaking sector. In other words, if one is
interested in the full loop corrections which include corrections of
the order $g$,
then we cannot relate these corrections (in powers of $g$)
entirely to the pure scalar sector.

The mass terms in Eq.~(\ref{eq4}) can be expanded as
\begin{eqnarray}
\frac{v^2}{4}{\cal W}_{\mu}^{+}{{\cal W}^{-}}^{\mu}
+\frac{v^2}{8}{\cal Z}_{\mu}{{\cal Z}}^{\mu}
&=&\partial_{\mu}\phi^{+}\partial^{\mu}\phi^{-}
+\frac{1}{2}\partial_{\mu}\phi^{3}\partial^{\mu}\phi^{3} \nonumber \\
&+&\frac{g^2v^2}{4}W_{\mu}^{+}{W^{\mu}}^{-}
+\frac{g^2v^2}{8c^2}Z_{\mu}Z^{\mu}+{\dots}\,\, .
\end{eqnarray}
At the tree level, the mass of $W^\pm$ boson is $M_W=gv/2$ and
the mass of $Z$ boson is $M_Z=gv/2c$.

Fermions can be included in this context by assuming that each flavor
transforms under G$ =SU(2)_L\times U(1)_{Y}$ as \cite{pecc}
\beq f\rightarrow {f}^{\prime}=e^{iyQ_f}f \, ,
\enq
where $Q_{f}$ is the electromagnetic charge of $f$.

 Out of the fermion fields $f_1$, $f_2$ (two different flavors),
and the Goldstone bosons matrix field $\Sigma$,
the usual linearly realized fields
$\Psi$ can be constructed. For example, the left-handed
fermions ($SU(2)_L$ doublet) are
\beq
\Psi_{L} = \Sigma F_{L} = \Sigma{f_1\choose {f_2}}_{L} \,
\enq
with $Q_{f_1}-Q_{{f_2}}=1$.
One can easily show that $\Psi_{L}$\,transforms under G linearly as
\beq
\Psi_{L}\rightarrow {\Psi}^{\prime}_{L}={\rm g} \Psi_{L}\, ,
\enq
where $ {\rm g}=
{{\rm {exp}}}(i\frac{\alpha^{a}\tau^{a}}{2})
{{\rm {exp}}}(i\frac{y}{2})\in {\rm{G}} $.
Linearly realized right-handed fermions
$\Psi_{R}$  ($SU(2)_L$ singlet) simply coincide with $F_{R}$, {\it i.e.},
\beq
\Psi_{R}= F_{R}={f_1\choose {f_2}}_{R}\, .
\enq
It is then straightforward to construct
a chiral Lagrangian containing both the bosonic and the fermionic
fields defined as above .

Our goal is to study the large Yukawa corrections to
the low energy data from the chiral Lagrangian formulated
electroweak theories.
We shall separate the radiative corrections as an expansion in
both the Yukawa coupling $g_t$ and the weak coupling $g$.
($g_t=\sqrt{2} m_t /v$, where $m_t$ is the mass
of the top quark.)
With this separation we can then consider the case of
ignoring the corrections of the order
$g$ as compared to that of $g_t$.
This kind of study has been done in Ref.~\cite{barxn2}
for the SM, where one can
concentrate on the pure scalar sector and treat the gauge bosons as
classical fields so that the full gauge invariance
of the SM Lagrangian is maintained
and a set of Ward identities can be derived to relate the Green's
functions of the Goldstone boson and the gauge boson sectors.
Hence, large $g_t$ corrections can be easily obtained from
calculating Feynman diagrams involving only fermions and
scalar bosons ({\it e.g.}, Goldstone bosons and Higgs boson) but not
gauge bosons.
In the chiral Lagrangian such approach is far more obvious and clear.

Why is the chiral Lagrangian formulation useful in finding large
$g_t$ corrections beyond tree level?
After constructing the gauge invariant bare Lagrangian, we can perform
the necessary loop calculations to any order and
organize all the loop corrections in a compact form,
which possesses a simple $U(1)_{em}$ invariance,
using the composite fields ${\cal W}^{\pm}_{\mu},
{\cal Z}_{\mu}$, and ${\cal A}_{\mu}$.
In principle one needs to
fix a gauge to perform loop calculations.
Fixing a gauge will however destroy the gauge invariance of the Lagrangian.
This is true because the gauge fixing term ({\it e.g.}, in $R_{\xi}$ gauge)
will explicitly break gauge invariance.
Since we are interested in  large $g_t$ corrections,
we do not need to consider gauge bosons in loops \cite{barxn2}. This
means that we do not need to fix a gauge and thus we can maintain the
full gauge invariance of the effective
 Lagrangian. This is obvious by observing that the
leading contributions enhanced by powers of $m_t$
are clearly  products of the SSB and have nothing to do with the weak
gauge coupling $g$.
The point is that because the final result can be organized in the
gauge invariant way as described above we can immediately notice
the equivalence between the two sets of
calculations, {\it i.e.}, using the Goldstone
bosons and using the gauge bosons.
{}From the expansion of the field
\beq
{\cal Z}_{\mu}=\frac{2}{v}\partial_{\mu}\phi^3 -\frac{g}{c}Z_{\mu}+...\, ,
\label{ex}\enq
one notices that each gauge boson field has a factor $g$ in front. Hence,
if we are interested in corrections independent of the gauge coupling
$g$, we need only to consider the pure scalar sector.

\subsection{\twelveit Effective Lagrangian}
\indent

To obtain the large contributions of the top quark
mass (in powers of $m_t$) to low energy data, we need only to
concentrate on the top-bottom fermionic sector
(with $f_1=t$ and $f_2=b$)
in addition to the bosonic sector.
The most general gauge invariant chiral Lagrangian can be written as
\begin{eqnarray}
{\cal L}_0&=&i\overline{t}\gamma^{\mu}\left ( \partial_{\mu}
 +i\frac{2s_0^2}{3}{\cal A}_{\mu}\right) t
+i\overline{b}\gamma^{\mu}\left (\partial_{\mu}-i\frac{s_0^2}{3}
{\cal A}_{\mu}\right ) b\nonumber \\
&-&\left (\frac{1}{2}-\frac{2s_0^2}{3}+\kln\right)
\overline{t_{L}}\gamma^{\mu} t_{L}{{\cal Z}_{\mu}}
 -\left ( \frac{-2s_0^2}{3}+\krn\right ) \overline{{t}_{R}}
\gamma^{\mu} t_{R}{{\cal Z}_{\mu}} \nonumber \\
&-&\left( \frac{-1}{2}+\frac{s_0^2}{3}\right)
\overline{b_{L}}\gamma^{\mu} b_{L}{{\cal Z}_{\mu}}
-\frac{s_0^2}{3}\overline{b_{R}}\gamma^{\mu} b_{R}
{{\cal Z}_{\mu}}\nonumber \\
&-&\frac{1}{\sqrt{2}}\left (1+\klc\right ) \overline{{t}_{L}}
\gamma^{\mu} b_{L}
{{\cal W}_{\mu}^+}-\frac{1}{\sqrt{2}}\left (1+{\klc}^{\dagger}\right)
\overline{{b}_{L}}\gamma^{\mu}t_{L}{{\cal W}_{\mu}^-} \nonumber \\
&-&\frac{1}{\sqrt{2}}\krc \overline{{t}_{R}}\gamma^{\mu} b_{R}
{{\cal W}_{\mu}^+}-\frac{1}{\sqrt{2}}{\krc}^{\dagger}
 \overline{{b}_{R}}\gamma^{\mu} t_{R}{{\cal W}_{\mu}^-} \nonumber \\
&-&m_t \overline{t} t +{\dots}
 \label{eq2_n} \,\, ,
\end{eqnarray}
where $\kln$, $\krn$, $\klc$, and $\krc$
parameterize possible deviations from the SM predictions~\cite{ehab},
and ${\dots }$ indicates possible
Higgs boson interactions and all possible higher dimensional operators.
Here we have assumed that new physics from the SSB sector might
modify the interactions of the top quark to gauge bosons due to the
heavy mass of the top quark, but the
bare $b$-$b$-$Z$ couplings are not modified in the limit of ignoring
the mass of the bottom quark \cite{ehab}.
The subscript $0$ denotes bare quantities and
all the fields in the  Lagrangian ${\cal L}_0$,
Eq.~(\ref{eq2}), are bare fields.

Needless to say, the composite fields are only used to
organize the radiative corrections in the chiral Lagrangian.
To actually calculate loop corrections one should expand these
operators in terms of the Goldstone boson and the gauge boson fields.
The gauge invariant result of loop calculations can be written
in a form similar to Eq.~(\ref{eq2}). Denoting this effective Lagrangian
as ${\cal L}_{eff}$, then its fermionic sector
takes the following form:
\begin{eqnarray}
{\cal L}_{eff} & = &i Z_b^{L} \gblbl +Z_1\frac{s_0^2}{3}\gblblaa
+\frac{1}{2}\left( Z_{v}^{L}-Z_2\frac{2s_0^2}{3}\right )
\gblblzz\nonumber \\
&+&iZ_b^{R}\gbrbr+Z_3\frac{s_0^2}{3}\gbrbraa
-Z_4\frac{s_0^2}{3}\gbrbrzz +{\dots}
 \label{eq3_n} \, ,
\end{eqnarray}
in which  the coefficient functions $Z_1$, $Z_2$, $Z_{3}$, and $Z_{4}$
contain all the loop results and as in ${\cal L}_0$ all the fields
in ${\cal L}_{eff}$ are bare fields.

In the case of ignoring the corrections of the order $g$ the
effective Lagrangian can be further separated into two
parts: one part has the explicit linear $U(1)_Y$ symmetry
in the unitary gauge, another contains all the radiative
corrections which do not vanish when taking the $g \ra 0$
limit. Namely, in this approximation, we can write
\begin{eqnarray}
{\cal L}_{eff} & = &i Z_b^{L} \gblbl -Z_1\frac{1}{3}\gblblbb
+\frac{1}{2} Z_{v}^{L} \gblblzz\nonumber \\
&+&iZ_b^{R}\gbrbr-Z_3\frac{1}{3}\gbrbrbb +{\dots}
 \label{eq3prime} \, ,
\end{eqnarray}
where
\beq
{\cal B}_{\mu}=s_0^2({\cal Z}_{\mu}-{\cal A}_{\mu})
\,\, ,
\enq
derived from Eqs.~(\ref{b1}) and (\ref{b2}).
Note that as shown in Eqs.~(\ref{b}) and (\ref{bb}) the
field ${\cal B}_{\mu}$ is not composite and
transforms exactly like $B_\mu$.
Comparing Eq.~(\ref{eq3_n}) with (\ref{eq3prime}), we conclude that
the coefficient functions
 $Z_1$, $Z_2$, $Z_{3}$, and $Z_{4}$ must be related and
\beq
Z_2=Z_1\,\, ,
\enq
\beq
Z_4=Z_3\,\, .
\enq
All the radiative corrections to the vertex $b$-$b$-$\phi^3$
in powers of $m_t$ are summarized by
the coefficient function $Z_{v}^{L}$ because, from Eq.~(\ref{ex}),
\beq
\frac{1}{2}Z_{v}^{L}\gblblzz=Z_{v}^{L}{\frac{1}{v}}
{\overline{b_L}}\gamma_{\mu}b_L \partial^{\mu}\phi^3 +{\dots }\,\,.
\enq

Since the effective Lagrangian ${\cal L}_{eff}$
must possess an explicit $U(1)_{em}$ symmetry and under G the
field ${\cal A}_\mu$ transforms as a gauge boson field while
the field ${\cal Z}_\mu$
transforms as a neutral vector boson field,
therefore, based upon the Ward identities in QED
we conclude that in Eq.~(\ref{eq3_n})
\beq
Z_1 = Z_b^{L}\,\, ,
\enq
and
\beq
Z_3=Z_b^{R}\,\, .
\enq
Hence, the effective Lagrangian ${\cal L}_{eff}$ can be rewritten as
\begin{eqnarray}
{\cal L}_{eff} & = &i Z_b^{L} \overline{b_{L}}\gamma^{\mu}\left(
\partial_{\mu}-i\frac{s_0^2}{3}{\cal A}_{\mu}\right ) b_{L}
+ iZ_b^{R} \overline{b_{R}}\gamma^{\mu}\left(
\partial_{\mu}-i\frac{s_0^2}{3}{\cal A}_{\mu}\right ) b_{R}\nonumber \\
&+&\frac{1}{2}\left( Z_{v}^{L}-Z_{b}^{L}\frac{2s_0^2}{3}\right )\gblblzz
-Z_b^{R}\frac{s_0^2}{3}\gbrbrzz +{\dots} \,\, .
\label{eq38}
\end{eqnarray}
This effective Lagrangian summarizes all the loop corrections
in powers of $m_t$ in
the coefficient functions $Z_{b}^{L}$, $Z_{b}^{R}$,
and $Z_{v}^{L}$. Recall that up to here
all the fields in ${\cal L}_{eff}$ are bare fields.
To make use of the effective Lagrangian to extract out the information
on low energy data we prefer to express ${\cal L}_{eff}$ in terms of
the renormalized fields.
After inspecting Eq.~(\ref{eq38}) it is clear that
the kinetic terms associated
with the $b_L$ and $b_R$ fields can be properly normalized
to make the residue of their propagators to be unity by
redefining (renormalizing) the fields $b_L$ and $b_R$ by
${(Z_{b}^L)}^{\frac{-1}{2}} b_{L}$ and
${(Z_{b}^R)}^{\frac{-1}{2}} b_{R}$, respectively.
In terms of the renormalized fields $b_L$ and $b_R$,
${\cal L}_{eff}$ can be rewritten as
\begin{eqnarray}
{\cal L}_{eff} &=& \overline{b_{L}}i\gamma^{\mu}\left(
\partial_{\mu}-i\frac{s_0^2}{3}{\cal A}_{\mu}\right ) b_{L}
+ \overline{b_{R}}i\gamma^{\mu}\left(
\partial_{\mu}-i\frac{s_0^2}{3}{\cal A}_{\mu}\right ) b_{R}\nonumber \\
&+&\frac{1}{2}\left( \frac{Z_{v}^{L}}{Z_b^L} -
\frac{2s_0^2}{3}\right )\gblblzz
-\frac{s_0^2}{3}\gbrbrzz +{\dots} \,\,\, .
\label{eq39}
\end{eqnarray}

Before considering the physical observables to low energy data
let us first examine the bosonic sector.
Similar to our previous discussions,
loop corrections to the bosonic sector can be organized by
the effective Lagrangian
\begin{eqnarray}
{\cal L}^{B}_{eff} &=&-\frac{1}{4g_0^2} {{\cal W}_{\mu \nu}^a}
{{\cal W}^{\mu \nu}}^{a}
 -\frac{1}{4{g_0^\prime}^2} {\cal B}_{\mu \nu}
{\cal B}^{\mu \nu}\nonumber \\
&+&Z^{\phi}\frac{v_0^2}{4}{\cal W}^{+}_{\mu}{{\cal W}^{-}}^{\mu}+
Z^{\chi}\frac{v_0^2}{8}\zz_{\mu}\zz^{\mu}+{\dots }
\label{eq5}\,\, .
\end{eqnarray}
We note that in the above equation we have explicitly
used the subscript
$0$ to indicate bare quantities.
The bosonic Lagrangian in Eq.~(\ref{eq4})
and the identity in Eq.~(\ref{eq05}) imply that
the Yang-Mills terms (the first two terms in ${\cal L}^B$)
are not directly related to the SSB sector.
Hence, any radiative corrections to the field ${W}_{\mu \nu}^a$
must know about the weak coupling $g$, {\it i.e.}, suppressed by $g$ in our
point of view. This also holds for operators, of dimension
four or higher, including ${W}_{\mu \nu}^a$ in the chiral Lagrangian
formulated electroweak theories
where these gauge invariant terms are all suppressed by the
weak coupling $g$ ~\cite{app,fer}.
The same conclusion applies to $B_{\mu \nu}$.
Therefore we conclude that the fields ${\cal W}_{\mu}^{\pm},
{\cal Z}_{\mu}$, and ${\cal A}_{\mu}$ in
${\cal L}_{eff}$ and ${\cal L}^B_{eff}$ do not get wavefunction corrections
(renormalization) in the limit of ignoring all the corrections
of the order $g$, namely the renormalized fields and the
bare fields are identical in this limit.

Expanding the mass terms in Eq.~(\ref{eq5}) we find
\begin{eqnarray}
Z^{\phi}\frac{v_0^2}{4}{\cal W}^{+}_{\mu}{{\cal W}^{-}}^{\mu}+
Z^{\chi}\frac{v_0^2}{8}\zz_{\mu}\zz^{\mu}
&=&Z^{\phi}\partial_{\mu}\phi^{+}\partial^{\mu}\phi^{-}
+\frac{1}{2}Z^{\chi}\partial_{\mu}\phi^{3}
\partial^{\mu}\phi^{3}\nonumber \\
&+& Z^{\phi}\frac{g_0^2v_0^2}{4}W_{\mu}^{+}{W^{-}}^{\mu}
+Z^{\chi}\frac{g_0^2v_0^2}{8 c_0^2}Z_{\mu}Z^{\mu}
+{\dots}\,\, .
\label{eq5prime}
\end{eqnarray}
It becomes clear that $Z^{\phi}$ denotes the self energy corrections of
the charged Goldstone boson $\phi^{\pm}$, and $Z^{\chi}$ denotes
the self energy corrections of the neutral Goldstone boson $\phi^3$.
Since $W^{\pm}_{\mu}$ and $Z_{\mu}$ do not get wavefunction correction in
powers of $m_t$, therefore the gauge boson masses are
\begin{eqnarray}
M_W^2 & = &Z^{\phi} \frac{g_0^2 v_0^2}{4} =Z^{\phi} {M_W}_{0}^2 \,\, ,
\nonumber \\
M_Z^2 & = & Z^{\chi} \frac{g_0^2 v_0^2}{4c_0^2}=Z^{\chi} {M_Z}_{0}^2 \,\, .
\label{wzmass}
\end{eqnarray}

In summary, all the loop corrections in powers of $m_t$ to low energy data
can be organized in the sum of ${\cal L}_{eff}$
(in Eq.~(\ref{eq39})) and ${\cal L}^B_{eff}$ (in Eq.~(\ref{eq5})).
Comparing them to the bare Lagrangian ${\cal L}_0$
in Eq.~(\ref{eq2}), we found that in the limit of taking $g \ra 0$ the
chiral Lagrangian ${\cal L}_0$ behaves as a renormalizable theory
although in general a chiral Lagrangian is non-renormalizable.
In other words, no higher dimensional operators (counterterms)
are needed to renormalize the theory in this limit.
The same feature was also found in another application of
a chiral Lagrangian with $1/N$ expansion \cite{largen}.

\subsection{\twelveit Renormalization}
\indent

Now we are ready to consider the large $m_t$ contributions
to low energy data. We choose our renormalization scheme
to be the $\alpha$, $G_F$, and $M_Z$ scheme. With
\beq
g_{0}^2=\frac{4\pi \alpha_0}{s_0^2}\,\,
\enq
and
\beq
s_0^2c_0^2=\frac{\pi \alpha_0}{\sqrt{2}{G_F}_{0}{M_Z^2}_{0}} \,\, ,
\enq
or,
\beq
s_0^2=\frac{1}{2}\left[1-\left (1-\frac{4\pi\alpha_0}
{\sqrt{2}{G_F}_{0}{M_Z^2}_{0}}\right )^{1/2}\right]\,\, .
\enq
Define the counterterms as
\begin{eqnarray}
\alpha &=& \alpha_0 + \delta \alpha \nonumber \,\, ,\\
G_F &=& {G_F}_0 + \delta G_F  \nonumber \,\, ,\\
M_Z^2 &=& {M_Z^2}_0 + \delta M_Z^2 \,\, ,
\end{eqnarray}
and
\begin{eqnarray}
s^2 &=& s_0^2 + \delta s^2 = s_0^2-\delta c^2
\nonumber \,\, ,\\
c^2 &=& c_0^2 + \delta c^2 \,\, ,
\end{eqnarray}
then
\beq
s^2c^2+(c^2-s^2) \,  \delta c^2=
\frac{\pi \alpha}{\sqrt{2}{G_F}{M_Z^2}} \left(
1 - \frac{\delta \alpha}{\alpha} + \frac{\delta G_F}{G_F}
 +\frac{\delta M_Z^2}{M_Z^2} \right)
\label{sinw} \,\, .
\enq
As shown in the above equation,
even after the counterterms $\delta \alpha$, $\delta G_F$,
and $\delta M_Z^2$
are fixed by data ({\it e.g.}, the electron (g-2), muon lifetime, and the
mass of the $Z$ boson), we still have the freedom to choose what
$\delta c^2$ is by defining differently the renormalized quantity $s^2c^2$.
In our case we would  choose the definition of the
renormalized $s^2$ such that there will be no large top quark mass
dependence (in powers of $m_t$) in the counterterm $\delta c^2$.
We shall show later that for this purpose our renormalized $s^2$
will satisfy\footnote{
If we define
$ s'^2c'^2 = \frac{\pi \alpha}{\sqrt{2}{G_F}{M_Z^2}} $, then
$s^2=s'^2 ( 1 + \Delta k')$ with
$ \Delta k'= \frac{- c'^2 \delta \rho }{c'^2 - s'^2} $,
and the counterterm of $s'^2$ will contain contributions in
powers of $m_t$.}
\beq
s^2c^2 = \frac{\pi \alpha}{\sqrt{2}{G_F}{M_Z^2} \rho} \,\, ,
\enq
where $\rho$ is defined from the partial width of $Z$ into
lepton pair, cf. Eq.~(\ref{widthmu}).
With this choice of $s^2$ and the definition of the renormalized
weak coupling
\beq
g^2=\frac{4 \pi \alpha}{s^2} \,\, ,
\enq
one can easily show that the counterterm $\delta g^2$
($=g^2-g_0^2$) does not contain large $m_t$ dependence either.
(Obviously, $\delta \alpha$ will not have contributions
purely in powers of $m_t$.)
Namely, in this renormalization scheme, $\alpha$, $g$,
and $s^2$ do not get renormalized after ignoring all the contributions
of the order $g$.
Hence, all the bare couplings $g_0$, $g_0'$, and $s_0^2$ in
the effective Lagrangians ${\cal L}_{eff}$ and ${\cal L}^B_{eff}$
do not get corrected when considering the contributions which
do not vanish in the limit of $g \ra 0$.
The only non-vanishing counterterm needs to be considered
in Eq.~(\ref{eq5}) is $\delta v^2$ ($=v^2-v_0^2$),
which can be obtained by noting that  neither $g$ nor ${\cal W}^\pm$
(or $W^\pm$) gets renormalized.
Therefore, from Eq.~(\ref{wzmass}),
\beq
Z^\phi v_0^2 = v^2 \,\, .
\enq
Thus
\beq
{G_F}_0=\frac{1}{\sqrt{2} v^2_0}=
Z^{\phi} \frac{1}{\sqrt{2} v^2} = Z^{\phi} G_F \,\, .
\enq
Consequently,
\beq
\frac{g_0^2}{c_0^2}=\frac{8{G_F}_{0}{M_Z^2}_{0}}{\sqrt{2}}
=\frac{8{G_F}{M_Z^2}}{\sqrt{2}}\frac{Z^\phi}{Z^{\chi}}\,\,
\enq
and the effective $Z$-$b$-$b$ coupling is
\beq
-\frac{g_0}{2c_0}\gamma^{\mu}\left[ \left(
\frac{Z_{v}^L}{Z_{b}^{L}} -\frac{2s_0^2}{3}
\right)P_L -\frac{2s_0^2}{3}P_R\right]=\\
-{\sqrt{\frac{G_FM_Z^2}{2\sqrt{2}}\frac{Z^{\phi}}{Z^{\chi}}}}
\gamma^{\mu}\left[ \left(\frac{Z_{v}^L}{Z_{b}^{L}}
-\frac{4s^2}{3}\right) -\frac{Z_{v}^L}{Z_{b}^{L}}\gamma_5\right] \,\, ,
\label{effzbb}
\enq
where $P_{L,R}=(1\mp\gamma_5)/{2}$.

\subsection{\twelveit Low Energy Observables}
\indent

In general, all the radiative corrections to low energy data can be
categorized in a model independent way into four parameters:
$S$, $T$, $U$ \cite{pesk}, and $R_b$ \cite{joserb};
or equivalently,
$\epsilon_1$, $\epsilon_2$, $\epsilon_3$, and $\epsilon_b$ \cite{bar}.

  These parameters are derived from four basic measured \mbox{observables},
$\Gamma_{\mu}$\,(the partial decay width of $Z$ into a
$\mu$ pair),
$A_{FB}^{\mu}$\,(the forward-backward asymmetry at the $Z$ peak for
the $\mu$ lepton), $ M_{W} / M_{Z} $ (the ratio of
$W^\pm$ and $Z$ masses),
and $\Gamma_{b}$\,(the partial decay width
of $Z$ into a $b\overline{b}$ pair).
The expressions of these observables in terms of
$\epsilon$'s are given in Ref.~\cite{bar}.
In this lecture we only give the relevant terms in $\epsilon$~'s
which might contain the leading effects in powers of $m_t$
from new physics.
The relations between these two sets of parameters are, to the order
of interest,
\begin{eqnarray}
S &=& \frac{4 s^2}{\alpha(M_Z^2)} \epsilon_3 \,\, , \nonumber \\
T &=& \frac{1}{\alpha(M_Z^2)} \epsilon_1 \,\, , \nonumber \\
U &=& \frac{-4 s^2}{\alpha(M_Z^2)} \epsilon_2 \,\, ,
\end{eqnarray}
and both $R_b$ ($=\Gamma_b / \Gamma_h$)
 and $\epsilon_b$ are measuring the
effects of new physics in the partial decay width of
$Z \ra b \bar b$. ($\Gamma_h$ is the hadronic width of $Z$.)

As shown in Lecture One, both
$\epsilon_1$ and $\epsilon_b$ gain corrections in powers of
$m_t$, and are sensitive to
new physics coming through the top quark~\cite{ehab}.
On the contrary, $\epsilon_2$\, and $\epsilon_3$ do not play any
significant role in our analysis because their dependence on the top mass is
only logarithmic.
Hence,
\begin{eqnarray}
\epsilon_1 &=& \delta \rho +
{\rm{corrections \,\, of \,\, the \,\, order }}\,\, g\, ,
\nonumber \\
\epsilon_b &=& \tau +
{\rm{corrections \,\, of \,\, the \,\, order }}\,\, g\, ,
\nonumber \\
\epsilon_2 &=&
{\rm{corrections \,\, of \,\, the \,\, order }}\,\, g\, ,
\nonumber \\
\epsilon_3 &=&
{\rm{corrections \,\, of \,\, the \,\, order }}\,\, g\, ,
\end{eqnarray}
where $\delta \rho = \rho -1$.
The parameters $\rho$ and $\tau$ are defined from
\begin{eqnarray}
\Gamma_{\mu} & \equiv & \Gamma(Z\ra \mu^{+}\mu^{-})=\rho
\frac{G_F M_Z^3}{6\pi\sqrt{2}}
\left ({g_{\mu}}_V^2+ {g_{\mu}^2}_A \right )
\label{widthmu} \,\, , \nonumber \\
\Gamma_b & \equiv & \Gamma(Z\ra \overline{b}b)=\rho
\frac{G_F M_Z^3}{2\pi\sqrt{2}}
\left ({g_{b}}_V^2+ {g_{b}^2}_A \right )\,\, ,
\label{gammamub}
\end{eqnarray}
where
\begin{eqnarray}
{g_{\mu}}_V & = & \frac{-1}{2} \left( 1-4s^2 \right)
,\,\,\, {g_{\mu}}_A=\frac{-1}{2} \,\, ,  \nonumber \\
{g_{b}}_V & = & \frac{-1}{2} \left(1-\frac{4}{3}s^2+\tau \right)
  ,\,\,\,  {g_{b}}_{A}=\frac{-1}{2} \left( 1+\tau \right) \,\, .
\end{eqnarray}
Hence, comparing to Eq.~(\ref{effzbb}) we conclude
\begin{eqnarray}
\delta\rho & = &\frac{Z^{\phi}}{Z^{\chi}}-1\,\, , \nonumber \\
\tau & =&\frac{Z_{v}^{L}}{Z_{b}^{L}}-1
\,\, .
\label{rhotauall}
\end{eqnarray}

\subsection{\twelveit One Loop Corrections in the SM}
\indent

\begin{figure}
\caption{
The relevant Feynman diagrams, which contribute to $\rho$ and
$\tau$ to the order ${\cal O}\,({m_t^2}\ln {\Lambda}^2)$.}
\label{fig1}
\end{figure}

The SM, being a linearly realized $SU(2)_L \times U(1)_Y$
gauge theory, can be formulated as a chiral Lagrangian
after non-linearly transforming the fields \cite{ehab}.
Applying the previous formalism, we can calculate one loop corrections
of order $m_t^2$ to $\rho$ and $\tau$ for the SM by
taking $\kln=\krn=\klc=\krc=0$ in Eq.~(\ref{eq2}).
These loop corrections can be summarized by the
coefficient functions
$Z^{\chi}$, $Z^{\phi}$, $Z^L_b$, and $Z^L_v$ which are calculated
from the Feynman diagrams shown in Figs.~\ref{fig1}(a),
\ref{fig1}(b), \ref{fig1}(c),
and the sum of \ref{fig1}(d) and \ref{fig1}(e), respectively.
We found
\begin{eqnarray}
Z^{\chi}&=&1+\frac{6m_t^2}{16\pi^2 v^2}\left (
\Delta-\ln m_t^2\right) \nonumber\,\, , \\
Z^{\phi}&=&1+\frac{6m_t^2}{16\pi^2 v^2}\left (\Delta+\frac{1}{2}-
\ln m_t^2\right)\nonumber  \,\, ,\\
Z_{b}^L&=&1+\frac{3m_t^2}{16\pi^2 v^2}\left (
-\Delta+\ln m_t^2 -\frac{5}{6}\right)\,\, ,\nonumber\\
Z_{v}^L&=&1+\frac{3m_t^2}{16\pi^2 v^2}\left (-\Delta+\ln m_t^2 -\frac{3}{2}
\right ) \,\, .
\end{eqnarray}
We note that Fig.~\ref{fig1}(e) arises
from the non-linear realization of the
gauge symmetry in the chiral Lagrangian approach.
Substituting the above results into Eq.~(\ref{rhotauall}), we obtain
\begin{eqnarray}
\delta \rho &=& \frac{3G_F m_t^2}{8\sqrt{2}\pi^2}\,\, , \nonumber \\
\tau &=& \frac{-G_F m_t^2}{4\sqrt{2}\pi^2}
\,\, ,
\end{eqnarray}
which are the established results~\cite{barxn2}.

\section{\twelvebf Updating the Top Quark Couplings to the EW Gauge Bosons}
\indent

In Lecture One we calculated the one-loop corrections
(of order $m_t^2\ln \Lambda^2)$ to $\rho$ and $\tau$
due to the non-standard couplings of the top quark to the EW gauge bosons
by considering a set of Feynman diagrams
with the external massive gauge bosons lines.
In this lecture we show how to reproduce those results
by considering a set of Feynman diagrams which include the pure
Goldstone boson and fermion lines as described in section 2.

Inserting these non-standard couplings in loop diagrams and keeping only
the linear terms in $\kappa$'s,
we found, at the oder of ${m_t}^{2}\ln {\Lambda}^{2}$,
\begin{eqnarray}
Z^{\chi}&=&1+\frac{6m_t^2}{16\pi^2 v^2}\left(2\kln-2\krn\right)
\ln \frac{\Lambda^2}{m_t^2} \nonumber\,\, , \\
Z^{\phi}&=&1+\frac{6m_t^2}{16\pi^2 v^2}\left(2\klc \right)
\ln \frac{\Lambda^2}{m_t^2} \nonumber\,\, ,\\
Z_{b}^L&=&1-\frac{6m_t^2}{16\pi^2 v^2}\klc\ln \frac{\Lambda^2}{m_t^2}
\nonumber \,\, ,\\
Z_{v}^L&=&1-\frac{m_t^2}{16\pi^2 v^2}\left(6\klc-4\kln+\krn\right)
\ln \frac{\Lambda^2}{m_t^2} \,\, .
\end{eqnarray}
Thus the nonstandard contributions to $\rho$ and $\tau$ are
\begin{eqnarray}
\delta \rho &=& \frac{3G_F m_t^2}{2\sqrt{2}\pi^2}\left (\klc-\kln+\krn\right)
\ln \frac{\Lambda^2}{m_t^2} \,\, ,  \nonumber \\
\tau &=& \frac{G_F m_t^2}{2\sqrt{2}\pi^2}\left (-\frac{1}{4}\krn+\kln\right)
\ln \frac{\Lambda^2}{m_t^2}\,\, ,
\label{rhotau}
\end{eqnarray}
which agree with our previous results obtained in Lecture One,
cf. Eqs.~(\ref{cal1}) and (\ref{cal2}).

Based upon the new LEP measurements \cite{lep94},
a global analysis indicates a SM top quark mass to be \cite{alt}
\beq
m_t=165 \pm 12\,\, {\rm{GeV\,\,\, for}}\,\,\, m_H=300 \,\,{\rm{GeV}}\,\,.
\enq
If the SLC measurement is included with LEP measurements, then
\beq
m_t=174 \pm 11\,\, {\rm{GeV\,\,\, for}}\,\,\, m_H=300 \,\,{\rm{GeV}}\,\,,
\enq
because the new SLC measurement of
$A_{LR}$ \cite{sld94} implied a heavier top quark.

Using the new LEP and SLC results we shall update the constraints on the
non-standard couplings of the top quark to the EW gauge bosons.
This can be done
by comparing the new experimental values for $\delta \rho$ and $\tau$ with
that predicted by the SM and the non-standard contributions combined.
In the limit of ignoring the contributions of the order $g$,
the observables $\Gamma_\mu$, $A_{FB}^\mu$, $M_W / M_Z$, and $\Gamma_b$
can all be expressed in terms of the two quantities
$\delta \rho$ and $\tau$.
In addition to Eq.~(\ref{gammamub}), we found
\beq
A_{FB}^{\mu}= \frac{ 3 {g_\mu}_V^2 {g_\mu}_A^2 }
{ \left( {g_\mu}_V^2 + {g_\mu}_A^2 \right)^2 } \,\,
\enq
and\footnote{
In terms of the quantity $\Delta r_w$ defined in Ref.~\cite{bar},
$\frac{M_W^2}{M_Z^2} \left( 1 - \frac{M_W^2}{M_Z^2} \right)
= \frac{ \pi \alpha(M_Z^2) }
{ \sqrt{2} G_F M_Z^2 ( 1-\Delta r_w) } $. For corrections in powers
of $m_t$, $ s^2 \Delta r_w = - c^2 \delta \rho $.
}
\beq
\frac{M_W^2}{M_Z^2} = \rho \, c^2 \,\, .
\enq

Using the minimum set of observables
($\Gamma_\mu$, $A_{FB}^\mu$, $M_W / M_Z$, and $\Gamma_b$),
we constrain the allowed space of $\kappa$'s in
a model independent way without specifying the explicit dynamics
for generating these non-standard effects.
One can also enlarge the set of observables used in the analysis by
including all the LEP measurements and
the measurement of the left-right cross section asymmetry $A_{LR}$
in $Z$ production with a longitudinally polarized electron beam
at the SLC, where
\beq
A_{LR} = \frac{2x}{1+x^2} \, ,
\enq
with
\beq
x=  \frac{ {g_e}_V }{ {g_e}_A } = 1-4 s^2  \,\, .
\enq

Following the same analyses carried out in the Lecture One, we
include both the SM and the non-standard contributions to low energy data.
The SM contributions to $\delta \rho$ and $\tau$ were given in
Ref.~\cite{bar} for various top quark and
Higgs boson masses. Our conclusions however are
not sensitive to the Higgs boson mass \cite{ehab},
as discussed in the previous lecture.

\begin{figure}
\caption{ Two-dimensional projection in the plane of $\krn$ and $\klc$, for
$m_t=175$\,GeV and $m_H=100$\,GeV.}
\label{fig2}
\end{figure}

\begin{figure}
\caption{ Two-dimensional projection in the plane of $\kln$ and $\krn$, for
$m_t=175$\,GeV and $m_H=100$\,GeV.}
\label{fig3}
\end{figure}

\begin{figure}
\caption{ Two-dimensional projection in the plane of $\kln$ and $\klc$, for
$m_t=175$\,GeV and $m_H=100$\,GeV.}
\label{fig4}
\end{figure}

Choosing $m_t=175$\,GeV and $m_H=100$\,GeV, we span the parameter space
defined by $-1.0\lsim\kln\lsim 1.0$, $-1.0\lsim \krn \lsim 1.0$, and
$-1.0\lsim\klc\lsim 1.0$, and compare with the values\footnote{
$\epsilon_1=\delta \rho$, $\epsilon_b=\tau$,
$\epsilon_2=(-9.2 \pm 5.1)\times 10^{-3}$, and
$\epsilon_3=(3.8 \pm 1.9)\times 10^{-3}$.
}\beq
\delta \rho =( 3.5\pm 1.8 )\times 10^{-3}\,\, ,
\enq
and
\beq
\tau=(0.9\pm 4.2)\times 10^{-3}\,\,
\enq
from a global fit \cite{alt} using all the new LEP and SLC data.
For reference, we listed in the following
some of the relevant data, taken from
\cite{alt},
\begin{eqnarray}
{\alpha}^{-1}(M_Z^2) &=&   128.87 \pm 0.12  \nonumber\, ,\\
G_F          & =  & 1.16637(2) \times {10}^{-5} \,\,\,\,
 {\rm{GeV}}^{-2} \nonumber\, ,\\
M_Z          & =  &  91.1899 \pm 0.0044  \,\,\, \rm{GeV} \nonumber \, ,\\
M_W/M_Z      & =  &  0.8814 \pm 0.0021   \nonumber \, ,\\
\Gamma_{\ell}& =  &  83.98 \pm 0.18  \,\,\, \rm{MeV} \nonumber \, ,\\
\Gamma_b     & =  &  385.9 \pm 3.4  \,\,\,\rm{MeV}   \nonumber \, ,\\
A^{\ell}_{FB}       & =  &  0.0170 \pm 0.0016 \nonumber \, ,\\
A^{b}_{FB}       & =  &  0.0970 \pm 0.0045  \nonumber \, ,\\
A_{LR} \, ({\rm SLC}) \,       & =  &  0.1668 \pm 0.079 \nonumber \, .
\end{eqnarray}
We found that within $2\sigma$ the allowed
region of these three parameters exhibits the same features as that
obtained using the old set of data (see Lecture One).
These features can be deduced from the two-dimensional projections of
the allowed parameter space shown in Figs.~\ref{fig2}, \ref{fig3},
and \ref{fig4}.
\begin{itemize}
\item[(1)]
As a function of the top quark mass, the allowed parameter space shrinks
as the top quark mass increases.
\item[(2)]
Data do not exclude possible new physics coming through
the top quark coupling to the EW gauge bosons.
As shown in Fig.~\ref{fig2},
 $\klc$ and $\krn$ are not yet constrained by the current data.
 Furthermore, no conclusion can be drawn regarding
$\krc$ since $\delta\rho$ and $\tau$ are independent of $\krc$ at the
order of $m_t^2$. Also we notice from
\item[(3)]
$\kln$ is almost constrained. New physics prefers
positive $\kln$, see Figs.~\ref{fig3} and \ref{fig4}. For example, $\kln$ is
constrained within ($-0.3$ to 0.5) for a 175 GeV top quark.
\item[(4)]
New physics prefers $\klc\approx -\krn$. This is clearly shown
in Fig.~\ref{fig2}.
\end{itemize}

As compared with the old set of data from LEP and SLC,  new data
tighten the allowed region of the non-standard parameters $\kappa$'s by
no more than a factor of two. This difference is due to the slightly smaller
errors on the new measurements as compared with the old ones.
The largest impact of these new data on our results
comes from the more precise measurement of
$\Gamma_b$  which turns out to be about
$2\sigma$ higher than the SM prediction and implies a lighter
top quark. For a much heavier top quark, new physics must come in because
all the $\kappa$'s cannot simultaneously vanish.
If the large discrepancy between LEP and SLC data persists, then
our model of having non-standard top quark couplings to the
gauge bosons is one of the candidates that can accommodate
such a difference.

If we restrict ourselves to the minimum set of observables,
which give \cite{alt}
\beq
\delta \rho =(4.8\pm 2.2)\times 10^{-3}\,\, ,
\enq
\beq
\tau=(5.0\pm4.8)\times 10^{-3}\,\, ,
\enq
we reach almost the same conclusion. The
main difference is that $\kln$ shifts slightly to the right,
due to the fact that the central value of
$\tau$ in this case is larger than its global fit value.

\begin{figure}
\caption{The allowed region of
$\kln$ and $\krn$ ($\kln =2\klc$), for
$m_t=150$\,GeV and $m_H=100$\,GeV.}
\label{fig5}
\end{figure}

\begin{figure}
\caption{ The allowed region
 of $\kln$ and $\krn$ ($\kln =2\klc$), for
$m_t=175$\,GeV and $m_H=100$\,GeV.}
\label{fig6}
\end{figure}

\begin{figure}
\caption{The allowed range of $\klc$ as a
function of the mass of the top quark. (Note that $\kln=2\klc$.)}
\label{fig7}
\end{figure}

\begin{figure}
\caption{The allowed range of $\kln - 2 \krn$ as a
function of the mass of the top quark. (Note that $\kln=2\klc$.)}
\label{fig8}
\end{figure}

In Lecture One we discussed an effective model  incorporated with
an additional approximate global custodial symmetry
(responsible for $\rho=1$ at the tree level), and
concluded that $\kln=2\klc$ as long as the tree-level vertex
{\mbox{$b$-$b$-$Z$\,}}is not modified.
{}From Eq.~(\ref{rhotau}), we found for this model
\beq
\delta \rho=\frac{3G_F m_t^2}{2\sqrt{2}\pi^2}\left (-\frac{1}{2}\kln
+\krn\right)\ln \frac{\Lambda^2}{m_t^2}\,\,
\enq
and
\beq
\tau =\frac{G_F m_t^2}{2\sqrt{2}\pi^2}\left (-\frac{1}{4}\krn+\kln\right)
\ln \frac{\Lambda^2}{m_t^2}\,\, .
\enq
Using this effective model, we span the plane
defined by $\kln$ and $\krn$ for top quark
mass 150 and 175 GeV, respectively.
Figs.~\ref{fig5} and \ref{fig6} show the allowed
range for those parameters within
$2\sigma$. As a general feature one observes that the allowed range forms
a narrow area aligned close to the line $\kln=2\krn$. For
$m_t=150$\,GeV (175 GeV) we see that $-0.05\lsim\kln\lsim 0.3$
($0.0\lsim \kln\lsim 0.25$). As the top quark mass increases
this range shrinks and moves downward to the right, away from the origin
$(\kln,\krn)=(0,0)$, although
positive $\kappa$'s remain preferred.
 The reason for this behavior is simply due to
the fact that as $m_t$ increases, the SM value for $\rho$ ($\tau$)
increases in the positive (negative) direction.
To summarize this behaviour, we show, respectively,
in Figs.~\ref{fig7} and \ref{fig8}
the allowed ranges for $\klc$ and $\kln - 2 \krn$ as a function of $m_t$.
An interesting point to mention is that in the global
fit analysis the SM ceases to be
a solution for $m_t \gsim 200$\,GeV. However, with new physics effects,
{\it e.g.}, $\klc\neq 0$, $m_t$ can be as large as 300 GeV.

In this analysis we concentrated on the physics at the $Z$
resonance, {\it i.e.}, at LEP and SLC.
Other lower energy observables may as well be used to constrain
the non-standard couplings of the top quark to the gauge bosons. In
Ref.~\cite{fuj} a constraint on the right-handed charged current
($\krc$) was set using the CLEO measurement of $b\ra s\gamma$. The authors
concluded that $\krc$ is well constrained to within a few percent
from its SM value ($\krc=0$).
This provides a complementary information to our result
because LEP and SLC data are
not sensitive to $\krc$ as compared to $\klc$, $\kln$, and $\krn$.

\section{\twelvebf Conclusions}
\indent

Because top quark is heavy, close to the symmetry-breaking scale,
it will be more sensitive than the other light fermions
to new physics from the SSB sector.
Concentrating on the effects to the low energy data directly
related to the SSB sector, we applied the chiral Lagrangian
formalism in Lecture One to examine whether
the non-standard couplings
($\kappa$'s) of the top quark to the gauge bosons ($W^\pm$ and $Z$).
were already strongly constrained by
the old (1993) data from LEP and SLC.
 Surprisingly, we found that to the order of $m_t^2 \ln \Lambda^2$
only the left-handed neutral current ($\kln$) is somewhat constrained
by the precision low energy data.
The precision data did impose some correlations among $\kln$, $\krn$,
and $\klc$. Since $\krc$ does not contribute to the LEP or SLC
observables in
the limit of taking $m_b = 0$, therefore
$\krc$ cannot be constrained by these data. It is however
strongly constrained by the complementary process
$b \ra s \gamma$ \cite{fuj}.

In Lecture One we obtained our results by considering a set
of Feynman diagrams, derived form the non-linear chiral Lagrangian,
whose  external lines were the gauge boson lines.
The leading corrections (in power of $m_t$) to the low energy
observables were found not to vanish in the limit of turning
off the weak coupling $g$ because they originated from being strongly
coupled to the SSB sector, {\it e.g.}, through large Yukawa coupling $g_t$.
Therefore, our previous results should be able to
be reproduced by considering an
effective Lagrangian in which all the gauge boson fields are treated
as classical fields, namely, they do not contribute to loop
calculations. All the loop corrections which do not vanish
after taking $g \ra 0$ can be obtained from calculating a set of
Feynman diagrams only involving the unphysical Goldstone bosons
(and probably the Higgs boson) and fermions.
In  sec. 2 of this lecture we discussed how to relate these two
sets of Green's functions for the low energy observables of interest.
We showed that by considering a completely different set of
Green's functions from that been discussed in Lecture One we
obtained exactly the same results.

In sec. 3 we used the new (1994) LEP and SLC data to constrain the
non-standard interactions of the top quark to the EW gauge bosons.
As compared with the old data from LEP and SLC, the new data
tighten the allowed region of the non-standard parameters $\kappa$'s by
no more than a factor of two. This difference is mainly due to
the more precise measurement of
$\Gamma_b$  which turns out to be about
$2\sigma$ higher than the SM prediction and favors a lighter top
quark.
If the large discrepancy between LEP and SLC data persists, then
our model of having non-standard top quark couplings to the
gauge bosons is one of the candidates that can accommodate
such a difference. Positive values for $\kappa$'s are preferred
in an effective model, as discussed in Lecture One, where an approximate
custodial symmetry is assumed.

\newpage

\setcounter{section}{0}
\setcounter{subsection}{0}
\centerline{\twelvebf LECTURE THREE:}
\vspace{0.2cm}
\centerline{\twelvebf Physics of Top Quark}
\baselineskip=14pt
\centerline{\twelvebf At Hadron Colliders}
\vspace{0.8cm}

\def\kln{\kappa_{L}^{NC}}
\def\krn{\kappa_{R}^{NC}}
\def\klc{\kappa_{L}^{CC}}
\def\krc{\kappa_{R}^{CC}}
\def\ttz{{\mbox {\,$t$-${t}$-$Z$}\,}}
\def\bbz{{\mbox {\,$b$-${b}$-$Z$}\,}}
\def\tta{{\mbox {\,$t$-${t}$-$A$}\,}}
\def\bba{{\mbox {\,$b$-${b}$-$A$}\,}}
\def\tbw{{\mbox {\,$t$-${b}$-$W$}\,}}
\def\tbW{{\mbox {\,$t$-${b}$-$W$}\,}}
\def\tltlz{{\mbox {\,$t_L$-$\overline{t_L}$-$Z$}\,}}
\def\blblz{{\mbox {\,$b_L$-$\overline{b_L}$-$Z$}\,}}
\def\brbrz{{\mbox {\,$b_R$-$\overline{b_R}$-$Z$}\,}}
\def\tlblw{{\mbox {\,$t_L$-$\overline{b_L}$-$W$}\,}}
\def\beq{\begin{equation}}
\def\enq{\end{equation}}
\def\pbarp{ \bar{{\rm p}} {\rm p} }
\def\pp{ {\rm p} {\rm p} }
\def\ipb{ {\rm pb}^{-1} }
\def\ifb{ {\rm fb}^{-1} }
\def\stds{\strut\displaystyle}
\def\SST{\scriptscriptstyle}
\def\TT{\textstyle}
\def\ra{\rightarrow}
\def\cro{\cropen{12pt}}
\def\mf{m_f}
\def\mb{m_b}
\def\mt{m_t}
\def\MW2{M^2_W}
\def\MZ{M_Z}
\def\Cw{C_w}
\def\Sw{S_w}
\def\mHn{m_{\SST H^{\SST 0}} }
\def\mh0{m_{\SST h^{\SST 0}} }
\def\mHp{m_{\SST H^{\pm}} }
\def\mA0{m_{\SST A^{\SST 0}} }
\def\mH{m_{\SST H} }
\def\mHs{m^2_{\SST H} }
\def\qq{q_1 \bar{q}_2}
\def\jj{j_1 j_2}
\def\ee{e^+ e^-}
\def\ptW{P_{\SST T}^{\SST W} }
\def\DR{\Delta R}
\def\fL{f_{\SST L}}
\def\yW{y_{\SST W}}
\def\qs{\theta^\ast}
\def\MWW{M_{\SST WW}}
\def\EWQCD{(1.1)~}
\def\eg{${\it e.g.}$}
\def\ie{${\it i.e.}$}
\def\etc{${\it etc}$}
\def\etal{${\it et al.}$}
\def\hatt{ \hat {\rm T} }
\def\ETslash{\not{\hbox{\kern-4pt $E_T$}}}
\def\mynot#1{\not{}{\mskip-3.5mu}#1}
\def\sss{\scriptscriptstyle}
\def\rts{\sqrt{S}}
\def\ra{\rightarrow}
\def\d{{\rm d}}
\def\M {{\cal M}}
\def\qgtb{q' g \ra q t \bar b}
\def\Wgtb{q' g (W^+ g) \ra q t \bar b}
\def\ubdt{q' b \ra q t}
\def\udbt{q' \bar q \ra W^* \ra t \bar b}
\def\Wbt{W^+ b \ra t}
\def\ggtt{q \bar q, \, g g \ra t \bar t}
\def\ttb{t \bar t}
\def\Wt{W t}
\def\gbtW{g b \ra W^- t}
\def\width{\Gamma(t \ra b W^+)}
\def\tevs{Di-TeV}
\def\flong{f_{\rm Long}}
\def\ltap{\;\centeron{\raise.35ex\hbox{$<$}}{\lower.65ex\hbox{$\sim$}}\;}
\def\gtap{\;\centeron{\raise.35ex\hbox{$>$}}{\lower.65ex\hbox{$\sim$}}\;}
\def\gsim{\mathrel{\gtap}}
\def\lsim{\mathrel{\ltap}}
\def\del{\partial }

\section{\twelvebf  Introduction }
\indent

The most important consequence of a heavy top quark
is that to a good approximation it decays as a free quark because its
lifetime is short and it does not have time to
bind with light quarks before it decays \cite{decay}.
Furthermore, because the heavy top quark
has the weak two-body decay $t \ra b W^+$, it will
analyze its own polarization.
Thus we can use the polarization
properties of the top quark
as additional observables to test the SM and
to probe new physics.
In the SM, the heavy top quark produced from the usual
QCD process, at the Born level, are unpolarized.
However, top quarks will have longitudinal
polarization if weak effects are present in their production \cite{qqttew}.
For instance, the top quark produced from the $W$-gluon
fusion process is left-handed polarized.
With a large number of top quark events, it
will be possible to test the polarization effects of the top quarks.

\begin{figure}[h]
\caption{ Diagrams contributing to the QCD production of $\ggtt$ }
\label{ttbAll}
\end{figure}
How to detect a SM top quark pair produced via the QCD processes
$\ggtt$, as shown in Fig.~\ref{ttbAll},
has been extensively studied in the literature \cite{argon}.
In this lecture we would concentrate on how to detect and study
the top quark produced from the single-top quark processes
$\Wgtb$, $\ubdt$, $\gbtW$, and $\udbt$.
For the single-top productions we will only consider the decay mode of
$t \ra b W^+ \ra b \ell^+ \nu$, with $\ell^+=e^+ \,{\rm or}\,\nu^+$.
(The branching ratio for this decay mode is
$\rm{Br}=\frac{2}{9}$.)

The rest of this lecture is organized as follows.
In section 2 we  discuss the production rates of top quarks at hadron
colliders. Following that, we will discuss in sections
3 and 4, respectively, how to measure the mass and
the width of the top quark.
In section 5 we discuss what we have learned about
the couplings of the top quark to the weak gauge bosons and show what
can be improved from measuring the production rate of single-top quark event.
We will also discuss in section 6 the potential of the Tevatron
as a $\pbarp$ collider
to probe CP properties of the top quark by simply measuring the
single-top quark production rate.
Section 7 contains our conclusions.
Throughout this lecture we will use $\mt = 140$\,GeV or $180$\,GeV
as an example of a light or a heavy top quark for our studies.

\noindent
\section{\twelvebf  The Single-Top Production Mechanism}
\indent

In this section we consider the production rate of
a single-top quark
at the Tevatron, the \tevs~(the upgraded Tevatron)
and the LHC (Large Hadron Collider) colliders.
In referring to single-top production, unless stated otherwise,
we will concentrate only on the positive charge mode
({\it i.e.}, only including single-$t$, but not single-$\bar t$).
The colliders we consider are the Tevatron
(a $\pbarp$ collider) with the Main Injector at
$\rts = 2\,$TeV, the Di-TeV (a $\pbarp$ collider) at $4\,$TeV
and the LHC (a $\pp$ collider) at $\rts = 14\,$TeV
with an integrated luminosity of
1\,$\ifb$, 10\,$\ifb$, and 100\,$\ifb$, respectively.\footnote{
In reality, the integrated luminosity can be higher than the
ones used here. For instance, with a couple of years of running
a $2\,$TeV Tevatron can accumulate, say, 10\,$\ifb$ luminosity.
Similarly, it is not out of question to have a $4\,$TeV Di-TeV
to deliver an integrated luminosity of about 100\,$\ifb$.}

\begin{figure}
\caption{ Diagrams for various single-top quark processes.}
\label{newdiag}
\end{figure}

A single-top quark signal can be produced from either the $W$-gluon
fusion process $\Wgtb$ (or $\ubdt$) \cite{sally,wgtb},
the Drell-Yan type process $\udbt$ (also known as
``$W^*$'' production) \cite{cortese},
or the $\Wt$ production via $\gbtW$ \cite{galwt}.
The corresponding Feynman diagrams for these processes are shown in
Fig.~\ref{newdiag}

\begin{figure}
\caption{ Rate in [pb] for $\ggtt$, $\Wgtb$, $\udbt$ and $\gbtW$ at
various energies of $\pbarp$ colliders.}
\label{fig1a}
\end{figure}

\begin{figure}
\caption{ Rate in [pb] for $\ggtt$, $\Wgtb$, $\udbt$ and $\gbtW$ at
various energies of $\pp$ colliders.}
\label{fig1b}
\end{figure}

In Figures~\ref{fig1a} and~\ref{fig1b}
we show the total cross sections of these processes
for the Tevatron, the \tevs~and the LHC energies referred to above.
For reference we include plots of the cross sections of top quarks
as a function of $m_t$ in both the $\pbarp$ collisions,
shown in Figure~\ref{fig1a}, and $\pp$ collisions, shown in
Figure~\ref{fig1b}.
The parton distribution function (PDF) used in our calculation is
the leading order set CTEQ2L \cite{pdf}.
We note that taking the $\Lambda_{\rm QCD}$
value given in CTEQ2L PDF we obtain $\alpha_s(M_Z)=0.127$
which is about 15\% larger than
the value of 0.110 in CTEQ2M PDF \cite{pdf}.
We found that if we rescale the $t \bar t$ production
rates obtained from CTEQ2L PDF with born level amplitudes by
the ratio of $\alpha_s^2(Q, \Lambda_{QCD})$ from CTEQ2M and
and that from CTEQ2L,
which yields 0.7 for $Q=M_Z$, then our total rates are
in good agreement with those obtained using
NLO PDF and NLO amplitudes \cite{qcdtt}, see, for
example, Ref.~\cite{smith}.
Hereafter we shall use the scaled results for our rates.
The constituent cross sections are all calculated at tree
level for simplicity
to study the kinematics of the top quark and its decay products.

To include the production rates for
both single-$t$ and  single-$\bar t$ events
at  $\pbarp$ colliders,
a factor of 2 should be multiplied to the single-$t$ rates shown in
Figures~\ref{fig1a} and~\ref{fig1b}
because the parton luminosity for single-$\bar t$
production is the same as that for single-$t$.
Similarly, at $\pp$ colliders the rates
should be multiplied by $\sim 1.5$
for the center-of-mass energy ($\rts$) of the collider
up to $\sim 4$ TeV,
but almost a factor of two at higher energies (say, $\rts \geq 8$ TeV
up to about 14 TeV)
because the relevant parton luminosities for
producing a single-$t$ and a single-$\bar t$ event in $\pp$
collisions are different.
As shown in Figures~\ref{fig1a} and~\ref{fig1b}
the total rate for single-top production is about the
same at $\pbarp$ and $\pp$ colliders for
 $\rts \geq 8$\,TeV  because the relevant
valence and sea quark parton distributions
are about equal for $100 \,{\rm GeV} \, < m_t < 300$\,GeV.
For smaller $\rts$, up to $\sim 4$ TeV, a $\pbarp$ collider
is preferred over
a $\pp$ collider for heavy top quark production because of
its larger parton luminosities.
Similarly, for $t \bar t$ pair productions at small $\sqrt{S}$,
 the quark initiated process $ q \bar q \ra t \bar t$
is more important than the gluon fusion process
$gg \ra t \bar t$.
At $\rts \sim 8$ to $14$ TeV the $\ttb$ rate
is about the same in $\pbarp$ and $\pp$ collisions
because the $gg \ra t \bar t$ subprocess becomes dominant.

\vspace{4mm}
\begin{table}
\caption{ Rates of the above processes
for $\mt = 140 (180)$ GeV. (Branching ratios are not included here.)
For $\protect\rts = 2\,$TeV and $4\,$TeV we include rates for a $\pbarp$
machine. At $\protect\rts = 14$ TeV the rates are for a $\pp$ machine.
For the single-top rates we only include single-$t$ production.}
\label{trates}
\begin{center}
\begin{tabular}{|l||l|l|l|l|r|}              \hline
            &   \multicolumn{4}{c|}{Cross Section (pb)}  \\ \hline
$\rts$(TeV) & $\ggtt$   & $\qgtb$ (or $\ubdt$)& $\udbt$  & $\gbtW$  \\ \hline
2           & 16(4.5)   & 2(1)     & 0.8(0.3) & 0.3(0.1) \\ \hline
4           & 88(26)    & 11(7)    & 2.1(0.8) & 2.9(1.3) \\ \hline
14          & 1300(430) & 140(100) & 11(4.6)  & 8.8(3.6) \\ \hline
\end{tabular}
\end{center}
\vspace{4mm}
\end{table}

For later reference in this lecture,
we show the rates of the above processes in Table~\ref{trates}
for $\mt = 140 (180)$ GeV. (Branching ratios are not included here.)
For $\rts = 2$ and 4 TeV we include only the rates for a $\pbarp$
machine, whereas at $\rts = 14$ TeV the rates are for a $\pp$ machine.
Again, for the single-top rates we only include $t$ production.

Both in Figures~\ref{fig1a} and~\ref{fig1b} and Table~\ref{trates},
we have given the cross section of single-top quark produced from either
the $\Wgtb$ or $\ubdt$ processes. From now on, we will refer to
this production rate as the rate of the $W$-gluon fusion process.
The single-top quark produced from the $W$-gluon fusion process
involves a very important and not
yet well-developed technique of handling the kinematics of
a {\it heavy} $b$ parton inside a hadron.
Thus the kinematics of the top quark produced from this process can
not be accurately calculated yet.
However, the total event rate of the
single-top quark production via this process
can be estimated using the method proposed in Ref.~\cite{wuki}.
The total rate for $W$-gluon fusion process
involves the ${\cal O}(\alpha^2)$
($2 \ra 2$) process $\ubdt$
plus the ${\cal O}(\alpha^2 \alpha_s)$ ($2 \ra 3$)
process $q' g (W^+ g) \ra q t \bar b $
(where the gluon splits to $b \bar b$)
minus the {\it splitting} piece $g \ra b \bar b \,\otimes\, \ubdt$
in which $b \bar b$ are nearly collinear.
\begin{figure}
\caption{ Feynman diagrams illustrating the subtraction procedure
for calculating the total rate for $W$-gluon fusion:
$\ubdt\,\oplus\, q' g (W^+ g) \ra q t \bar b \,
\ominus\, (g \ra b \bar b \,\otimes\, \ubdt)$. }
\label{feynm}
\end{figure}
These processes are shown diagrammatically in Figure~\ref{feynm}.
The helicity amplitudes and the cross sections
for these processes were given in Ref.~\cite{wgtbnew}.

The splitting piece is subtracted to avoid double counting the regime in
which the $b$ propagator in the ($2 \ra 3$) process closes
to on-shell. This procedure is to resum the large logarithm
$\alpha_s \ln (m_t^2/m_b^2)$ in the $W$-gluon fusion process
to all orders in $\alpha_s$ and include part of the higher order
${\cal O}(\alpha^2 \alpha_s)$ corrections to
its production rate.
($m_b$ is the mass of the bottom quark.)
We note that to obtain the complete ${\cal O}(\alpha^2 \alpha_s)$
corrections beyond just the leading log contributions
one should also include
virtual corrections to the ($2 \ra 2$) process, but
we shall ignore these non-leading contributions in this work.
Using the prescription described as above
we found that the total rate of the $W$-gluon fusion process
is about a $25\%$ decrease as compared to the ($2 \ra 2$) event rate for
$\mt = 140\,(180)$ GeV regardless of
the energy or the  type ({\it i.e.}, $\pp$ or $\pbarp$)
of the machine.
\begin{figure}
\caption{ Rate in [pb] for single-$t$ production: $\ubdt$
$(2 \ra 2)$, $\qgtb$ $(2 \ra 3)$ and
the {\it splitting} piece $g \ra b \bar b \,\otimes\, \ubdt$
in which $b \bar b$ are collinear.  The rates are for
$\pbarp$ colliders. }
\label{fig2a}
\end{figure}
\begin{figure}
\caption{ Rate in [pb] for single-$t$ production: $\ubdt$
$(2 \ra 2)$, $\qgtb$ $(2 \ra 3)$ and
the {\it splitting} piece $g \ra b \bar b \,\otimes\, \ubdt$
in which $b \bar b$ are collinear.  The rates are for
$\pp$ colliders. }
\label{fig2b}
\end{figure}
In Figures~\ref{fig2a} and~\ref{fig2b}
we show the total rate of $W$-gluon fusion  versus $\mt$ with scale
$Q = \mt$ as well as a breakdown of the
contributing processes at the Tevatron, the \tevs~and the LHC.

\begin{figure}
\caption{ Rate of $W$-gluon fusion process versus scale $Q$
for $m_t = 180\,$GeV and $\protect\rts=2$\,TeV. }
\label{scale}
\end{figure}

To estimate the uncertainty in the production rate
due to the choice of the scale $Q$ in evaluating
the strong coupling constant $\alpha_s$ and the
parton distributions, we show
in Figure~\ref{scale} the scale dependence
of the $W$-gluon fusion rate.
As shown in the figure, although the individual rate from
either ($2 \ra 2$), ($2 \ra 3$), or the splitting piece
is relatively sensitive to the choice of the scale,
the total rate as defined by $(2 \ra 2)\, + \, (2 \ra 3) \, - \,
({\rm splitting\,piece})$ only varies by about 30\%
for $M_W/2 < Q < 2 m_t$ at the Tevatron. At the \tevs~and the LHC,
it varies by about 30\% and 10\%, respectively.
Based upon the results shown in Figure~\ref{scale}, we argue
that $Q < M_W/2$ probably is not a good choice of the relevant
scale for the production of the top
quark from the $W$-gluon fusion
process because the total rate rapidly increases
by about a factor of 2 in the low $Q$ regime.
In view of the prescription adopted in calculating the total rate,
the only relevant scales are the top quark mass $m_t$ and the
virtuality of the $W$-line in the scattering amplitudes.
Since the typical transverse momentum
of the quark ($q$) which comes from the initial quark ($q'$)
after emitting the $W$-line is about half of the $W$-boson mass,
the typical virtuality of the $W$-line is about
$M_W/2 \sim 40$\,GeV.
$m_b \sim 5$\,GeV is thus not an appropriate scale
to be used in calculating the $W$-gluon fusion rate using
our prescription.
We note that in the ($2 \ra 2$) process the
$b$ quark distribution effectively contains sums to order
$[\alpha_s \ln(Q/{\mb})]^n$ from $n$-fold collinear
gluon emission, whereas the subtraction term (namely, the splitting piece)
contains only first order
in $\alpha_s \ln({\it Q}/{\mb})$.  Therefore, as
${\it Q} \ra {\mb}$ the
($2 \ra 2$) process picks up only the leading order in
$\alpha_s \ln({\it Q}/{\mb})$ and so gets largely
cancelled in calculating the total rate.
Consequently, as shown in Figure~\ref{scale}, the total rate is about the
same as the ($2 \ra 3$) rate for $Q \ra m_b$.
We also note that at $Q \sim  {M_W}/2$,
the ($2 \ra 2$) and ($2 \ra 3$) processes have about the same rates,
and  as $Q$ increases the ($2 \ra 2$) rate, which effectively contains
sums of $[\alpha_s \ln(Q/{\mb})]^n$, gradually increases
while the ($2 \ra 3$) rate decreases such that the total rate is
not sensitive to the scale $Q$.
It is  easy to see also that the total rates calculated via this prescription
will not be sensitive to the choice of PDF although each individual
piece can have different results from different PDF's, based upon the
factorization of the QCD theory \cite{wuki}.

Another single-top quark production mechanism is the Drell-Yan type process
$\udbt$. As shown in Figures~\ref{fig1a} and~\ref{fig1b},
for top quarks with mass
on the order of 180 GeV the rate for $W^*$ production is about one fourth
that of $W$-gluon fusion at $\sqrt{S}=2$\,TeV. The $W^*$ process becomes
much less important for  heavier top quark.
This is because at higher invariant masses $\hat s$
(for producing heavier top quark) of the $t \bar b$ system,
$W^*$ production suffers the usual $1/{\hat s}$ suppression
in the constituent cross section. However, in the $W$-gluon fusion
process the constituent cross section does not fall off as $1/{\hat s}$
but flattens out asymptotically to $1/M_W^2$.
For colliders with higher energies,
therefore with large range of $\hat s$,
the $W^*$ production mechanism for
heavy top quarks becomes much less important. However, the kinematics of the
top quarks produced from this process are different from those in the
$W$-gluon fusion events.
Moreover, possible new physics may introduce a high mass state
(say, particle $V$) to couple strongly with the $t \bar b$ system such that
the production rate from $q' \bar q \ra W^* \ra V \ra t \bar b$
can largely deviate from the SM $W^*$ rate.\footnote{
This is similar to the speculations made in Ref.~\cite{hill}
for having some high mass resonants in the $t \bar t$ productions.}
We will however not discuss it in details here because
its rate is highly model dependent.

The $W$-gluon fusion process becomes more important for heavier top quark.
Why?
Effectively,  the $W$-gluon fusion process
can be viewed as the scattering of a longitudinal
$W$-boson ($W_L$) with gluon to produce a top quark and a bottom
anti-quark ($W^+_L g \ra t \bar b$) after applying the effective-$W$
approximation \cite{dawson}. For large $\hat s$ this scattering
process is equivalent to ($\phi^+ g \ra t \bar b$) where $\phi^+$
is the corresponding Goldstone boson of the gauge boson $W^+$ due
to the Goldstone Equivalence Theorem \cite{equiv,equivrad}.
Since the coupling of
$t$-$b$-$\phi$ is proportional to the mass of the top quark, the
constituent cross section of the $W$-gluon fusion process
grows like $m_t^2/M_W^2$ when $m_t$ increases.
This explains why the $W$-gluon fusion rate only decreases
slightly as the mass of the top quark increases even
though both the parton luminosity and the available phase space
decrease for a heavier top quark.
In contrast, the $t \bar t$ pair production rate from the QCD processes
decreases more rapidly as $m_t$ increases because the constituent
cross section of $\ggtt$ goes as $1/{\hat s}$ and the phase space
for producing a $t \bar t$ pair is smaller than that for producing
a single-$t$. Therefore, the $W$-gluon fusion process becomes
more important for the production of a heavy top quark.

Before closing this section, we note that
the Effective-$W$ approximation has been the essential tool
used in studying the strongly interacting longitudinal $W$ system to probe the
symmetry breaking sector at the supercolliders such as the LHC \cite{wwww}.
By studying the single-top production from the
$W$-gluon fusion process at the Tevatron,
one can learn about the validity of the
Effective-$W$ approximation prior
to the supercolliders.

\noindent
\section{\twelvebf Measuring the Top Quark Mass}
\indent

By the year 2000, we expect results from the Tevatron (with 10\,$\ifb$)
and results from LEP-200, giving error of $\sim 50$\,MeV on $M_W$.
Due to Veltman's screening theorem,
the low energy data are not sensitive to the mass of the Higgs boson
\cite{veltman}.
For a heavy Higgs boson, they can at most depend on $m_H$ logarithmically
up to the one loop level.
Therefore, within the SM one needs to also know the mass of the
top quark to within $\sim 5$\,GeV to start getting
useful information on $m_H$, with an uncertainty less than
a few hundred GeV, through studying  radiative
corrections to the low energy
data which include LEP, SLC, and neutrino experiments
\cite{mele,alta,lep94,sld94}.
(Of course, $m_H$ will be measured to some better
precision if it is detected in direct production at colliders.)

How accurate can the mass of the top quark be measured at hadron colliders?
At hadron colliders, $m_t$ can be measured in the $\ttb$ events
by several methods \cite{argon}.
\begin{figure}
\caption{ The lepton+jet decay mode of $\ttb$ production. }
\label{decay2}
\end{figure}
The first method is to use the lepton+jet decay mode of
the $\ttb$ pair, as shown in Fig.~\ref{decay2},
 by reconstructing the invariant mass of the three jets in
the opposite hemisphere
from the isolated lepton $\ell$ ($= e \,{\rm or}\, \mu$)
in $t \ra bW (\ra \ell \nu)$,
and requiring that two of the three jets reconstruct to a
$W$ and the third be tagged as a $b$-jet.
\begin{figure}
\caption{ The di-lepton decay mode of $\ttb$ production. }
\label{decay1}
\end{figure}
The second method is to use the di-lepton decay mode of the $\ttb$
pair, as shown in Fig.~\ref{decay1}, by requiring both $W$'s to decay
leptonically and for one of the $b$'s to decay semileptonically
to measure the mass distribution of the non-isolated
lepton $\ell_{b}$ (from $b$ decay) and one of the two
isolated leptons ($\ell_1$ and $\ell_2$ from $W^\pm$ decay)
which is closer to $\ell_{b}$.
The third method is to measure the cross section of the
di-lepton decay mode of the $\ttb$ pair.
At the LHC, there will be about $10^8$ $\ttb$ pairs produced in one
year of running for $\mt < 200$\,GeV.
With such a large number of events, ATLAS and CMS
concluded that $m_t$ can be measured with a precision
of $\leq 5$\,GeV using the first method described above,
and about a factor of 2 improvement using the second method
\cite{atlas,cms}.
A similar conclusion was also drawn by the CDF and the D0 collaborations
for the Tevatron with Main Injector after the upgrade of their
detectors \cite{chip}.  This is remarkable given that
the $\ttb$ cross section at the Tevatron is smaller
by about two orders of magnitude as compared with that at the
LHC, as shown in Figures~\ref{fig1a} and~\ref{fig1b}.

Next, we would like to discuss how to measure the mass of the top
quark in the $W$-gluon fusion process. Why do we care?
After $m_t$ is measured in the $\ttb$ events, we would like to
test whether this is a SM top quark. Thus we have to verify
its mass measured from other processes,
such as in the single-top quark events.
Suppose that $m_t$ is measured from the  methods described above in
$\ttb$ events,
and the coupling of \tbw~is not of the SM nature, then
we would find that the single-top quark production rate of
the $W$-gluon fusion process is different from the SM prediction
because its production rate is directly proportional to the square
of this coupling. (We will discuss more on this point in section 5.)
Hence, without knowing the nature of the \tbw~interactions
one can not use the production rates of the single-top
quark events to measure $m_t$.

Alternatively, we propose two methods to measure $m_t$ in the single-top
quark events. We will refer to them as the fourth and the fifth method.
The fourth method is a slight variation of the second method.
Instead of measuring the invariant mass of the leptons, we propose to
directly measure the invariant mass ($m_{b\ell}$) of the $\ell$ and $b$
in $t \ra b W (\ra \ell \nu)$.
We expect that the efficiency of $b$ tagging
using the displaced vertex is higher  for detecting a heavier top quark,
and the $b$ jet energy measurement
is better for $b$ having  larger transverse momentum from
a heavy top quark decay.
Thus $m_{b\ell}$ can be used to measure
the mass of a SM top quark.
A detailed Monte Carlo study on the detection of
a single-top quark event in hadron collisions was performed
in Refs.~\cite{wgtb,wgtbnew}, in which
various unique features of the kinematics of the
single-top quark and methods in suppressing
backgrounds were discussed. We shall however
not reproduce that study here.
In the $\ttb$ event there are two $b$'s,
therefore this method may not work as well as in the single-top
event which only contains one $b$.
\begin{figure}
\caption{ Distributions of $m_{b\ell}$ (solid) and
$m_{{\bar b}\ell}$ (dash) in $t \bar t$ events for a 180 GeV top quark. }
\label{figttb}
\end{figure}
However it is not entirely impossible to use this method because,
as shown in Figure~\ref{figttb},
the sum of the invariant mass distributions of
$b\ell$ and $\bar{b} \ell$ for a 180 GeV top quark
still show a bump near the region that the distribution of
$m_{b\ell}$ peaks.
(With a larger sample of $t \bar t$ events one might be able to afford
using the electric charge of the soft-lepton from $b$-decay to
separate $b$ from $\bar b$ on an event-by-event basis at the cost of
the small branching ratio of $b \ra \mu + X$, of about 10\%.)
We will explain in more details how to use $\flong$
(the fraction of longitudinal $W$-boson from top quark decay),
derived from the distribution of
$m_{b\ell}$, to measure $m_t$ in section 5.

The fifth method is to reconstruct the invariant mass of the top quark
in the $t \ra b W (\ra \ell \nu)$ decay mode by
measuring  the missing transverse momentum
and  choosing a two-fold solution
of the longitudinal momentum of the neutrino from the mass constraint of
the $W$ boson.
In Refs.~\cite{wgtb,wgtbnew} we concluded that it
is possible to measure $m_t$ using
either of these last two methods to
a precision of 5\,GeV at the Tevatron
($\sqrt{S}=2\,$TeV) with 1\,$\ifb$ integrated luminosity.
We also find that after applying all the kinematical
cuts to suppress the dominant
background  $W+ b \bar b$,
at most $10\%$ of $W^*$ events contribute to the
single-top production for a 140 (180)\,GeV top quark.
The SM $W^*$ production rate is already much smaller than
the $W$-gluon fusion rate for a heavier top quark,
therefore the contribution from the $W^*$ is not important in our
study.

\noindent
\section{\twelvebf Measuring the Top Quark Width}
\indent

As shown in Ref.~\cite{steve} the intrinsic width of the top quark
can not be measured at the high energy hadron collider such as the LHC
through the usual QCD processes.\footnote{
In Ref.~\cite{steve} we studied the effects of the QCD
radiation in top quark decay (at one loop level) to the
measurement of $m_t$ in $t \bar t$ events produced
in hadron collisions. We concluded that
the peak position of the $m_t$ distribution remains
about the same as the tree level result, but the shape
is different. We also found that the $m_{b\ell}$ distribution
is not sensitive to the QCD radiations in top decay.
} For instance,
the intrinsic width of a 150 GeV Standard Model top quark is
about 1 GeV, and the full width at half maximum of the reconstructed
top quark invariant mass (from $t \ra b W (\ra \,jets)$ decay mode)
is $\sim 10$ GeV after including the detector resolution
effects by smearing the final state parton momenta.
Here, the ratio of the measured width and the intrinsic width for a 150 GeV
top quark is about a factor of 10. For a heavier top quark,
this ratio may be slightly improved because the jet energy can be better
measured. (The detector resolution $\Delta E/E$ for a QCD jet with
energy $E$ is proportional to $1/\sqrt{E}$.)
A similar conclusion was
also given from a hadron level analysis presented in the SDC Technical
Design Report which concluded that reconstructing the top quark invariant
mass gave a width of 9 GeV for a 150 GeV top quark \cite{sdc}.
Is there a way to measure the top quark width $\width$, say, within
a factor of 2 or better, at hadron colliders?
Yes, it can in principle be measured in single-top events.

The width $\width$ can be measured by counting the production rate of top
quarks from the $W$--$b$ fusion process which is {\it equivalent}
to the $W$-gluon fusion process by a proper treatment of the bottom
quark and the $W$ boson as partons inside the hadron.
The $W$-boson which interacts with the $b$-quark to produce the top
quark can be treated as an on-shell boson
in the leading log approximation \cite{dawson,effw}.
The result is that even
under the approximations considered,
a factor of 2 uncertainty in the production
rate for this process gives a
factor of 2 uncertainty in the measurement of $\width$.
This is already  much better than what can be measured
from the invariant mass distribution of the jets from
the decay of top quarks in the $t \bar t$ events produced via
the usual QCD processes.
More precisely, as argued in section 2 that the production rate of
single-top at the Tevatron can probably be known within about 30\%,
thus it implies $\width$ can be measured to
about the same accuracy.\footnote{
Strictly speaking, from the production rate of single-top event,
one measures the sum of all the possible partial decay widths, such as
$\Gamma(t \ra b W^+) +\Gamma(t \ra s W^+) +\Gamma(t \ra d W^+) + \cdots$,
therefore, this measurement is really measuring the width of
$\Gamma(t \ra XW^+)$ where $X$ can be more than one particle state
as long as it originates from the partons inside proton (or anti-proton).
In the SM, $\Gamma(t \ra b W^+)$ is about equal to
the total width of the top quark.}
Therefore, this is an extremely important measurement
because it directly tests the couplings of~\tbw.

$W$-gluon fusion can also tell us about the
 Cabibbo-Kobayashi-Maskawa (CKM) matrix element
$|V_{tb}|$. Assuming only three generations of quarks, the constraints
from low energy data together with unitarity of the CKM matrix
require $|V_{tb}|$ to be in 0.9988 to 0.9995 at the 90\% confidence
level~\cite{databook}.
As noted in Ref.~\cite{databook}
the low energy data do not preclude there being
more than three generations of quarks (assuming the same
interactions as described by the SM). Moreover, the entries deduced
from unitarity might be altered when the CKM matrix is expanded to
accommodate more generations. When there are more than three generations
the allowed ranges (at 90\% CL) of the matrix element
$|V_{tb}|$ can be anywhere  between 0 and 0.9995~\cite{databook}.
Since  $|V_{tb}|$  is directly involved in
single-top production via $W$-gluon fusion,
any deviation from SM value in $|V_{tb}|$ will
either enhance or suppress the production rate of single-top events.
It can therefore be measured by simply counting the single-top
event rates. For instance, if the single-top production rate is
measured to within 30\%, then $|V_{tb}|$ is determined to within 15\%.

In conclusion, after the top quark is found,
the branching ratio of $t \ra b W^+(\ra \ell^+\nu)$ can be measured
from the ratio of $(2\ell+\,jets)$ and $(1\ell+\,jets)$
 rates in $t \bar t$ events.
Because the measured single-top quark event rate is equal to the single-top
production rate multiplied by the branching ratio of
$t \ra b W^+(\ra \ell^+\nu)$ for the  $(1 \ell+\,jets)$ mode, and
the same $t$-$b$-$W$ couplings appearing in the decay of $t$
in this process appear also in the production of $t$.
Thus, a model independent measurement of the decay
width~$\width$ can  be made by simply counting the production rate
of $t$ in the $W$-gluon fusion process.
Should the top quark width be found to be different from the SM expectations,
we would then have to look for
non-standard decay modes of the top quark.
We note that it is important to measure at least one
partial width (say, $\width$) precisely in order to discriminate between
different models of new physics, if any. In the SM, the
partial width $\width$
is about the same as the total width of the top quark at the tree level
because of the small CKM matrix element
$|V_{ts}|$, thus measuring the
single-top quark production rate measures the
lifetime of the top quark.

\noindent
\section{\twelvebf Top Quark Couplings to $W$ Gauge Boson }
\indent

It is equally important to ask what kind of interactions the~\tbw
vertex might involve \cite{toppol}.
For instance, one should examine the form factors of~\tbw
which result from  higher order corrections due to
SM strong and/or electroweak interactions.
It is even more interesting to examine these form factors to test the
plausibility of having  {\it nonuniversal} gauge couplings
of~\tbw due to some dynamical symmetry breaking scenario \cite{pecc,ehab}.

The QCD~\cite{qcd} and the electroweak~\cite{elecw} corrections to the
decay process $t \ra b W^+$ in the SM have
been done in the literature.
The most general operators for this coupling are
described by the interaction lagrangian
\begin{eqnarray}
L&=&\ {g\over \sqrt{2}}\left[ W^-_\mu\bar{b}\gamma^\mu
                (f_1^L P_-+f_1^R P_+)t -{1\over M_W}
\del_\nu W^-_\mu\bar{b}\sigma^{\mu\nu}(f_2^LP_-+f_2^RP_+)t \right] \nonumber \\
 & &+ {g\over \sqrt{2}}\left[W^+_\mu\bar{t}\gamma^\mu
      ({f_1^L}^* P_-+{f_1^R}^* P_+)b -{1\over M_W}
      \del_\nu W^+_\mu\bar{t}\sigma^{\mu\nu}
      ({f_2^R}^*P_-+{f_2^L}^*P_+)b \right] \, , \nonumber \\
 & & \qquad \,\,
\label{eqlag}
\end{eqnarray}
where $P_\pm ={1\over 2}(1\pm \gamma_5)$,
$i\sigma^{\mu\nu}=-{1\over 2}[\gamma^\mu,\gamma^\nu]$ and
the superscript $*$ denotes the complex conjugate.
In general, the form factors $f_{1}^{L,R}$ and $f_{2}^{L,R}$ can be complex.
Note that in Eq.~(\ref{eqlag}), if there is a relative phase between
$f^L_1$ and $f^R_2$ or between $f^R_1$ and $f^L_2$, then CP is violated.
For instance, in the limit of $m_b=0$ ,
a CP-violating observable will have a coefficient proportional to
${\rm Im}(f^L_1 {f^R_2}^*)$ for a left-handed bottom quark,
and ${\rm Im}(f^R_1 {f^L_2}^*)$ for a right-handed
bottom quark~\cite{toppol}.
(We will discuss  more on CP violation in section 6.)
If the $W$-boson can be off-shell then there are additional form factors
such as
\beq
\del^\mu W^-_\mu \bar{b}(f_3^LP_-+f_3^RP_+)t
+\del^\mu W^+_\mu \bar{t}({f_3^R}^* P_- + {f_3^L}^* P_+)b \, ,
\label{eqlag2}
\enq
which vanish for an on-shell $W$-boson or when the off-shell $W$-boson
couples to massless on-shell fermions.
Here, we only consider on-shell $W$-bosons for
 $m_t> M_W + m_b$.
At tree level in the SM the form factors are
$f_1^L = 1$ and $f_1^R = f_2^L = f_2^R = 0$.
These form factors will in general affect the
experimental observables related to the top quark, such as
the fraction of longitudinal $W$'s produced in top quark decays.

The fraction ($\flong$) of longitudinally polarized $W$-bosons produced
in the rest frame of the decaying top
quarks strongly depends on the form factors
$f_1^{L,R}$ and $f_2^{L,R}$ \cite{toppol}.
Hence, $\flong$ is a useful observable for
measuring these form factors.
The definition of $\flong$ is simply the
ratio of the number of longitudinally
polarized $W$-bosons produced with respect to the total number of
$W$-bosons produced in top quark decays:
\beq
\flong={\Gamma(\lambda_W=0)
\over{\Gamma(\lambda_W=0)+\Gamma(\lambda_W=-)+\Gamma(\lambda_W=+)}}
\enq
where we use $\Gamma(\lambda_W)$ to refer to the decay rate for a
top quark to decay into a $W$-boson with polarization $\lambda_W$.
($\lambda_W=-, +, 0$ denotes a left-handed, right-handed,
and longitudinal $W$-boson.)
Clearly, the polarization of the $W$-boson depends on the form factors
$f_1$ and $f_2$.\footnote{
$f_1^R$ and $f_1^L$ contribute the same amount of
longitudinal $W$'s in top quark decays \cite{toppol}.}
 Therefore, one can measure the polarization of the
$W$-boson to measure these form factors.
The polarization of the
$W$-boson can be determined by the
angular distribution of the lepton, say, $e^+$ in the rest frame of $W^+$ in
the decay mode $t \ra b W^+ (\ra e^+ \nu)$.
However, the reconstruction of the $W$-boson rest frame (to measure
its polarization) could be a non-trivial matter due to the missing
longitudinal momentum ($P_{\SST Z}$) (with a two-fold ambiguity)
of the neutrino ($\nu$) from $W$ decay.
Fortunately, as shown in Eq.~(\ref{mbe1}),
one can determine the polarization
of the $W$-boson without reconstructing its rest frame by using the
Lorentz-invariant observable $m_{be}$, the invariant mass of
$b$ and $e$ from $t$ decay.

The polar angle $\theta^*_{e^+}$
distribution of the $e^+$ in the rest frame
of the $W^+$ boson whose z-axis is defined to be the moving direction of
the $W^+$ boson in the rest frame of the top quark can be written in terms of
$m_{be}$ through the following derivation:
\begin{eqnarray}
\cos \theta^*_{e^+} &=& {{ E_e E_b - p_e \cdot p_b }\over
{|\vec{\bf p}_e| |\vec{\bf p}_b| }}       \nonumber \\
&\simeq & 1-{p_e \cdot p_b \over E_e E_b}
= 1-{2 m_{be}^2 \over m_t^2 - M_W^2}.
\label{mbe1}
\end{eqnarray}
The energies $E_e$ and $E_b$ are evaluated in the rest frame of
the $W^+$ boson from the top quark decay and are given by
\begin{eqnarray}
E_e &=& {M_W^2+m_e^2-m_\nu^2 \over 2 M_W}, \qquad |\vec{\bf p}_e|=
\sqrt{E_e^2-m_e^2}, \nonumber \\
E_b &=& {m_t^2-M_W^2-m_b^2 \over 2 M_W}, \qquad |\vec{\bf p}_b|=
\sqrt{E_b^2-m_b^2}.
\label{mbe2}
\end{eqnarray}
$m_e$ ($m_\nu$) denotes the mass of $e^+$ ($\nu_e$) for the
sake of bookkeeping.
The first line in Eq.~(\ref{mbe1}) is exact when using Eq.~(\ref{mbe2}),
while the second line of Eq.~(\ref{mbe1}) holds in the limit of $m_b=0$.
It is now trivial to find $\flong$ by first calculating the
$\cos\theta^*_{e+}$ distribution
then fitting it according to the decay amplitudes
of the $W$-boson from top quark decay \cite{toppol}.
In what follows we will show how to use the distribution of
$m_{be}$ to measure the mass of the top quark and
its couplings to the $W$-boson.

In the previous lectures we considered the effective couplings
\beq
 W-t_{L}-b_{L}:\,\, \frac{g}{2\sqrt{2}}\frac{ 1 + \kappa^{CC}_{L}}{2}
 \gamma_{\mu}(1-\gamma_{5})\,
\enq
and
\beq
 W-t_{R}-b_{R}:\,\, \frac{g}{2\sqrt{2}}\frac{ \kappa^{CC}_{R}}{2}
 \gamma_{\mu}(1+\gamma_{5})\,
\enq
derived from an electroweak chiral lagrangian with the symmetry
$SU(2)_{L}\times U(1)_Y$ broken down to $U(1)_{em}$.
(Here, $\kappa_L^{CC}=f_1^L - 1$, and $\kappa_R^{CC}=f_1^R$.)
At the Tevatron and the LHC, heavy top quarks are
predominantly produced from the QCD process
$gg, q \bar q \ra t \bar t$ and the $W$-gluon fusion process
$qg (Wg) \ra t \bar{b}, \bar{t} b$.
In the former process, one can probe $\klc$ and $\krc$ from the decay of the
top quark to a bottom quark and a $W$ boson. In the latter process,
these non-standard couplings can also
be measured by simply counting the production
rates of signal events with a single $t$ or $\bar t$.
Let us discuss them in more details as follows.

\noindent
\subsection{\twelveit From the Decay of Top Quarks}
\indent

To probe $\klc$ and $\krc$ from the decay of the
top quark to a bottom quark and a $W$ boson, one needs to measure the
polarization of the $W$ boson which can be measured from
the distribution of the invariant mass $m_{b\ell}$.
 For a massless $b$, the $W$ boson from top
quark decay can only be either longitudinally or left-handed polarized for
a left-handed charged current ($\krc=0$). For a right-handed
charged current ($\klc=-1$) the $W$ boson can only be either longitudinally
or right-handed polarized.
(Note that the handedness of the $W$ boson is reversed for a massless
$\bar b$ from $\bar t$ decays.)
\begin{figure}
\caption{ For a left-handed $\tbw$ vertex. }
\label{left}
\end{figure}
\begin{figure}
\caption{ For a right-handed $\tbw$ vertex. }
\label{right}
\end{figure}
This is the consequence of helicity conservation, as diagrammatically
shown in Figures~\ref{left} and ~\ref{right} for a polarized
top quark. In these figures we show the preferred
moving direction of the lepton from a polarized $W$-boson in the rest
frame of a polarized top quark for
either a left-handed and a right-handed $\tbw$ vertex.
As indicated in these figures, the invariant mass
$m_{b \ell}$ depends on the polarization of the $W$-boson from the decay
of a polarized top quark.
\begin{figure}
\caption{ $m_{b{\ell}}$ distribution for SM top quark
(solid) and for pure right-hand~$\tbW$ coupling of
$tbW$(dash).}
\label{mbe}
\end{figure}
\begin{figure}
\caption{ $\cos \theta^*_{\ell}$ distribution for SM top quark
(solid) and for pure right-hand~$\tbW$ coupling of
$tbW$(dash).}
\label{thesta}
\end{figure}
Also, $m_{b \ell}$ is preferentially larger for a pure
right-handed $\tbw$ vertex than a pure left-handed one.
This is clearly shown in Figure~\ref{mbe}, in which
the peak of the $m_{b{\ell}}$
distribution is shifted to the right and the distribution falls
off sharply at the upper mass limit for a pure right-handed $\tbw$ vertex.
In terms of  $\cos \theta^*_{\ell}$, their difference is shown in
 Figure~\ref{thesta}.
However, in both cases the fraction ($\flong$) of longitudinal $W$
from top quark decay is enhanced by ${m_t}^2/{2{M_W}^2}$ as compared
to the fraction of transversely polarized $W$ \cite{toppol}, namely,
\beq
\flong = { {m_t^2 \over 2 M_W^2 } \over { 1 + {m_t^2 \over 2 M_W^2 } } } \,.
\enq
 Therefore, for a heavier
top quark, it is more difficult to untangle the $\klc$ and $\krc$
contributions.
On the other hand, because of the very same reason, the mass of a
heavy top quark can be accurately measured from $\flong$
irrespective of the nature of the $\tbw$ couplings
(either left-handed or right-handed).

The QCD production rate of $t \bar t$ is
obviously independent of $\klc$ and $\krc$. (Here
we assume the electroweak production rate of $q\bar{q}\ra A,Z \ra t\bar t$
remains small as in the SM.)
Let us estimate how well the couplings  $\klc$ and $\krc$
can be measured at the Tevatron, the \tevs, and the LHC.
First, we need to know the production rates of the top
quark pairs from the QCD processes.
As shown in Table~\ref{trates},
 the QCD production rate of $gg,q\bar{q}\ra t \bar t$
for a 180\,GeV top quark
is about 4.5\,pb, 26\,pb and 430\,pb at the Tevatron, the \tevs, and the LHC,
 respectively.
For simplicity, let's consider the $\ell^\pm \, + \geq 3\, {\rm jet}$
decay mode whose
branching ratio is $\rm{Br}=2 {\frac{2}{9}} {\frac{6}{9}} = \frac{8}{27}$,
where the charged lepton $\ell^\pm$ can be either $e^\pm$ or $\mu^\pm$.
We assume the experimental detection
efficiency ($\epsilon$), which includes
both the kinematic acceptance and the efficiency of $b$-tagging,
to be 15\% for the signal event \cite{CDF}.
Let's further assume that there is
no ambiguity in picking up the right $b$ ($\bar b$)
to combine with the charged lepton $\ell^+$ ($\ell^-$)
to reconstruct $t$ (or $\bar t$), then in total there are
$4.5\,{\rm pb}\,\times \, 10^3\,{\rm pb}^{-1}\,
\times \,{\frac{8}{27}}\,\times \,0.15=200$
reconstructed $t \bar t$ events to be used in measuring
$\klc$ and $\krc$ at $\sqrt{S}= 2$\,TeV.
The same calculation at
the \tevs~and the LHC yields 1100 and 19000 reconstructed
$t \bar t$ events, respectively.
Given the number of reconstructed top quark events,
one can fit the $m_{b\ell}$ distribution to measure
$\klc$ and $\krc$.
For example we have done a study for the Tevatron.
Let us assume the effects of
new physics only modify the SM results ($f_1^L=1$ and $f_1^R=0$
at Born level) slightly and the form factors
$f_2^{L,R}$ are as small as expected from the usual dimensional
analysis~\cite{geor2,dimens}.\footnote{
The coefficients of the form factors $f_2^{L,R}$, assumed to be
induced through loop effects,  will be a factor of
${1 \over 16 \pi^2}$ smaller than that of the form factors
$f_1^{L,R}$.}
We summarize our results on the accuracy of measuring
$f_1^{L,R}$ for various luminosities
in Table~\ref{formfac} \cite{danf1}.
(Only statistical errors are included at the 95\% confidence level.)
\vspace{4mm}
\begin{table}
\caption{
Results on the accuracy of measuring
$f_1^{L,R}$ for various luminosities.
(Only statistical errors are included at the 95\% confidence level.)
}
\label{formfac}
\begin{center}
\begin{tabular}{|l|l|l|l|l|r|}                                    \hline
Integrated  & Number of            & &            &            \\
Luminosity  & reconstructed        & ${\Delta f_1^L}\over f_1^L$
& $\Delta f_1^R$ & ${\Delta m_t} \over m_t$                    \\
$\ifb$      & $t \bar t $ events   & &            &            \\ \hline
1           & 200       & $8\%  $    & $\pm 0.5 $ & $4\%  $    \\ \hline
3           & 600       & $4\%  $    & $\pm 0.3 $ & $2\%  $    \\ \hline
10          & 2000      & $2\%  $    & $\pm 0.2 $ & $1\%  $    \\ \hline
\end{tabular}
\end{center}
\vspace{4mm}
\end{table}

In the same table (\ie\,Table~\ref{formfac})
we also show our estimate on how well the mass of the top
quark $m_t$ can be measured from $\flong$.
By definition of $\flong$, for a SM top quark
({\it i.e.}, $f_1^L=1$ and $f_1^R=0$), the distribution of
  $\cos \theta^*_{\ell}$ has the functional form, in shape, as
\beq
F(\cos \theta^*_{\ell}) \sim
\left( { 1- \cos \theta^*_{\ell} \over 2 } \right)^2
+ \flong \left( { \sin \theta^*_{\ell} \over \sqrt{2} } \right)^2 \, .
\enq
Therefore, $\flong$ can be calculated by fitting with the distribution
of $\cos \theta^*_{\ell}$, or equivalently with
the distribution of $m_{b\ell}$.
We prefer to measure $\klc$ and $\krc$ using the distributions of
$m_{b\ell}$ than of $\cos \theta^*_{\ell}$ because
the former can be directly calculated from the
measured momenta of $b$ and $\ell$. However,
to convert from the distributions of $m_{b\ell}$ to
$\cos \theta^*_{\ell}$, as given in Eq.~(\ref{mbe1}), the effects
from the widths of $W$-boson and top quark might slightly
distort the distribution of $\cos \theta^*_{\ell}$.
(Notice that in the full calculation of the scattering amplitudes
the widths of the $W$-boson and the top quark have to be included
in the Breit-Wigner form to generate a finite event rate.)

However, in reality, the momenta of the bottom quark and the
charged lepton will be smeared by
detector effects and another problem in this analysis is
the identification of the right $b$ to reconstruct $t$.
There are three possible strategies to improve the efficiency of identifying
the right $b$.  One is to demand a large invariant mass of the $t \bar t$
system so that $t$ is boosted and its decay products are collimated.
Namely, the right $b$ will be moving closer to the lepton from $t$ decay.
This can be easily enforced by demanding leptons with a larger transverse
momentum.
Another is to identify the soft (non-isolated) lepton from $\bar b$ decay
(with a branching ratio ${\rm Br}(\bar b \ra \mu^{+} X) \sim 10\%$).
The other is to statistically determine the electric charge of the
$b$-jet (or $\bar b$-jet) to be $1/3$ (or $-1/3$) \cite{lepjet}.
All of these methods may further reduce the reconstructed signal rate by
an order of magnitude. How will these affect our conclusion on
the determination of the non-universal couplings $\klc$ and $\krc$?
It can only be answered by detailed Monte Carlo studies
which are yet to be done.

\noindent
\subsection{\twelveit From the Production of Top Quarks}
\indent

Here we propose another method to measure
the couplings $\klc$ and $\krc$
from the production rate of the single-top quark process.

For $m_t=180$ GeV, the sum of the production rates of single-$t$ and
single-$\bar t$
events is about 2\,pb and 14\,pb for $\sqrt{S}= 2$\,TeV and
$\sqrt{S}= 4$\,TeV respectively. The branching ratio of interest is
$\rm{Br}=\frac{2}{9}$. The kinematic acceptance
of this event at $\sqrt{S}= 2$\,TeV is about $0.55$ \cite{wgtbnew}.
Assuming the efficiency of $b$-tagging is about 30\%, then
there will be $2\,{\rm pb}\,\times \,10^3\,{\rm pb}^{-1}\,
\times \,{\frac{2}{9}}\,\times \,0.55\, \times \,0.3=75$
events reconstructed for a 1\,$\ifb$ integrated
luminosity.  At $\sqrt{S}= 4$\,TeV,
 the kinematic acceptance of this event is about
$0.40$ \cite{wgtbnew} which, from the above calculation, yields
about $3700$ reconstructed events for 10\,$\ifb$ integrated luminosity.
Based on statistical
error alone, this corresponds to a 12\% and 2\%
measurement on the single-top cross section.
A factor of 10 increase in the luminosity of
the collider can improve the measurement by a factor of 3 statistically.
Taking into account the theoretical uncertainties,
as discussed in section 2, we
examine two scenarios: 20\% and 50\% error on
the measurement of the cross section for single-top production.
\begin{figure}
\par
\caption{ Constraint on $|\klc|$ and $\krc$ given $20\%$ and $50\%$
error in measurement of Standard Model rate for $W$-gluon fusion.
Curves are identical for $m_t = 140$ GeV and $m_t = 180$ GeV.
}
\label{klkr}
\par
\end{figure}
The results, which are not sensitive to
the energies of the colliders considered here (either 2\,TeV or 4\,TeV),
 are shown in Figure~\ref{klkr} for a 180 GeV top quark
at the Tevatron.
We found that $\klc$ and $\krc$ are well constrained inside
the region bounded by two (approximate) ellipses.
To further determine the sizes of $\klc$ and $\krc$ one needs to
study the kinematics of the decay products, such
as the charged lepton $\ell$, of the top quark.
Since the top quark produced from the $W$-gluon fusion process
is almost one hundred percent left-handed (right-handed) polarized
for a left-handed (right-handed) $\tbw$ vertex,
the charged lepton $\ell^+$ from $t$ decay has a harder momentum
for a right-handed $\tbw$ coupling than for a left-handed coupling.
(Note that the couplings of
light-fermions to $W$ boson have been well tested
from the low energy data to be left-handed as described in the SM.)
As shown in Figures~\ref{left} and~\ref{right},
this difference becomes smaller when the top quark is much heavier because
the $W$ boson from the top quark decay tends to be more
longitudinally polarized.

A right-handed charged current is absent in a
linearly $SU(2)_L$ invariant gauge
theory with massless bottom quark.
In this case,  $\krc=0$,
then $\klc$ can be constrained to within
about $-0.08 < \klc < 0.03$ ($-0.20 < \klc < 0.08$)
with a 20\% (50\%) measurement on the production rate
of single-top quark at the Tevatron
\cite{ehab}. (Here we assume the experimental data
agrees with the SM prediction within 20\% (50\%).) This means that if
we interpret {\mbox {($1+\klc$)}} as the CKM matrix element $|V_{tb}|$,
then $|V_{tb}|$ can be bounded as $|V_{tb}| > 0.9$ (or 0.75) for a 20\%
(or 50\%) measurement on the single-top production rate.

Before closing this section, we would remark that
in the previous lectures and in the Refs.~\cite{ehab} and \cite{fuj}
some bounds on the
couplings of $\klc$ and $\krc$ were obtained by
studying the low energy data with the assumption that
the effects of new physics at low energy can only modify the couplings
of $\klc$ and $\krc$ but not introduce any other light fields
in the effective theory. However, nature might not behave exactly in this
way. It is possible that some light fields may exist just below the
TeV scale, then  the bounds
obtained from Refs.~\cite{ehab} and \cite{fuj} may no longer hold.
Thus, it is important to have the direct measurement on all
the form factors listed in Eq.~(\ref{eqlag}) from the  production of
top quarks, in spite of the present bounds on $\kappa$'s derived from
radiative corrections to low energy data.

\noindent
\section{\twelvebf Probing CP Properties in Top Quarks}
\indent

It is known that explicit CP violation requires the presence of both the CP
non-conserving vertex and the complex structure of the physical amplitude.
Due to the origin of this complex structure, the possible CP-violating
observables can be separated into two categories.
In the first category, this complex structure comes from the absorptive
part of amplitude due to the final state interactions.
In the second category, this complex structure does not arise from the
absorptive phase but from the correlations in the kinematics of the
initial and final state particles involved in the physical process.
Hence, it must involve a triple product
correlation ({\it i.e.}, a Levi-Civita tensor).

To distinguish the symmetry properties between these two cases,
we introduce the transformation $\hatt$, as defined in Ref.~\cite{hatt},
which is simply the application of time reversal to all momenta and
spins without interchanging initial and final states.
The CP-violating observables in the first category are CP-odd and
CP$\hatt$-odd, while those in the second category are
CP-odd and CP$\hatt$-even. Of course, both of them are CPT-even.

As an illustration of the above two categories, we consider the CP-violating
observables for the decay of the top quark.
Consider the partial rate asymmetry
\begin{eqnarray}
{\cal A}_{bW} & \equiv &
    { \Gamma( t \ra b W^+) - \Gamma({\bar t} \ra {\bar b} W^-)
    \over \Gamma( t \ra b W^+) + \Gamma({\bar t} \ra {\bar b} W^-) }.
\label{abw}
\end{eqnarray}
This observable clearly violates CP and CP$\hatt$ and therefore belongs to
the first category.
We note that
because of CPT invariance, the total decay width of the top quark
$\Gamma(t)$ has to equal the total decay width of the top anti-quark
$\Gamma(\bar t)$.
Thus, any non-zero ${\cal A}_{bW}$ implies that
there exists a state (or perhaps more than one state) $X$ such that
$t$ can decay into $X$, and ${\bar t}$ into ${\bar X}$.
The absorptive phase
of $t \ra b W^+$ is therefore generated
by re-scattering through state $X$, \ie\,,
$t \ra X \ra b W^+$, where $X \neq b W^+$ because
the final state interaction should be off-diagonal \cite{wolf}.

Next, let's consider the observable of the second category.
In the decay of $t \ra b W^+ (\ra \ell^+ \nu_{\ell})$, for a polarized $t$
quark, the time-reversal invariance (T) is violated if
the expectation value of
\begin{eqnarray}
\vec{\sigma}_t \times \vec{p}_b \cdot \vec{p}_{\ell^+}
\label{tripro}
\end{eqnarray}
is not zero \cite{toppol}. Assuming CPT invariance, this
implies CP is violated.
Therefore, this observable is CP-odd but CP$\hatt$-even.
A non-vanishing triple product observable,
such as that in Eq.~(\ref{tripro}), from the decay of the top quark
violates T, however it may be entirely due to final state interaction
effects without involving any CP-violating vertex.
To construct a truly CP-violating observable, one must combine
information from both the $t$ and $\bar t$ quarks.
For instance, the difference in the expectation values of
$\vec{\sigma}_t \times \vec{p}_b \cdot \vec{p}_{\ell^+}$
and
$\vec{\sigma}_{\bar t} \times \vec{p}_{\bar b} \cdot
\vec{p}_{\ell^-}$
would be a true measure of an intrinsic CP violation.

There have been many studies
on how to measure the CP-violating effects in the $t \bar t$ system
produced in either electron or hadron collisions.
(For a review, see a recent paper in Ref.~\cite{cpcpy}.)
At hadron colliders, the number of
$t \bar t$ events needed to
measure a CP-violating effect of the order of $10^{-3} - 10^{-2}$
is about $10^7 - 10^8$.
To examine the potential of various current and future
hadron colliders in measuring the CP-violating asymmetries,
we estimate the total event rates of $t \bar t$ pairs
for a 180 GeV SM top quark produced at these colliders.
At the Tevatron, the Di-TeV, and the LHC, an integrated luminosity of
10, 100, and 100 ${\rm fb}^{-1}$ will produce about
$4.5 \times 10^4$, $2.6 \times 10^6$, and $4.3 \times 10^7$
$t \bar t$ pairs, respectively, as given in Table~\ref{trates}.
Therefore, the LHC
would be able to probe the CP asymmetry  of the top
quark at the level of a few percent. A similar
number of the $t \bar t$ pairs
is required in electron collision to probe the CP asymmetry
at the same level.
Thus, for a $\sqrt{S}=500\,$GeV $e^-e^+$ collider,
an integrated luminosity of about $10^4 - 10^5$ ${\rm fb}^{-1}$
has to be delivered. This luminosity is
at least a factor of $100$ higher than the
planned next linear colliders.
We note that although the initial state in a pp collision
(such as at the LHC)
is not an eigenstate of a CP transformation,
these CP-odd observables can still be defined as long as
the production mechanism is dominated by $gg$ fusion. This is
indeed the case for $t \bar t$ pair productions at the LHC.

In the SM, the top quark  produced via the $W$-gluon fusion
process is about one hundred percent
left-handed (longitudinally) polarized.
Given a polarized top quark, one can use
the triple product correlation, as defined in Eq.~(\ref{tripro}),
to detect CP violation of the top quark.
For  a polarized top quark, one can either use
$\vec{\sigma}_t \times \vec{p}_b$
or $\vec{p}^{\rm Lab}_t \times \vec{p}_b$
to define the decay plane of $t \ra b W (\ra \ell^+ \nu) $.
Obviously, the latter one is easier to implement experimentally.
Define the asymmetry to be
\begin{eqnarray}
{\cal A}_{io} & \equiv &
      { {   N(\ell^+ \, {\rm out \, of \, the \, decay \, plane})
         -  N(\ell^+ \, {\rm into \, the \, decay \, plane }) } \over
           {   N(\ell^+ \, {\rm out \, of \, the \, decay \, plane})
           +  N(\ell^+ \, {\rm into \, the \, decay \, plane }) } }~.
\end{eqnarray}
If ${\cal A}_{io}$ is not zero, then the time-reversal T is not conserved,
therefore CP is violated for a CPT invariant theory.
Due to the missing momentum of the neutrino from the decay of the $W$-boson,
it is difficult to reconstruct the azimuthal angle ($\phi_W$)
of the $W$-boson from the decay of the top quark. Once the angle
$\phi_W$ is integrated over, the transverse polarization of the top
quark averages out, and only the longitudinal polarization of the top
quark contributes to the asymmetry ${\cal A}_{io}$.
Thus, the asymmetry ${\cal A}_{io}$ can be used to study the
effects of CP violation in the top quark,
which in the SM is about one
hundred percent left-handed (longitudinally) polarized
as produced from the $W$-gluon fusion process.
To apply the CP-violating observable
${\cal A}_{io}$, one needs to reconstruct the directions of both the $t$
and $b$ quarks.
It has been shown in Ref.~\cite{onetcp} that it takes about
$10^7-10^8$ single-top events
to detect CP violation at the order of $\sim 10^{-3} - 10^{-2}$.

For $m_t=180\,$GeV at the Tevatron, the Di-TeV, and the LHC,
an integrated luminosity of
10, 100, and 100 ${\rm fb}^{-1}$ will produce about
$2 \times 10^4$, $1.4 \times 10^6$, and $2 \times 10^7$
single-$t$ or single-$\bar t$ events, respectively,
Table~\ref{trates}.
At the NLC, the single top quark production rate is
much smaller. For a $2\,$TeV electron collider,
the cross sections for
$e^-e^+ \ra e^- {\bar \nu_e} t {\bar b}$
and
$e^+ \gamma \ra {\bar \nu}_e t {\bar b}$
are 8 fb and 60 fb, respectively \cite{eeonetop}.
Hence, it will be extremely difficult
to detect CP violation effects at the order of $\leq 10^{-2}$
in the single-top events produced in electron collisions.

A few comments are in order. First, to extract the {\it genuine}
CP-violating effects, we need to study the difference in the asymmetry
${\cal A}_{io}$ measured in the single-$t$ and single-$\bar t$
events because the time-reversal violation in
${\cal A}_{io}$ of the $t$ (or $\bar t$) alone could be
generated by final state interactions without CP-violating
interactions.
Second, the detection efficiency for this method is not close
to one, so a good understanding of the kinematics of the decay products and
how the detector works are needed to make this method useful.

The asymmetry ${\cal A}_{io}$ belongs to the second category of
CP-violating observables, and is CP-odd and CP$\hatt$-even. Here, let's
consider another asymmetry ${\cal A}_t$ which belongs to the first
category of CP-violating observables, and is CP-odd and CP$\hatt$-odd.
Using ${\cal A}_t$ for detecting CP-violating effects is to make use of
the fact that $\pbarp$ is a CP eigenstate; therefore, the difference in the
production rates for $\pbarp \ra t X$ and $\pbarp \ra \bar t X$
is a signal of CP violation.
This asymmetry is defined to be
\begin{eqnarray}
{\cal A}_t & \equiv &
    { \sigma(\pbarp \ra t X) - \sigma(\pbarp \ra \bar t X) \over
               \sigma(\pbarp \ra t X) + \sigma(\pbarp \ra \bar t X) } ~~.
\end{eqnarray}
As discussed in section 5,
the production rate of $\pbarp \ra t X$ is proportional to
the decay rate of $t \ra b W^+$, and the rate of
$\pbarp \ra \bar t X$ is proportional to
the rate of $\bar t \ra \bar b W^-$.
This implies that ${\cal A}_t = {\cal A}_{bW}$, cf. Eq.~(\ref{abw}).
There have been quite a few models studied in the literature about the
asymmetry in ${\cal A}_{bW}$. For instance, in the Supersymmetric
Standard Model where a CP-violating phase may occur in the
left-handed and right-handed top-squark, ${\cal A}_{bW}$ can
be as large as a few percent
depending on the details of the parameters in the model \cite{stopcp}.

Next, let's examine how many top quark events are needed to
detect a few percent effect in the CP-violating asymmetry ${\cal A}_t$.
Consider $t \ra b W^+ \ra b \ell^+ \nu$, where
$\ell= e \, {\rm or} \, \mu$.
Define the branching ratio $B_W$ as the product of
${\rm Br}(t \ra bW^+)$ and  ${\rm Br}(W^+ \ra \ell^+ \nu)$, where
${\rm Br}(W^+ \ra \ell^+ \nu)$ is $2/9$.
(${\rm Br}(t \ra bW^+)$ depends on the details of a model, and is
almost 1 in the SM.)
Let us assume that the efficiency of $b$-tagging ($\epsilon_{\rm btag}$)
is about 30\%, and the kinematic acceptance
($\epsilon_k$) of reconstructing the single-top event,
$\pbarp \ra t X \ra b W^+ X \ra b \ell^+ \nu X$,
is about 50\%.\footnote{
This was obtained from a Monte Carlo study performed
in Ref.~\cite{wgtbnew}.}
The number of single-$t$ and single-$\bar t$
events needed to measure ${\cal A}_t$ is
\beq
{\cal N}_t = { 1 \over B_W \epsilon_{\rm btag} \epsilon_k}
  \left(1 \over {\cal A}_t \right)^2 ~~.
\enq
Thus, to measure ${\cal A}_t$ of a few percent,
${\cal N}_t$ has to be as large as $\sim 10^6$, which corresponds
to an integrated luminosity of 100\,$\ifb$ at the Di-TeV.

\newpage
\section{\twelvebf Discussions and Conclusions}
\indent

We discussed the physics of top quark production and decay
at hadron colliders, such as the Tevatron, the \tevs~and the LHC.
We showed how to measure the mass and the width of the
top quark, produced from either a single-top or a $t \bar t$ pair
process, using the invariant mass distribution of $m_{b\ell}$.
It has been shown in Ref.~\cite{steve} that
the distribution of $m_{b\ell}$ is not sensitive to
radiative corrections from  QCD interactions, thus
it can be reliably used to test the polarization of the
$W$-boson from $t$ decay, therefore test the polarization of
the top quark from the production mechanism.
We also discussed how well the couplings of $t$-${b}$-$W$ vertex
can be studied to probe new physics, and how well the
CP properties of the top quark can be tested in
electron or hadron colliders.

In Ref.~\cite{wgtb} we showed that an almost perfect efficiency
for ``kinematic $b$ tagging'' can be achieved
due to the characteristic features of the transverse momentum and
rapidity distributions of the spectator quark which emitted the virtual $W$.
In addition, the ability of performing $b$-tagging using a
vertex detector increases the detection efficiency
of a heavy top quark
produced via the $W$-gluon fusion process.
A detailed Monte Carlo study was performed in Ref.~\cite{wgtbnew}
to show that this process is extremely useful at the Tevatron
with Main Injector.
For an integrated luminosity of 1 $\ifb$,
 there will be about 105 (75)
single-$t$ or single-$\bar t$  events reconstructed
in the lepton+jet mode
for $\mt = 140 \,(180)$ GeV at $\rts = 2$ TeV.
(The branching ratio of $W \ra e, \, {\rm or} \, \mu$ is included, and
The $b$-quark tagging efficiency is assumed to be 30\% for $P^b_t > 30\,$GeV
with no misidentifications of a $b$-jet from other QCD jets.)
The dominant background process is the electroweak-QCD process
$W+b\bar b$ whose rate is about $60\%$($80\%$) of the signal rate in
the end of the analysis. The $t \bar t$ events are not as important
to our study.
The results for $\rts = 4$\,TeV at the Di-TeV
and for $\rts = 14$\,TeV at the LHC were also given in Ref.~\cite{wgtbnew}.

\section*{ Acknowledgments }
\indent

I would like to thank Douglas O. Carlson and Ehab Malkawi
for helping me in preparing these lecture notes.
To them and Gordon L. Kane, Chung Kao, Glenn A. Ladinsky , and
Steve Mrenna I thank for the fruitful collaborations.
I also thank the organizers of the School for a pleasant
setup.
This work was supported in part by an NSF grant No. PHY-9309902.

\newpage


\begin{thebibliography}{99}

\bibitem{mele}The LEP Collaborations
ALEPH, DELPHI, L3, OPAL and The LEP Electroweak
Working Group, CERN/PPE/93--157 (1993); \\
W. Hollik, in {\it Proceedings of the
$XVI$ International Symposium on Lepton--Photon Interactions},
Cornell University, Ithaca, N.Y., Aug. 10--15, 1993; \\
M. Swartz, in {\it Proceedings of the
$XVI$ International Symposium on Lepton--Photon Interactions},
Cornell University, Ithaca, N.Y., Aug. 10--15, 1993; \\
Barbara Mele, in {\it $XIV$ Encontro
Nacional de Fisica de Campos e Particulas},
 Caxambu, Brazil, 29 Sept.--3 Oct., 1993;\\
J. Lefrancois, in {\it Proceedings of the EPS Conference on
Higs Energy Physics}, Marseille, France, 1993.
\bibitem{alta}G. Altarelli,
in {\it Proceedings of International University School of Nuclear and
Particle Physics: Substructures of Matter as Revealed with Electroweak
Probes}, Schladming, Austria, 24 Feb - 5 Mar 1993.
\bibitem{lep94}
M. Koratsinos and S. de Jong, in {\it Proceedings
of the Rencontres de la Vallee d'Aoste}, La Thuile, 1994 ;\\
P. Clarke, P. Siegrist and B. Pietrzyk, in
{\it Proceedings of the Rencontres de Moriond}, Meribel, 1994.
\bibitem{sld94}
M. Woods, in {\it Proceedings of the Rencontres de Moriond},
Meribel, 1994.
\bibitem{alt}
G. Altarelli, CERN-TH-7319/94,
talk at 1st International Conference on Phenomenology of
Unification: from Present to Future, Rome, Italy, 23-26 Mar 1994.
\bibitem{pc}See, for example, \\
R.D. Peccei, in {\it 1993 Scottish
Summer School}, St. Andrews, Scotland, August 1993, and {\it 1993
Escuela Latino Americano de Fisica}, Mar del Plata, Argentina, July 1993.
\bibitem{kane}G.L. Kane, in {\it Proceedings of
the Workshop on High Energy Phenomenology}, Mexico City, July
1--10, 1991.
\bibitem{D0}S. Abachi {\it et al.}, {\it Phys. Rev. Lett.}
 {\bf 72}, 2138 (1994).
\bibitem{CDF}F. Abe {\it et al.}, {\it Phys.~Rev.} {\bf D50}, 2966 (1994).
\bibitem{pczh}R.D. Peccei, S. Peris and X. Zhang, {\it Nucl. Phys.}
 {\bf B349}, 305 (1990).
\bibitem{sekh}R.S. Chivukula, E. Gates, E.H. Simmons and J. Terning,
{\it Phys. Lett.} {\bf B311}, 157 (1993);\\
R.S. Chivukula, E.H. Simmons and J. Terning, {\it Phys. Lett.}
{\bf B331}, 383 (1994).
\bibitem{lopez}J.L. Lopez, D.V. Nanopoulos, G.T. Park, Xu Wang
and A. Zichichi, {\it Phys. Rev.} {\bf D50}, 2164 (1994).
\bibitem{pesk}
M.E. peskin and T. Takeuchi, {\it Phys. Rev. Lett.}
{\bf 65}, 964 (1990); {\it Phys. Rev.} {\bf D46}, 381 (1991); \\
D.C. Kennedy and P. Langacker, {\it Phys. Rev. Lett.}
 {\bf 65}, 2967 (1990);\\
W. Marciano and J. Rosner, {\it Phys. Rev.} {\bf D44}, 1591 (1991).
B. Holdem, {\it Phys. Lett.} {\bf B259}, 329 (1991);\\
\bibitem{bar}R. Barbieri, in {\it Proceedings of the Symposium
on Particle Physics at the Fermi scale}, Beijing, China, May
27 --June 4, 1993.
\bibitem{bar1}G. Altarelli, R. Barbieri and
 F. Caravaglios, {\it Nucl. Phys.} {\bf B405}, 3 (1993).
\bibitem{bar2}G. Altarelli and R. Barbieri, {\it Phys. Lett.} {\bf B253},
161 (1990); \\
G. Altarelli, R. Barbieri and S. Jadach, {\it Nucl. Phys.}
 {\bf B369}, 3 (1992).
\bibitem{geor}H. Georgi, {\it Nucl. Phys.} {\bf B361}, 339 (1991).
\bibitem{gold}M. Golden and L. Randall,
{\it Nucl. Phys.} {\bf B362}, 3 (1991);\\
R.D. Peccei and S. Peris, {\it Phys. Rev.} {\bf D44}, 809 (1991); \\
A. Dobado {\it et al.}, {\it Phys. Lett.} {\bf B255}, 405 (1991); \\
M. Dugan and L. Randall, {\it Phys. Lett.} {\bf B264}, 154 (1991).
\bibitem{buch}W. Buchm\"{u}ller and D. Wyler, {\it Nucl. Phys.}
{\bf B268}, 621 (1986).
\bibitem{wwww} See, for example, \\
 C.--P. Yuan, in {\it Perspectives
on Higgs Physics}, edited by G. Kane (World Scientific, 1992), pp 415--428.
\bibitem{tana}Y. Nambu, in {\it Proceedings of the 1988 International
Workshop on New Trends in Strong Coupling gauge Theories},
eds. by M. Bando, T. Muta and K. Yamawaki
(World Scientific, Singapore, 1989);\\
W.A. Bardeen, C.T. Hill and M. Lindner, {\it Phys. Rev.}
 {\bf D41}, 1647 (1990);\\
R. B\"{o}nisch and A. Leike, DESY 93--111, August 1993;\\
B.A. Kniehl and A. Sirlin, DESY 93--194, NYU--TH--93/12/01,
December 1993.
\bibitem{wein}S. Weinberg, {\it Physica} {\bf 96A}, 327 (1979).
\bibitem{geor2}H. Georgi, {\it Weak Interactions and Modern Particle
Theory} (The Benjamin/Cummings Publishing Company, 1984).
\bibitem{cole}S. Coleman, J. Wess and Bruno Zumino, {\it Phys. Rev.}
{\bf D177}, 2239 (1969);\\
C.G. Callan, S. Coleman, J. Wess and Bruno Zumino, {\it Phys. Rev.}
{\bf D177}, 2247 (1969).
\bibitem{fer}F. Feruglio, {\it Int.J. Mod. Phys.} {\bf A8}, 4937 (1993).
\bibitem{nlc}
See, for example,\\
P. Chen, {\it Phys. Rev.} {\bf D46}, (1992) 1186; and the references therein.
\bibitem{pecc}R.D. Peccei and X. Zhang, {\it Nucl. Phys.}
 {\bf B337}, 269 (1990).
\bibitem{hol}B. Holdom and J. Terning, {\it Phys. Lett.}
 {\bf 247B}, 88 (1990).
\bibitem{how1}H. Georgi, {\it Nucl. Phys.} {\bf B363}, 301 (1991).
\bibitem{fer1}F. Feruglio, A. Masiero and L. Maiani,
{\it Nucl. Phys} {\bf B387}, 523 (1992).
\bibitem{app}T. Appelquist and Guo--Hong Wu, {\it Phys. Rev.} {\bf D48},
3241 (1993).
\bibitem{burg}C.P. Burgess and D. London, {\it Phys. Rev.} {\bf D48},
4337 (1993).
\bibitem{mart}M.B. Einhorn, UM--TH--93--12, Apr 1993.
\bibitem{ehab}
D. Carlson, Ehab Malkawi and C.--P. Yuan,
{\it Phys. Lett.} {\bf B337} (1994) 145; \\
Ehab Malkawi and C.--P. Yuan. {\it Phys. Rev.}
{\bf D50}, 4462 (1994);\\
Ehab Malkawi and C.--P. Yuan. MSUHEP-50107, January 1995.
\bibitem{velt}M. Veltman, {\it Nucl. Phys.} {\bf B123}, 89 (1977).
\bibitem{long}A. C. Longhitano, {\it Phys. Rev.} {\bf D22}, 1166 (1980).
\bibitem{gal}
G.A. Ladinsky and C.--P. Yuan, {\it Phys. Rev.}
{\bf D49}, 4415 (1994).
\bibitem{barxn2}R. Barbieri {\it et al.}, {\it Nucl. Phys.} {\bf B409},
105 (1993);\\
R. Barbieri {\it et al.}, {\it Nucl. Phys.} {\bf B288}, 95 (1992).
\bibitem{largen}
S. Naculich and C.--P. Yuan, {\it Phys. Rev.} {\bf D48}, 1097 (1993).
\bibitem{joserb}
J. Bernabeu, A. Pich and A. Santamaria, {\it Phys. Lett.} {\bf B200},
569 (1988); \\
W. Beenakker and W. Hollik, {\it Z. Phys.} {\bf C40}, 141 (1988); \\
A.A. Akhundov, D.Yu. Bardin and T. Riemann, {\it Nucl. Phys.}
{\bf B276}, 1 (1988).
\bibitem{fuj}
K. Fujikawa and A. Yamada, {\it Phys. Rev.} {\bf D49}, 5890 (1994).


\bibitem{decay}
I.I.Y. Bigi, Yu L. Dokshitzer, V.A. Khoze, J.H. Kuhn and P. Zerwas,
{\it Phys. Lett.} {\bf 181B} (1986) 157;\\
L.H. Orr and J.L. Rosner, {\it Phys. Lett.} {\bf 246B} (1990) 221;
{\bf 248B} (1990) 474(E).

\bibitem{qqttew}
F.~Berends et al., in the report of the Top Physics Working
Group from the {\it Proceedings of the Large Hadron
Collider Workshop}, 4-9 October 1990, Aachen, ed. G.~Jarlskog and D.~Rein,
CERN publication CERN 90-10, pg.~310, Fig.~7/8; \\
A.~Stange and S.~Willenbrock, {\it Phys.~Rev.} {\bf D48} (1993) 2054; \\
C.~Kao, G.A.~Ladinsky, C.--P.~Yuan,
MSUHEP-93/04, 1993;  MSUHEP-94/04, 1994; \\
W.~Beenakker, {\it et al.}, {\it Nucl.~Phys.} {\bf B411} (1994) 343; \\
Chung Kao, FSU-HEP-940508, June 1994.



\bibitem{argon}
See, for example, \\
C.--P. Yuan, {\it et al.}, {\it Report of the subgroup on the Top Quark}, in
{\it Proceedings of Workshop on Physics at
Current Accelerators and Supercolliders},
eds. by J. Hewett, A. White and D. Zeppenfeld, 1993, pp 495-505.

\bibitem{sally}
S. Dawson,  {\it Nucl.~Phys.} {\bf B249}, 42 (1985); \\
S. Dawson and S. Willenbrock,  {\it Nucl.~Phys.} {\bf B284},
449 (1987); \\
S. Willenbrock and D.A. Dicus, {\it Phys.~Rev.} {\bf D34},
155 (1986); \\
F. Anselmo, B. van Eijk and G. Bordes, {\it  Phys.~Rev.}
{\bf D45}, 2312 (1992); \\
T. Moers, R. Priem, D. Rein and H. Reithler, in {\it Proceedings of Large
Hadron Collider Workshop}, preprint CERN 90-10, 1990; \\
R.K. Ellis and S. Parke,  {\it Phys.~Rev.} {\bf D46}, 3785 (1992).

\bibitem{wgtb}
C.--P. Yuan, {\it Phys.~Rev.} {\bf D41}, 42 (1990); \\
D. Carlson and C.--P. Yuan, {\it Phys. Lett.} {\bf B306} (1993) 386.

\bibitem{cortese}
S. Cortese and R. Petronzio, {\it Phys. Lett.} {\bf B253} (1991) 494.

\bibitem{galwt}
G.A. Ladinsky and C.--P. Yuan, {\it Phys.~Rev.} {\bf D43}, 789 (1991);

\bibitem{pdf}
J. Botts, J. Huston, H. Lai, J. Morfin, J. Owens, J. Qiu,
W.-K. Tung, H. Weerts, Michigan State University preprint MSUTH-93/17.

\bibitem{qcdtt}
P. Nason, S. Dawson and R.K. Ellis,
{\it Nucl.~Phys.} {\bf B303}, 607 (1988); {\bf B327}, 49 (1989); \\
W. Beenakker, H. Kuijf, W.L. van Neerven and J. Smith,
{\it  Phys.~Rev.} {\bf D40}, 54 (1989); \\
R. Meng, G.A. Schuler, J. Smith and W.L. van Neerven,
{\it Nucl.~Phys.} {\bf B339}, 325 (1990).

\bibitem{smith}
E. Laenen, J. Smith and W.L. van Neerven,
Nucl. Phys. {\bf B369} (1992) 543

\bibitem{wuki}
J. C. Collins and Wu-Ki Tung, {\it Nucl. Phys.} {\bf B278} (1986) 934;\\
F. Olness and Wu-Ki Tung, {\it Nucl. Phys.} {\bf B308} (1988) 813;\\
M.  Aivazis, F. Olness and Wu-Ki Tung, {\it Phys. Rev. Lett.}
{\bf 65} (1990) 2339; {\it Phys. Rev.} {\bf D50}, 3085 (1994); \\
M.  Aivazis, J.C. Collins, F. Olness and Wu-Ki Tung,
{\it Phys. Rev.} {\bf D50}, 3102 (1994).

\bibitem{wgtbnew}
D. Carlson and C.--P. Yuan, preprint MSUHEP-40903.

\bibitem{hill}
C.T. Hill and S. Parke, {\it Phys. Rev.} {\bf D49} 4454 (1994); \\
E. Eichten and K. Lane, {\it Phys. Lett.} {\bf B327} 129 (1994).

\bibitem{dawson}
R. N. Cahn and S. Dawson, {\it Phys. Lett.}
{\bf B136}, 196 (1984), {\it Phys.
Lett.} {\bf B138}, 464(E) (1984); \\
M. S. Chanowitz and M. K. Gaillard, {\it Phys. Lett.} {\bf B142}, 85 (1984);\\
G. L. Kane, W. W. Repko and W. R. Rolnick, {\it Phys. Lett.}
 {\bf B148}, 367 (1984); \\
S. Dawson, {\it Nucl. Phys.} {\bf B249}, 42 (1985);
{\it Phys. Lett.} {\bf B217}, 347 (1989); \\
J. Lindfors, {\it Z. Phys.} {\bf C28}, 427 (1985);\\
W. B. Rolnick, {\it Nucl. Phys.} {\bf B274}, 171 (1986);\\
P. W. Johnson, F. I. Olness and Wu--Ki Tung, {\it Phys. Rev.}
 {\bf D36}, 291 (1987);\\
Z. Kunszt and D. E. Soper, {\it Nucl. Phys.} {\bf B296}, 253 (1988);\\
A. Abbasabadi, W. W. Repko, D. A. Dicus and R. Vega,
{\it Phys. Rev.} {\bf D38}, 2770 (1988).


\bibitem{equiv}
J. M. Cornwall, D. N. Levin, and G. Tiktopoulos,
{\it Phys. Rev.} {\bf D10}, 1145 (1974); \\
C.~Vayonakis, Lett. Nuovo Cimento {\bf 17}, 383 (1976); \\
B.~W.~Lee, {\hbox {C. Quigg}}, and H.~Thacker, {\it Phys. Rev.}
 {\bf D16}, 1519 (1977); \\
M.~S.~Chanowitz and M.~K.~Gaillard, {\it Nucl. Phys.}
 {\bf B261}, 379 (1985);\\
G.~J.~Gounaris, R.~Kogerler, and H.~Neufeld, {\it Phys. Rev.}
 {\bf D34}, 3257 (1986); \\

\bibitem{equivrad}
Y.--P. Yao and C.--P. Yuan, {\it Phys. Rev.} {\bf D38}, 2237 (1988); \\
J. Bagger and C. R. Schmidt, {\it Phys. Rev.} {\bf D41}, 264 (1990); \\
H. Veltman, {\it Phys. Rev.} {\bf D41}, 2294 (1990); \\
H.-J. He, Y.-P. Kuang, and X. Li, {\it Phys. Rev. Lett.}
{\bf 69} (1992) 2619; {\it Phys.  Rev.} {\bf D49} (1994) 4842;
{\it Phys. Lett.} {\bf B329} (1994) 278;\\
A. Dobado and J.R. Pelaez, {\it Phys. Lett.} {\bf B329} (1994) 469
(Addendum, {\it ibid}, {\bf B335} (1994) 554);
{\it Nucl. Phys.} {\bf B425} (1994) 110; \\
H.-J. He, Y.-P. Kuang, and C.--P. Yuan, VPI-IHEP-94-04, MSUHEP-40909,
1994.

\bibitem{veltman}
M. Veltman, {\it Acta. Phys. Pol.} {\bf B8} (1977) 475;\\
M. Einhorn, J. Wudka, {\it Phys. Rev.} {\bf D39} (1989) 2758; \\
M. Veltman, UM-TH-94-28, June 1994,  Lectures given at
{\it 34th Cracow School of Theoretical Physics},
Zakopane, Poland, 31 May - 10 June 1994.

\bibitem{atlas}
ATLAS Letter of Intent, CERN/LHCC/92-4, October 1992.
ATLAS Internal Note, A.Mekki and L.Fayard, PHYS-NO-028, 1993.
ATLAS Internal Note, Nicholas Andell, PHYS-NO-037, 1994.

\bibitem{cms}
CMS Letter of Intent, CERN/LHCC/92-3, October 1992.

\bibitem{chip}
Raymond Brock; private communication.

\bibitem{steve}
S. Mrenna and C.--P. Yuan, {\it Phys.~Rev.} {\bf D46} (1992) 1007.

\bibitem{sdc}
SDC technical Design Report, preprint SDC-92-201, 1992.

\bibitem{effw}
For a review, see, \\
 S. Cortese and R. Petronzio,  {\it Phys. Lett.} {\bf B276} (1992) 203.

\bibitem{databook}
Review of Particle Properties, {\it Phys.~Rev.}
 {\bf D50}, Aug. 1994, pp. 1315-1317.

\bibitem{toppol}
G. L. Kane, G. A. Ladinsky and C.--P. Yuan,
{\it Phys. Rev.} {\bf D45} (1992) 124.

\bibitem{qcd}
J.~Liu and Y.--P. Yao, University of Michigan preprint UM-TH-90-09;  \\
C.S. Li, R.J. Oakes and T.C. Yuan, {\it Phys. Rev.} {\bf D43} (1991) 3759; \\
M. Jezabek and J.H. Kuhn, {\it Nucl.~Phys.} {\bf B314} (1989) 1;
{\it ibid} {\bf B320} (1989) 20.

\bibitem{elecw}
A. Denner and T. Sack, {\it Nucl. Phys.} {\bf B358} 46 (1991);
{\it Z. Phys.} {\bf C46}, 653 (1990); \\
G. Eilam, R. R. Mendel, R. Migneron,
and A. Soni, {\it Phys. Rev. Lett.} {\bf 66} (1991) 3105; \\
B. A. Irwin, B. Margolis, and H. D. Trottier,
{\it Phys. Lett.} {\bf B256}, 533 (1991); \\
J. Liu and Y.--P. Yao, {\it Int.J.Mod.Phys.} {\bf A6} (1991) 4925; \\
T.C. Yuan and C.--P. Yuan, {\it Phys. Rev.} {\bf D44} (1991) 3603.


\bibitem{dimens}
A. Manohar and H. Georgi, {\it Nucl. Phys.} {\bf B234} (1984) 189.

\bibitem{danf1}
A similar analysis was also performed by D. Amidei and D. Winn; private
 communication.

\bibitem{lepjet}
This is the kind of analysis that has curently been performed at LEP for
seperating a quark jet from a gluon jet.

\bibitem{hatt}
D. Chang, W.-Y. Keung, I. Phillips, {\it Nucl. Phys.} {\bf B408} (1993) 286.

\bibitem{wolf}
L. Wolfenstein, {\it Phys. Rev.} {\bf D46} (1992) 256.

\bibitem{cpcpy}
C.--P. Yuan, preprint MSUHEP-41009, November 1994.

\bibitem{onetcp}
B. Grzadkowski and J.F. Gunion, {\it Phys. Lett.} {\bf B287} (1992)
237.

\bibitem{eeonetop}
R. Kauffman, {\it Phys. Rev.} {\bf D41} (1990) 3343.

\bibitem{stopcp}
B. Grzadkowski and W.-Y. Keung, {\it Phys. Lett.} {\bf B319} (1993) 526; \\
M. Nowakowski and A. Pilaftsis, {\it Mod. Phys. Lett.} {\bf A4} (1989)
   821; {\it Z. Phys.} {\bf C42} (1989) 449; \\
A. Pilaftsis, {\it Z. Phys.} {\bf C47} (1990) 95; \\
M. Nowakowski and A. Pilaftsis, {\it Mod. Phys. Lett.} {\bf A6}
   (1991) 1933; {\it Phys. Lett.} {\bf B245} (1990) 185.

\end{thebibliography}
\end{document}